\documentclass[%
reprint,
twocolumn,
superscriptaddress,
showpacs,preprintnumbers,
nofootinbib,
aps,
prd,
]{revtex4-2}

\usepackage[compat=1.1.0]{tikz-feynman} 
\usetikzlibrary{decorations.pathmorphing}
\usetikzlibrary{calc, intersections}

\newcommand{\sunset}[2]{
	\vertex (a) at (0,0);
	\vertex (b) at (1.5,0);
	\vertex (d) at (0.75,-0.75);
	\diagram*{
		(a) -- [#1] (b)
	};
	\draw[#2] (d) arc [start angle=-90, end angle=270, radius=0.75cm];
}

\newcommand{\daisydiagram}{
\begin{tikzpicture}[baseline=(current bounding box.center),scale=0.5, transform shape]
\node[circle,draw,minimum size=1cm] (center) at (0,0) {};
\foreach \angle in {0,72,...,288} {
\draw[double] (\angle:1cm) circle (0.48cm);
}
\end{tikzpicture}
}

\newcommand{\diagramA}{
\begin{tikzpicture}[baseline=(current bounding box.center)]
\draw (0,0) -- (1,0);
\end{tikzpicture}
}

\newcommand{\diagramAA}{
\begin{tikzpicture}[baseline=(current bounding box.center)]
\draw[dashed] (0,0) -- (1,0);
\end{tikzpicture}
}

\newcommand{\diagramB}{
\begin{tikzpicture}[baseline=(current bounding box.center)]
\draw (0,0) -- (1,0);
\draw[double] (0.5,0.25) circle (0.25cm);
\end{tikzpicture}
}

\newcommand{\diagramC}{
\begin{tikzpicture}[baseline=(current bounding box.center)]
\draw (0,0) -- (1,0);
\draw (0.5,0.25) circle (0.25cm);
\end{tikzpicture}
}

\newcommand{\diagramD}{
\begin{tikzpicture}[baseline=(current bounding box.center)]
\draw (0,0) -- (1,0);
\draw[double] (0.5,0.25) circle (0.25cm);
\draw[double] (0.5,0.75) circle (0.25cm);
\end{tikzpicture}
}

\newcommand{\diagramE}{
\begin{tikzpicture}[baseline=(current bounding box.center)]
\draw (0,0) -- (1,0);
\draw[double] (0.5,0) circle (0.25cm);
\end{tikzpicture}
}

\newcommand{\diagramX}{
\begin{tikzpicture}[baseline=(current bounding box.center)]
\draw[dashed] (0.5,0.25) circle (0.25cm);
\end{tikzpicture}
}

\newcommand{\diagramY}{
\begin{tikzpicture}[baseline=(current bounding box.center)]
\draw[dashed] (0.5,0.25) circle (0.25cm);
\draw[dashed] (0.5,0.75) circle (0.25cm);
\end{tikzpicture}
}

\newcommand{\diagramZ}{
\begin{tikzpicture}[baseline=(current bounding box.center)]
\draw[dashed] (0.25,0.25) -- (0.75,0.25);
\draw[dashed] (0.5,0.25) circle (0.25cm);
\end{tikzpicture}
}

\newcommand{\diagramXX}{
\begin{tikzpicture}[baseline=(current bounding box.center)]
\draw (0,0) -- (0.25,0.25);
\draw (0,0.5) -- (0.25,0.25);
\draw (1.0,0) -- (0.75,0.25);
\draw (1.0,0.5) -- (0.75,0.25);
\draw[double] (0.5,0.25) circle (0.25cm);
\end{tikzpicture}
}

\newcommand{\diagramYY}{
\begin{tikzpicture}[baseline=(current bounding box.center)]
\def\radius{0.25cm}
\draw[double] (0,0) circle[radius=\radius];
\draw (0.125, 0.2165) -- (0.425, 0.39);
\draw (0.125, 0.2165) -- (0.125, 0.5825);
\draw (-0.25, 0) -- (-0.525, 0.1665);
\draw (-0.25, 0) -- (-0.525, -0.1665);
\draw (0.125, -0.2165) -- (0.425, -0.39);
\draw (0.125, -0.2165) -- (0.125, -0.5825);
\end{tikzpicture}
}

\newcommand{\diagramZZ}{
\begin{tikzpicture}[baseline=(current bounding box.center)]
\draw[dashed] (0,0) -- (1,0);
\draw (0.5,0) circle (0.25cm);
\draw[dashed] (0.5,0.25)  -- (0.25,0.5);
\draw[dashed] (0.5,0.25)  -- (0.75,0.5);
\end{tikzpicture}
}

\usepackage{amssymb}
\usepackage{amsfonts}
\usepackage{bbm,bm}
\usepackage[breaklinks,urlbordercolor={1 1 1}]{hyperref}
\usepackage[utf8]{inputenc}
\usepackage{graphicx}

\graphicspath{ {./figures-final/} }

\usepackage{comment}
\usepackage{amsmath}
\usepackage{subcaption}
\usepackage{ragged2e}

\makeatletter
\renewcommand{\@makecaption}[2]{
	\vskip\abovecaptionskip
	\sbox\@tempboxa{#1: #2}%
	\ifdim \wd\@tempboxa >\hsize
	\RaggedRight{#1:} #2\par
	\else
	\global \@minipagefalse
	\hb@xt@\hsize{\hfil\box\@tempboxa\hfil}%
	\fi
	\vskip\belowcaptionskip}
\makeatother

\usepackage{enumerate}
\usepackage{multirow}
\usepackage{ulem} 

\newcommand\MSbar{$\overline{\text{MS}}$ } 
\newcommand\Veff{V_\text{eff}} 
\newcommand\he[1]{#1^\dagger}
\newcommand\gr[1]{\mathrm{#1}}
\newcommand\SU[1]{\mathrm{\gr{SU(#1)}}} 
\newcommand\grU[1]{\mathrm{\gr{U(#1)}}} 
\newcommand\sumint[1]{\int\kern-1.5em\sum\nolimits_{#1}}

\newcommand{\nn}{\nonumber \\}

\newcommand{\Tc}{T_{\rm c}}
\newcommand{\vphys}{v_\mathrm{phys}}

\newcommand{\der}[2][]{\frac{d#1}{d#2}}
\newcommand{\pder}[2][]{\frac{\partial#1}{\partial#2}} 

\newcommand{\DRALGO}{{\tt DRalgo} }

\usepackage{xcolor}

\begin{document}

\newcommand{\HEL}{\affiliation{%
Department of Physics and Helsinki Institute of Physics,
PL 64, FI-00014 University of Helsinki,
Finland }}

\newcommand{\NOR}{\affiliation{
Nordita,
KTH Royal Institute of Technology and Stockholm University,\\
Hannes Alfv\'ens v\"ag 12,
SE-106 91 Stockholm,
Sweden }}

\newcommand{\TSU}{\affiliation{Tsung-Dao Lee Institute \& School of Physics and Astronomy, Shanghai Jiao Tong University, Shanghai 200240, China }}

\newcommand{\SHA}{\affiliation{Shanghai Key Laboratory for Particle Physics and Cosmology, Key Laboratory for Particle Astrophysics and Cosmology (MOE), Shanghai Jiao Tong University, Shanghai 200240, China}}

\title{Investigating two-loop effects for first-order electroweak phase transitions}

\preprint{HIP-2024-10/TH}

\author{Lauri~Niemi} \email{lauri.b.niemi@helsinki.fi} \HEL \TSU

\author{Tuomas V. I.~Tenkanen} \email{tuomas.tenkanen@helsinki.fi} \HEL \NOR \TSU \SHA

\begin{abstract}

We study first-order electroweak phase transitions in the real-singlet extended Standard Model, for which non-zero mixing between the Higgs and the singlet can efficiently strengthen the transitions. We perform large-scale parameter space scans of the model using two-loop effective potential at next-to-next-to leading order in the high-temperature expansion, greatly improving description of phase transition thermodynamics over existing one-loop studies.
We find that
1) two-loop corrections to the effective potential lead to narrower regions of strong first-order transitions and significantly smaller critical temperatures,
2) transitions involving a discontinuity in the singlet expectation value are significantly stronger at two-loop order,
3) high-temperature expansion is accurate for a wide range of parameter space that allows strong transitions, 
although it is less reliable for the very strongest transitions. 
These findings suggest revisiting past studies that connect the possibility of a first-order electroweak phase transition with future collider phenomenology.

\end{abstract}

\maketitle

\section{Introduction}

Recent years have a seen an influx of articles discussing the cosmological electroweak phase transition (EWPT) in various extensions of the Standard Model (SM). The SM only has a smooth crossover transition \cite{Kajantie:1996mn,Csikor:1998eu}, while with new field content the transition can be first order, generating a stochastic background of gravitational waves \cite{Hogan:1986dsh, Kamionkowski:1993fg, Caprini:2019egz,LISACosmologyWorkingGroup:2022jok} with prospects for observation in next-generation experiments \cite{LISA:2017pwj, TianQin:2015yph, Ruan:2018tsw}, and resurrecting the possibility that matter-antimatter asymmetry could be generated by electroweak baryogenesis \cite{Kuzmin:1985mm, Morrissey:2012db,Bodeker:2020ghk}. 

A central tool in analytical modeling of the EWPT is the thermal effective potential for the Higgs field and possible beyond the SM (BSM) scalars, used to study the general phase structure of a given theory and to obtain the free energy from which thermodynamic information can be extracted. Among others, the effective potential enters in computations of the critical temperature, thermal bubble nucleation rate \cite{Enqvist:1991xw,Ignatius:1993qn,Moore:2000jw,Moore:2001vf,Gould:2021ccf}, transition ``strength'' or the amount of energy released \cite{Espinosa:2010hh,Giese:2020rtr,Giese:2020znk}, and the speed at which nucleating bubbles expand in the hot early universe \cite{Bodeker:2009qy,Kozaczuk:2015owa,Bodeker:2017cim, Hoche:2020ysm,Azatov:2020ufh, Gouttenoire:2021kjv,Laurent:2022jrs}. One may also extract various more theoretical, possibly renormalization-dependent quantities from the effective potential, such as the Higgs condensate $\langle \he\phi\phi \rangle$ that is related to the rate of anomalous baryon number violation in the Higgs phase \cite{DOnofrio:2014rug,Annala:2023jvr}.

Much effort has been put into understanding subtle aspects of the thermal effective potential. These include the need to resum certain thermal corrections \cite{Carrington:1991hz,Arnold:1992rz}, residual dependence on gauge-fixing details \cite{Buchmuller:1994vy,Laine:1994zq,Bodeker:1996pc,Patel:2011th} and the generally slow convergence of perturbation theory near the EWPT critical temperature \cite{Farakos:1994kx,Fodor:1994bs,Kajantie:1995kf,Laine:2017hdk,Gould:2021oba,Gould:2022ran}. The fact that loop corrections are relatively more important at finite temperature than in the vacuum is hardly surprising, given that thermal corrections to field theory only manifest themselves at loop level in the first place. Understanding and accounting for these complications is crucial for reliable modeling of phase transitions and related cosmological phenomena in realistic particle physics models.

In this article, our focus is on the real-singlet extended SM (sometimes called ``xSM'') which adds a new scalar excitation to the SM.  The EWPT in this model has been studied for example in Refs.~\cite{Profumo:2007wc,Kehayias:2009tn,Espinosa:2011ax,Curtin:2014jma,Profumo:2014opa,Kozaczuk:2015owa,Vaskonen:2016yiu,Beniwal:2017eik,Chen:2017qcz,Kurup:2017dzf,Chiang:2018gsn,Kozaczuk:2019pet}. These previous analyses include thermal effects at leading orders, ie. utilize one-loop thermal effective potential with so-called daisy resummation \cite{Carrington:1991hz, Arnold:1992rz}. Computation of the effective potential in a high-temperature approximation to two-loop order was carried out in our previous work~\cite{Niemi:2021qvp}. See \cite{Farakos:1994kx,Bodeker:1996pc,Niemi:2020hto,Croon:2020cgk} for analogous calculations in other models, and the \DRALGO package \cite{Ekstedt:2022bff} for automated computations in generic models.

The calculation of our effective potential is split in two parts: (i) incorporation of thermal corrections from high-momentum modes to a low-energy effective description, and (ii) calculation of remaining loops in resummed perturbation theory. Resummations of thermal loops are automatic in this approach due to step (i), often referred to as ``dimensional reduction'' because the imaginary Euclidean time direction decouples from the resulting low-energy effective theory (EFT) \cite{Ginsparg:1980ef, Appelquist:1981vg}. The method systematically incorporates thermal resummations beyond the leading daisy diagrams but is only valid at high temperatures; specifically, it assumes that mass-to-temperature ratios are small for all excitations relevant for the phase transition. Because the EWPT is associated with a Higgs mechanism, ratios like $m/T$ typically grow with the Higgs condensate $\langle \he\phi\phi\rangle / T^2$, ie. as one moves deeper in the Higgs phase of the theory. This can pose a problem for high-temperature approximations when describing very strong, possibly supercooled, phase transitions where particle masses before and after the EWPT can be vastly different.

The purpose of this paper is twofold. First, we perform large-scale scans over singlet-extended SM parameter space and compare predictions of EWPT thermodynamics at one- and two-loop orders. This allows us to make generic statements about the validity of existing one-loop analyses \cite{Profumo:2007wc,Espinosa:2011ax,Curtin:2014jma,Profumo:2014opa,Kozaczuk:2015owa,Vaskonen:2016yiu,Beniwal:2017eik,Chen:2017qcz,Kurup:2017dzf,Chiang:2018gsn,Kozaczuk:2019pet} and identify regions of the parameter space where two-loop corrections to the potential become important. A similar analysis recently appeared in \cite{Lewicki:2024xan} that focused on a narrow part of the $Z_2$ symmetric limit of the singlet model. Here we study the more general model without $Z_2$ symmetry in which the physical scalars can mix.

Second, we assess the accuracy of the high-temperature EFT approach by including a set of higher-dimensional thermal operators to the EFT and estimating their effect on the EWPT. These operators correspond to subleading terms in high-$T$ expansions \cite{Bodeker:1996pc, Laine:2000kv} and are often neglected in studies based on dimensional reduction. For instance, a recent lattice analysis \cite{Niemi:2024axp} of the singlet-extended model utilized the high-$T$ EFT to nonperturbatively study the problematic IR physics of the EWPT, and the applicability of these results for model phenomenology hinges on validity of the EFT constructed in \cite{Niemi:2021qvp}. 

Our scans suggest that
\begin{itemize}
	
	\item  One-loop approximations vastly overestimate the critical temperature, and the two-loop potential generically predicts narrower regions of strong first-order EWPT compared to the conventional one-loop potential. However, in the regions that survive at two-loop, transitions tend to be much stronger than at one-loop, characterized by larger latent heat and Higgs condensate discontinuity.
	
	\item Many of the strongest transitions feature a significant jump also in the singlet vacuum expectation value (VEV). These transitions appear near the boundary where the electroweak vacuum stops being the global energy minimum at zero temperature, ie. another vacuum in the singlet direction becomes favored.
	
	\item We find plenty of moderate and strong transitions in regions where the dimensionally-reduced EFT is accurate at the few-percent level and high-temperature expansions are under control. However, we also observe clear breakdown of high-$T$ approximations particularly in parameter space where the running couplings are close to being nonperturbatively large. This includes some of the strongest EWPT found in the scans, suggesting those points should not be trusted as physically viable benchmarks of a strong cosmological phase transition.

\end{itemize}

The remainder of this article is organized as follows: In section \ref{sec:model} we introduce the singlet-extended model at zero temperature and our parametrization for it.  Section \ref{sec:EFT} summarizes the high-temperature EFT for the model, and in section \ref{sec:limitations} we further discuss limitations of this approach. Section \ref{sec:Veff} reviews the structure of the thermal effective potential. In section \ref{sec:IR-div} we discuss subtleties related to non-analytic behavior of the effective potential, in particular, unphysical IR divergences that one encounters at two-loop level. Details of our numerical scanning routine are given in \ref{sec:analysis}. Section \ref{sec:conclusions} is a summary of our conclusions and outlook. The following appendices are included: Appendix \ref{sec:anatomy} reviews the generic structure of thermal effective potentials, appendix \ref{sec:impact} is dedicated to a technical exploration of two-loop contributions in a simplified setting, and appendix \ref{sec:more-plots} collects additional plots for our error estimators in several parameter planes.

\section{Model and parametrization}
\label{sec:model}

The most general renormalizable scalar potential for singlet-extended SM is 
\begin{align}
\label{eq:scalar-pot}
V(\phi, S) =& m^2_\phi \he\phi\phi + \lambda (\he\phi\phi)^2 + b_1 S + \frac12 m^2_S S^2 \nonumber \\
& + \frac13 b_3 S^3 + \frac14 b_4 S^4  + \frac12 a_1 S \he\phi\phi + \frac12 a_2 S^2 \he\phi\phi.
\end{align}
Here $\phi$ is the electroweak Higgs doublet and $S$ is a gauge singlet. We write, in unitary gauge, 
\begin{align}
\phi = \frac{1}{\sqrt{2}} \begin{pmatrix} 
0 \\ v_0 + h 
\end{pmatrix}
\end{align}
where $v_0$ is the Higgs VEV.\footnote{The subscript in $v_0$ is used to distinguish between the $T=0$ VEV and the background field $v$ introduced later for the effective potential.}
As usual, we require that there is Higgs mechanism, ie. the classical potential has its minimum at nonzero VEV, $v_0\neq 0$. Furthermore, the singlet field can be freely shifted without affecting physics, allowing us to remove one free parameter from the theory \cite{Profumo:2007wc,Espinosa:2011ax}. We choose a parametrization in which the singlet VEV vanishes in the EW minimum, fixing $b_1 = -a_1 v_0^2 / 4$ at tree level.

In general the mass matrix of $h, S$ is non-diagonal and the scalars mix. We denote the mass eigenstates by $h_1, h_2$ and the mixing angle as $\theta$:
\begin{align}
\begin{pmatrix} 
h_1 \\ h_2
\end{pmatrix}
= 
\begin{pmatrix} 
\cos\theta & -\sin\theta \\
\sin\theta & \cos\theta
\end{pmatrix}
\begin{pmatrix} 
h \\ S
\end{pmatrix}.
\end{align}
Masses of $h_1, h_2$ (denoted $m_1, m_2$) along with $\sin\theta$ appear frequently in phenomenological studies, see eg. \cite{Profumo:2007wc,Profumo:2014opa,Robens:2016xkb,Kozaczuk:2019pet}.
It is useful to express parameters in the potential \eqref{eq:scalar-pot} in terms of these and the Higgs VEV, giving at tree level 
\begin{align}
\label{eq:params-tree1}
m^2_\phi &= -\frac14 \left( m^2_1 + m^2_2 - (m^2_2 - m^2_1)\cos2\theta \right)  \\
m^2_S &= \frac12 \left( m^2_1 + m^2_2 + (m^2_2 - m^2_1)\cos2\theta - a_2 v_0^2 \right) \\
b_1 &= -\frac14 v_0 (m^2_2 - m^2_1) \sin2\theta \\
a_1 &= v_0^{-1} (m^2_2 - m^2_1) \sin2\theta \\
\label{eq:params-tree2}
\lambda &= \frac14 v_0^{-2} \left( m^2_1 + m^2_2 - (m^2_2 - m^2_2)\cos2\theta \right).
\end{align}
The remaining parameters $a_2, b_3, b_4$ are treated as input parameters. The range of $\theta$ is, without loss of generality, $(-\frac{\pi}{4}, \frac{\pi}{4})$ \cite{Profumo:2007wc}. For small $|\sin\theta| \ll 1$, the scalar state with larger couplings to gauge fields and fermions is $h_1$, which is identified with the observed Higgs boson.

However, these relations obtain quantum corrections, and at one-loop the effects are parametrically of same order as two-loop corrections to the effective potential at high temperature \cite{Niemi:2021qvp}. Therefore we replace Eqs~\eqref{eq:params-tree1} - \eqref{eq:params-tree2} by their loop corrected counterparts according to the renormalization procedure in appendix A of \cite{Niemi:2021qvp} (equations (A29) through (A35) in the reference). The main feature here is matching experimentally measured masses of EW particles to one-loop pole masses in the \MSbar scheme.
In the scalar sector we treat $a_2, b_3, b_4$ as direct inputs at \MSbar scale $\bar\mu = M_Z = 91.1876$ GeV and the input $\theta$ is defined to diagonalize the mass matrix at tree-level. For masses we input the pole masses $M_1, M_2$ of $h_1, h_2$. In the following we fix $M_1 = 125.1$ GeV, and $M_2$ will denote the mass of the new scalar excitation. For further details of our renormalization scheme see Ref.~\cite{Niemi:2021qvp}.

\section{High-$T$ effective description}
\label{sec:EFT}

For calculating 
equilibrium thermodynamics of the model
we turn to the imaginary time formalism. The generating functional (canonical partition function) at temperature $T$ is 
\begin{align}
\label{eq:partition-function}
Z = \int D\phi \; \exp\Big( -\int\limits_0^{1/T} d\tau \int d^3\mathbf{r} \; \mathcal{L}_E \Big),
\end{align}
where $\mathcal{L}_E$ is the 
model
Lagrangian in Euclidean 
space,
$\tau$ is the imaginary time and $D\phi$ denotes collective functional integration over all fields. The Euclidean momentum is $P = (\omega_n, \mathbf{p})$ where the Matsubara frequency is $\omega_n = 2\pi n T$ for bosons and $\omega_n = (2n+1)\pi T$ for fermions.

It is well known that finite-temperature perturbation theory requires resummation of 
certain 
thermal corrections \cite{Arnold:1992rz}. 
The underlying reason is a scale hierarchy between bosonic zero-Matsubara modes ($\omega_n = 0$) and other modes which always carry momentum $P^2 \geq (\pi T)^2$, 
ie. the ``hard'' modes with $\omega_n \neq 0$. 
For light fields with $m \lesssim T$, the zero mode propagator gets a correction of order $g^2 T^2$ from hard momenta ($\omega_n \neq 0$) loops, $g^2$ denoting a generic quartic coupling. At weak coupling this loop correction can be of same order of magnitude or larger as the tree-level mass, suggesting that a perturbative expansion around the free theory is not justified. The solution is to perform resummation for zero modes, absorbing the thermal mass corrections into a new ``tree level'' mass, frequently called the thermal mass.

Of course, implementing this resummation in a consistent manner while avoiding double counting of thermal loops and other spurious effects is a nontrivial task, especially beyond the leading one-loop order. Much has been said about this in the literature \cite{Carrington:1991hz,Arnold:1992rz,Laine:2017hdk}, and a self-consistent method for higher-order calculations was developed in \cite{Arnold:1992rz,Farakos:1994kx,Braaten:1995cm,Kajantie:1995dw}: one constructs a low-energy effective theory (EFT) for the zero Matsubara modes.
That an EFT description solves the problem of resummation should not come as a surprise, given that the scale hierarchy between zero- and non-zero modes is why resummation is needed in the first place. 

The process can be compactly described as follows. A formal integration over momentum modes with $\omega_n \geq \pi T$ in the path integral (\ref{eq:partition-function}) gives
\begin{align}
\label{eq:partition-EFT}
Z = \int_{\omega_n = 0} D\phi \; \exp\Big(\frac{1}{T} \int d^3 \mathbf{r} \; \mathcal{L}_\mathrm{3D} + \text{const.} \Big),
\end{align}
where $\mathcal{L}_\mathrm{3D}$ is an effective Lagrangian containing only fields that carry no momentum in the imaginary time direction. Consequently, the EFT is fully bosonic and effectively three dimensional. This dimensional reduction is a generic feature of high-temperature field theory \cite{Appelquist:1981vg,Ginsparg:1980ef}. Parameters appearing in $\mathcal{L}_\mathrm{3D}$ differ from those in the original Lagrangian by modifications from high-momentum loops, precisely the corrections needed for resummation. The ($T$-dependent) constant is irrelevant for phase transition quantities that depend on free energy difference.\footnote{The $T$-dependent constant must be included if one is interested in hydrodynamical quantities such as the speed of sound \cite{Giese:2020rtr,Giese:2020znk,Tenkanen:2022tly}.} 

In practice the EFT is most conveniently constructed by matching low-momentum Green's functions to a desired order in perturbation theory \cite{Kajantie:1995dw}. For SM + singlet this has been performed in \cite{Brauner:2016fla,Schicho:2021gca,Niemi:2021qvp}. The resulting EFT has action\footnote{This EFT includes additional (small) corrections from integrating out gauge field temporal components that are screened in the plasma and obtain a Debye mass of order $gT$. See \cite{Niemi:2021qvp} for details.}
\begin{align}
\label{eq:EFT-action}
S_\mathrm{3D} = \; \frac{1}{T} \int d^3 \mathbf{r} \; \Big\{ \frac14 F^a_{ij} F^a_{ij} + &\frac14 B_{ij} B_{ij} + |D_i \phi|^2 \nn 
& + \frac12 (\partial_i S)^2 + \bar{V}(\phi, S) \Big\}.
\end{align}
Here $F^a_{ij}$ and $B_{ij}$ are 3D field strengths for $\SU2$ and $\grU1$ gauge fields, respectively. The scalar potential is as in Eq.~(\ref{eq:scalar-pot}) but with resummed masses and couplings:
\begin{widetext}
\begin{align}
\label{eq:potential-3d}
\bar{V}(\phi, S)  = \bar{m}^2_\phi(T) \he\phi\phi + \bar{\lambda}(T) (\he\phi\phi)^2 + \bar{b}_1(T) S + \frac12 \bar{m}^2_S(T) S^2 + \frac13 \bar{b}_3(T) S^3 + \frac14 \bar{b}_4(T) S^4  + \frac12 \bar{a}_1(T) S \he\phi\phi + \frac12 \bar{a}_2(T) S^2 \he\phi\phi.
\end{align}
\end{widetext}
In particular, masses in the potential $\bar{V}$ differ from those in the vacuum potential \eqref{eq:scalar-pot} by two-loop thermal corrections, corresponding to perturbative order $\mathcal{O}(g^4)$ (more generally, second order in quartic couplings).
Expression for the parameters can be read from sections V and VI in Ref.~\cite{Niemi:2021qvp}. Unlike in the reference \cite{Niemi:2021qvp}, we have kept the overall $1/T$ factor explicit for the sake of presentation; with this convention all fields and couplings have same mass dimensions as in the original 4D theory.

\section{Limitations of the high-$T$ approach}
\label{sec:limitations}

The effective theory described above is accurate only when the scale hierarchy between zero and non-zero modes is manifest.
For EWPT applications this means that the relevant phase transition dynamics should occur at length scales $\gg (\pi T)^{-1}$, which is a reasonable expectation for transitions driven by thermal fluctuations (see eg. \cite{Gould:2021ccf} for discussion of this point). Integration over the heavy Matsubara modes produces higher-dimensional operators that are suppressed by appropriate powers of $\pi T$: examples include $(\he\phi\phi)^2 S/T$ and $(\he\phi\phi)^3/T^2$  at field dimension five and six, respectively. Effects of these operators are suppressed and dimensional reduction is valid when only light masses appear in the EFT, ie. $m^2 \ll (\pi T)^2$. Note that this coincides with the condition that high-$T$ expansion of thermal sum-integrals is valid \cite{Laine:2000kv,Laine:2017hdk}.

It is commonplace to neglect higher-dimensional operators and truncate the EFT at field dimension four
as we have done in (\ref{eq:EFT-action}). In many cases this truncation is perfectly reasonable as coefficients of these operators are perturbatively suppressed and, in a formal power counting sense, are same order of magnitude as three-loop corrections to the thermal masses \cite{Ghisoiu:2015uza}.\footnote{See however the discussion in \cite{Niemi:2021qvp} regarding dimension five operators and cubic interactions in the singlet extension.} 
Hence, operators beyond dimension four are unnecessary for calculations aiming at two-loop precision. As an aside, we note that high-$T$ EFTs also provide a useful starting point for nonperturbative lattice studies of thermal phase transitions, and in this context the truncation of higher-dimensional operators may be a practical necessity \cite{Niemi:2024axp}.

This discussion is complicated by the fact that particle masses are different before and after the EWPT. The EFT needs to accurately describe not only the symmetric phase at $\phi = 0$, but also the broken phase at least down to temperatures relevant for bubble nucleation dynamics.\footnote{We use the terms ``symmetric'' and ``broken'' phase sloppily to label regimes of the theory before and after the EWPT. In reality there is only one thermodynamical phase with a possible first-order transition somewhere in the phase diagram \cite{Fradkin:1978dv, Kajantie:1996mn}, ie. the phase diagram is continuously connected and no symmetry breaking occurs in the strict sense \cite{Elitzur:1975im}.}
In the broken phase, Higgs mechanism generates field-dependent masses of order $m(v) \sim \mathrm{(coupling)} \times v$, where $v$ denotes the Higgs VEV in some gauge, and ratios of type $m(v) / (\pi T)$ are not necessarily small for large field values. For $v \gtrsim \pi T$ the suppression of higher-dimensional operators is lifted and the accuracy of our truncated EFT (\ref{eq:EFT-action}) becomes questionable. This poses a potential problem for studying strong phase transitions that are characterized by $v \gtrsim T$ after the EWPT.
Similar concerns hold for the singlet field whose VEV modifies masses of scalar excitations. 

A simple way of estimating the validity of the high-$T$ EFT is to explicitly include higher-dimensional operators in the calculations and then compare results to those obtained from the truncated theory.
For EWPT the dominant operators are those that modify the effective Higgs potential directly at tree level, ie. pure scalar operators without derivatives. At dimensions five and six these are 
\begin{widetext}
\begin{align}
\label{eq:dim6}
S_\mathrm{3D}^{5,6} \supset  \; \frac{1}{T} \int d^3\mathbf{r} \; \Big\{ c_{0,5} S^5 + c_{2,3} \he\phi\phi S^3 + c_{4,1} (\he\phi\phi)^2 S + c_{6,0} (\he\phi\phi)^3 + c_{0,6} S^6 + c_{4,2} (\he\phi\phi)^2 S^2 + c_{2,4} \he\phi\phi S^4 \Big\}.
\end{align}
\end{widetext}
Matching relations for the Wilson coefficients are given in Eqs (43) - (49) of \cite{Niemi:2021qvp} (but note the difference in field scaling). For simplicity we only include contributions from scalar and top quark loops as corrections from gauge loops are subdominant \cite{Kajantie:1995dw}.

Our parameter-space scans and analysis of the EWPT will be performed without these operators included in the EFT. Their relevance for our results will be estimated in section~\ref{sec:quantities}.

\section{Thermal effective potential}
\label{sec:Veff}

Using the EFT (\ref{eq:partition-EFT}), it is now straightforward to calculate the resummed effective potential for $S$ and the Higgs. For this, we shift 
\begin{align}
\label{eq:field-shifts}
\phi \rightarrow \phi +  \frac{1}{\sqrt{2}} \begin{pmatrix} 0 \\ v \end{pmatrix} \quad \text{ and } \quad S \rightarrow S + x,
\end{align} 
where $v$ and $x$ are constant background fields (zero-momentum modes). The effective potential is, formally, 
\begin{align}
\label{eq:Veff-def}
\Veff(v,x) = - \frac{T}{V} \ln \int\limits_{P \neq 0} D\phi \; \exp\Big( -\frac{1}{T} \int d^3\mathbf{r} \; \mathcal{L}'_\mathrm{3D} \Big)
\end{align}
where the integration is restricted to field modes with non-zero momentum and $\mathcal{L}'_\mathrm{3D}$ is the EFT Lagrangian after field shifts (\ref{eq:field-shifts}). Infinite volume limit $V\rightarrow \infty$ is to be taken at the end. 

The two-loop expression for $\Veff(v, x)$ can be read from appendix~B of \cite{Niemi:2021qvp}; for a simplified and pedagogical review of the structure of $\Veff$ see Appendix~\ref{sec:anatomy} of the present paper.
In terms of quartic couplings ($g^2, g'^2, \lambda, a_2, b_4$) this means $\mathcal{O}(g^4)$ accuracy and next-to-next-to-leading (NNLO) order accuracy in the high-$T$ expansion, $(m/T)^4$. However, given that the overwhelming majority of EWPT studies are based on the one-loop approximation to $\Veff$, it will be interesting to explore how our two-loop results differ from those obtained with the one-loop potential. Our ``one-loop'' refers to an $\mathcal{O}(g^3)$ calculation at next-to-leading order in the high-$T$ expansion. In other words, our one-loop accuracy corresponds to the familiar one-loop high-$T$ effective potential with daisy resummation using one-loop thermal masses. The high-$T$ expansion of one-loop $\Veff$ produces terms of type $m^4 \ln (\bar\mu / T)$ at NNLO that are sometimes included in one-loop analyses; here we include these terms only in our two-loop results. For both calculations we use the one-loop corrected matching between \MSbar parameters and physical observables as described in section~\ref{sec:model}. These zero-temperature corrections are formally of order $g^4$ and exceed the accuracy of our one-loop description, but using the same $T=0$ theory for both one- and two-loop analyses allows for easier comparison of thermal effects in the high-$T$ approximation.

The background fields $v,x$ act as effective order parameters for distinguishing phases of the theory, and minimizing $\Veff$ is equivalent to finding the free energy density of the system in a thermodynamically (meta-)stable phase. We therefore wish to know how the local minima evolve with the temperature. The extremal conditions 
\begin{align}
\label{eq:min_cond}
\pder[\Veff]{v} = 0, \quad \pder[\Veff]{x} = 0 \quad \text{at} \quad (v, x) = (v_\text{min}, x_\text{min})
\end{align}
can be solved either with a numerical minimization algorithm, or perturbatively by expanding around leading order minima \cite{Laine:1994zq,Farakos:1994xh,Kajantie:1995kf,Karjalainen:1996rk,Kajantie:1997hn,Patel:2011th} (sometimes called the ``$\hbar$ expansion''). 
In terms of Feynman diagrams the minimization corresponds to appending the $\Veff$ with one-particle-reducible tadpole insertions \cite{Fukuda:1975di}. 
Numerical minimization inserts tadpoles to all diagrams of the effective potential while the order-by-order method inserts them only to a subset of low-order diagrams required for a consistent result in loop-counting sense.

In this paper we will use the former, numerical approach to minimization. Consequently, our method necessarily incorporates some corrections that are higher-order in loop and power counting sense, some of which can contain dependence on unphysical parameters such as the gauge-fixing parameter $\xi$ and the renormalization scale $\bar{\mu}$. 
In more technical terms and using the power counting from above, we obtain correct results for physical quantities at $\mathcal{O}(g^4)$ (which are of course independent of gauge and RG artefacts), but we also obtain partial $\mathcal{O}(g^5)$ and $\mathcal{O}(g^6)$ corrections whose gauge- and RG-dependent parts do not necessarily cancel out, and in the direct minimization method it is not possible to disentangle these from the $\mathcal{O}(g^4)$ part. In practice this is not a big issue as long as the residual corrections are small compared to the $\mathcal{O}(g^4)$ results. Indeed, our choice of Landau gauge ($\xi = 0$) is known to be good for minimizing residual contributions from gauge loops \cite{Chiang:2018gsn,Schicho:2022wty}.

Minimizing higher-order logarithms by an optimal choice of $\bar{\mu}$ is the goal of RG improvement programs for the effective potential and is not limited to the direct minimization method. RG improvements in high-$T$ contexts have been discussed in \cite{Arnold:1992rz,Farakos:1994kx,Gould:2021oba}. We discuss our treatment of RG effects more in section \ref{sec:analysis}; the summary is that $\bar{\mu}$-dependent logarithms from hard-scale thermal loops are automatically optimized through the dimensional reduction step \cite{Farakos:1994kx, Kajantie:1995dw}, and for the remaining loops over Matsubara zero modes we fix $\bar{\mu} = T$. See \cite{Niemi:2021qvp, Lewicki:2024xan} for studies in the singlet model with varying RG scale at two-loop.

In either minimization scheme one has to pay special attention to non-analytic behavior of the potential due to light scalar fields. We devote the next section for discussion of these effects.

\section{Infrared divergences at two-loop}
\label{sec:IR-div}

Loops over the Matsubara zero modes are not well behaved in the IR. Already in the one-loop high-$T$ effective potential, a bosonic field of mass $m$ produces a term proportional to $T (m^2)^{3/2}$ which is non-analytic at $m^2 = 0$. In general the masses depend on background fields that appear in the effective potential, in our case $m^2 = m^2(v,x)$, and for resummed potentials these masses are also temperature-dependent. The non-analytic terms become problematic if, for some background-field configurations, any thermally-corrected (bosonic) mass vanishes. In such points the potential is not smooth, and for $m^2 < 0$ the effective potential develops an imaginary part. 

This IR issue becomes much more severe at two-loop order where an explicit logarithmic divergence occurs \cite{Jackiw:1980kv}.
In the singlet model, the diagrams responsible for IR divergences are scalar ``sunset'' diagrams with three $h_1$ or $h_2$ propagators. The loop integral associated with such diagrams is, here in $d=3-2\epsilon$ dimensions,
\begin{widetext}
\begin{align}
	\label{eq:sunset-diagram}
	\begin{tikzpicture}[scale=0.75, transform shape, baseline={([yshift=-.5ex]current bounding box.center)}]
	\begin{feynman}
	\sunset{plain}{plain}
	\end{feynman}
	\end{tikzpicture} \;
	\propto T^2 H_3(m, m, m) \equiv T^2 \mu^{4\epsilon} \int \frac{d^d p}{(2\pi)^d} \frac{d^d k}{(2\pi)^d} \frac{1}{(p^2 + m^2)(k^2 + m^2)((p+k)^2 + m^2)}
\end{align}
\end{widetext}
where $\mu$ is the scale associated with dimensional regularization (related to the \MSbar scale by $\bar\mu^2 = 4\pi e^{-\gamma_E}\mu^2$), and we show only the problematic 3D (ie. zero-mode) part of the full Matsubara sum.
In the SM, the (dimensionful) cubic interactions needed for these diagrams are generated by the field shift in Eq.~(\ref{eq:field-shifts}). In the general singlet extension without $Z_2$ symmetry, a diagram of this type is present even in the ``symmetric'' phase ($v=0$) due to the $S^3$ term in the action. To reiterate, $m$ here refers to the tree-level mass of a resummed Matsubara zero-mode of $h_1$ or $h_2$ and is generally dependent on the background fields introduced by field shifts.

For $m^2 \neq 0$, the integral can be evaluated as \cite{Arnold:1994ps}
\begin{align}
\label{eq:sunset-integral}
H_3(m,m,m) = \frac{1}{16\pi^2} \Big( \frac{1}{4\epsilon} + \frac12 + \ln\frac{\bar\mu}{3\sqrt{m^2}} \Big) + \mathcal{O}(\epsilon).
\end{align}
The $1/\epsilon$ pole is removed by UV counterterms as usual. However, the logarithm is divergent for $m^2=0$ and renders the perturbative calculation useless for massless scalar fields. To avoid confusion, we note that for $m^2 = 0$ the integral in (\ref{eq:sunset-diagram}) vanishes in dimensional regularization and a finite result for $\Veff$ can be obtained by simply omitting the diagram in this special case. However, quantities such as the latent heat and various condensates (see section \ref{sec:quantities}) that depend on derivatives of the potential are ill defined when crossing the hypersurface in $(v, x, T)$ space that separates $m^2 > 0$ from $m^2 < 0$. As an example, consider the the leading-order high-$T$ potential in the minimal SM, which predicts a second-order phase transition, ie. the tree-level resummed Higgs mass vanishes at the critical temperature. If a perturbative expansion is performed with this mass in the Higgs propator, one encounters an IR divergence at two-loop level precisely at $T = \Tc$, due to the above logarithmic term \cite{Laine:1994zq}. 

That such divergences arise in perturbation theory is not surprising: For Matsubara zero modes the expansion parameter for scalar self-interactions must, on dimensional grounds, be of form $\lambda T / m$, invalidating the use of perturbative expansions in the IR region $m \ll \lambda T$. The full, nonperturbatively defined theory is nevertheless free of any IR divergences \cite{Jackiw:1980kv, Cossu:2020yeg}. We note in passing that the divergences described here are distinct from IR issues in the 4D Coleman-Weinberg potential in $\xi = 0$ gauge, caused by massless but gauge-dependent ``Goldstone'' modes in the broken minimum \cite{Andreassen:2014eha,Elias-Miro:2014pca,Martin:2014bca,Espinosa:2016uaw,Espinosa:2017aew,Ekstedt:2018ftj}.

\def\hackspace{\hspace{-0.65cm}}
\def\figscale{0.365}
\begin{figure*}[t]
	\hackspace
	\begin{subfigure}[b]{\figscale\textwidth}
		\includegraphics[width=\textwidth]{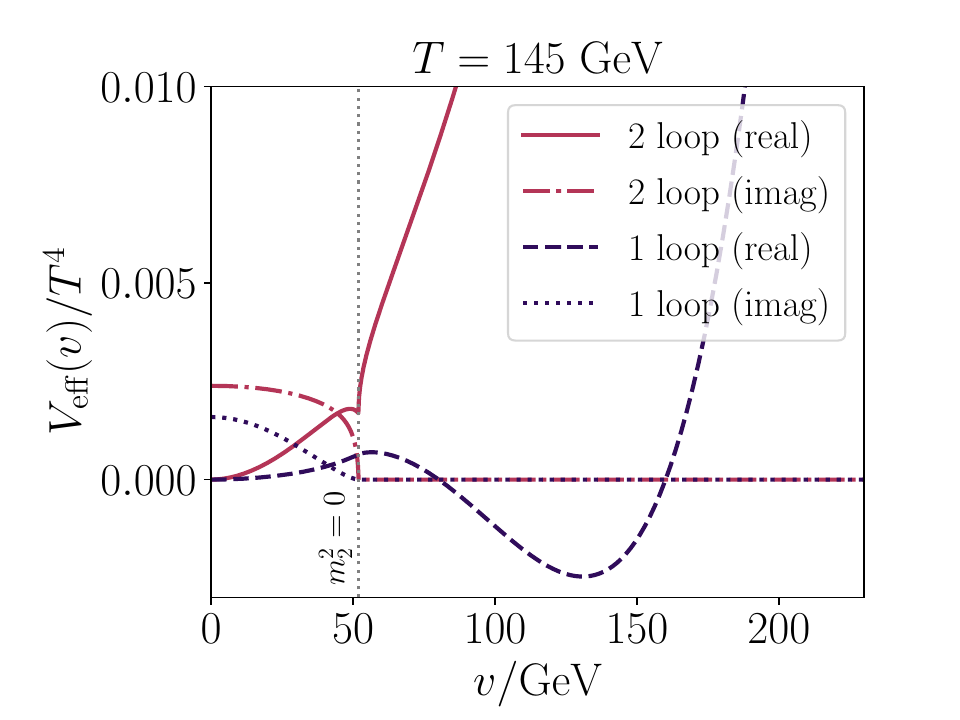}
	\end{subfigure}%
\hackspace
	\begin{subfigure}[b]{\figscale\textwidth}
		\includegraphics[width=\textwidth]{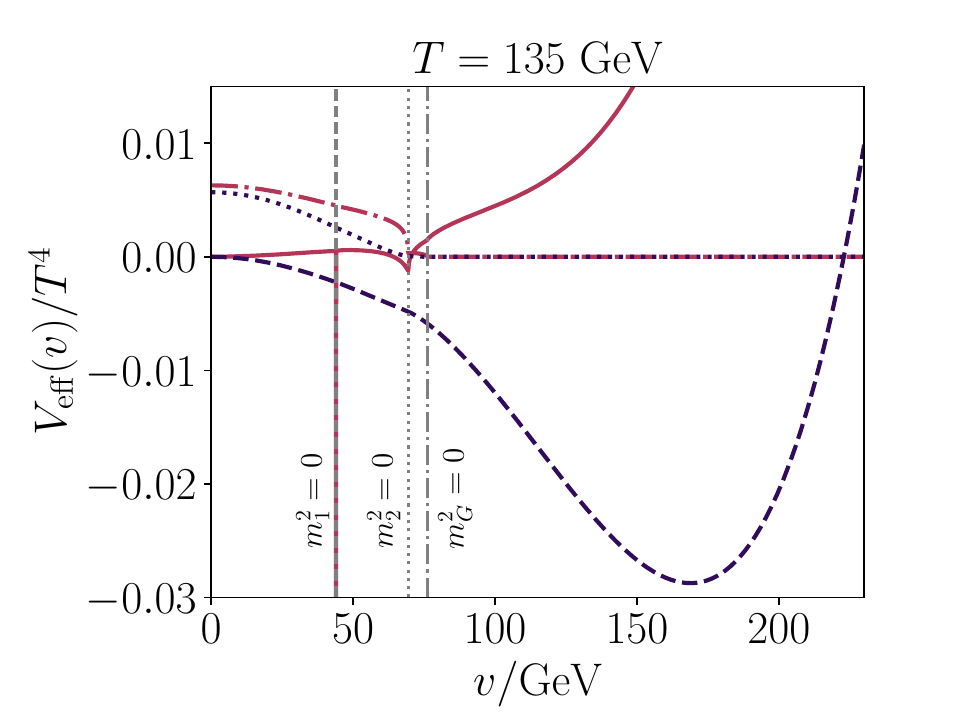}
	\end{subfigure}%
\hackspace\hspace{0.1cm}
	\begin{subfigure}[b]{\figscale\textwidth}
		\includegraphics[width=\textwidth]{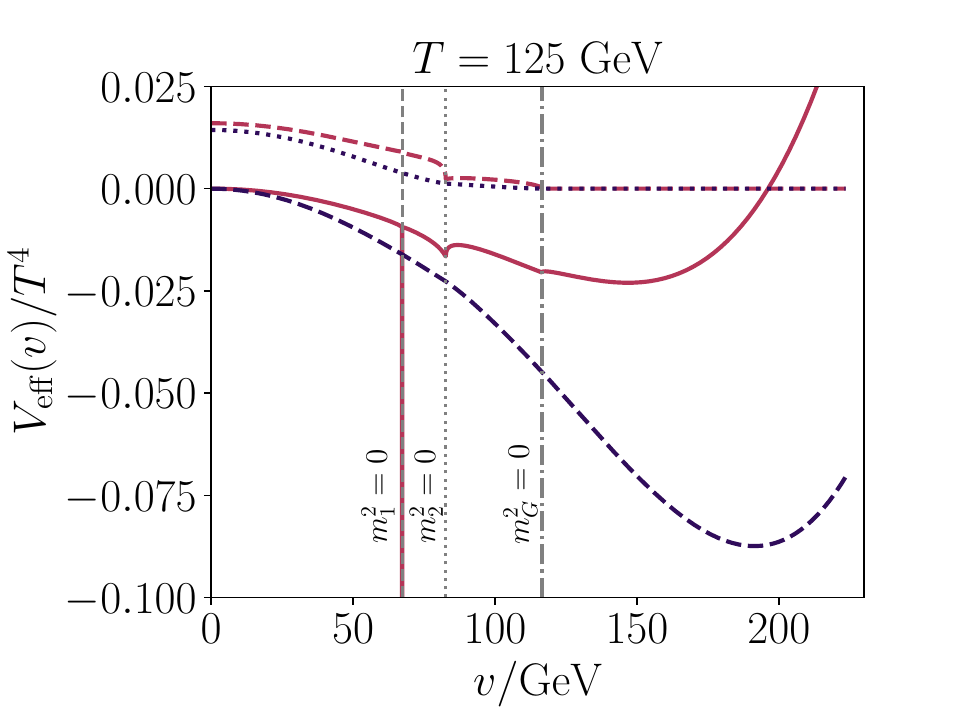}
	\end{subfigure}%
	\caption{
		Display of IR sensitivities in the resummed two- and one-loop thermal Higgs potentials, showing both real and imaginary parts at three fixed temperatures. The input parameters for this benchmark point are $M_2 = 189$ GeV, $a_2 = 1.66$, $b_4 = 1.7$ and $\sin\theta = b_3 = 0$.
		Vertical dotted, dashed and dot-dashed lines show values of the Higgs background field $v$ at which a resummed mass eigenvalue $m_1, m_2$ or $m_G$ (respectively) cross zero.
	}
	\label{fig:veff-IR-div} 
\end{figure*}

The emergence of a massless scalar excitation in the physical spectrum would imply a broken continuous symmetry, or a second order phase transition. As such, in the singlet-extended model we do not expect to find massless $h_1$ or $h_2$ in (meta-)stable minima of the effective potential.\footnote{In the $Z_2$-symmetric model there can be two-step phase transitions in which the singlet undergoes a symmetry breaking transition before the EWPT takes place. If the $Z_2$-breaking transition occurs at high enough temperatures such that the Higgs field can be decoupled (see eg. \cite{Gould:2023ovu} for how this works), the IR behavior of such transition is described by a $S^2 + S^4$ type theory which has a second order transition. In this paper we focus solely on the actual EWPT, ie. phase transitions in the Higgs field, so possible IR issues associated with second-order transitions in the singlet direction are not a concern.}
Despite this, there are still situations in which vanishing (or very small) scalar masses enter the analysis of also first-order EWPT:
\begin{enumerate}[(i)]
	\item As already mentioned above, for resummed potentials that \textit{at tree level} describe a second order transition, there is exact cancellation between thermal and vacuum masses at the critical temperature, leading to $m^2 = 0$ at tree level. This includes models like the minimal SM, our singlet extension in the $Z_2$ symmetric limit, and in fact we believe this to be generally the case for any BSM theory in which the potential barrier needed for first-order EWPT is generated at loop level. In these situations the order-by-order solution to the minimization conditions (\ref{eq:min_cond}) is increasingly unreliable when approaching the critical temperature and diverges at $T=\Tc$, ie. the ``$\hbar$ expansion'' discussed in \cite{Laine:1994zq,Farakos:1994xh,Kajantie:1995kf,Karjalainen:1996rk,Patel:2011th} cannot be used without modifications (see below). 
	
	This IR issue with radiatively-generated phase transitions does not prevent the use of $\hbar$ expansion in studies that limit themselves to one-loop approximations of the effective potential. However, given that already the next order in the expansion is divergent, it is unclear how reliable the one-loop approximation is. 

	\item For direct numerical minimization of $\Veff(v,x)$ one varies the background fields according to some minimization algorithm until a solution to (\ref{eq:min_cond}) is found. During this process, it can happen that for some values of $v,x$, either $h_1$ or $h_2$ becomes approximately massless. The potential then obtains a spuriously large contribution from the diagram in (\ref{eq:sunset-integral}) and may prevent the minimization algorithm from converging to a true minimum. For example, the minimization algorithm may find a local divergence instead of a physical minimum of the potential. This is illustrated in Fig.~\ref{fig:veff-IR-div} which we discuss in more detail shortly.
	
	Furthermore, the whole calculation breaks down if $m^2\approx0$ occurs close to a local minimum $(v_\text{min}, x_\text{min})$: in this case the value of $\Veff$ in the minimum is ``contaminated'' by the IR sensitive logarithm in (\ref{eq:sunset-integral}). The latter case is expected to occur at least near the endpoint of first-order transitions, ie. when approaching a second order EWPT.
	
\end{enumerate}

We illustrate the non-analytic behavior using the $Z_2$ symmetric limit of the model for simplicity ($\sin\theta = b_3 = 0$) and focusing on the Higgs direction of the potential (singlet background field $x = 0$). In this limit the mass-squares of resummed scalar eigenstates are given by
\begin{align}
\label{eq:m1sq}
m_1^2(T, v) &= \bar{m}_\phi^2(T) + 3 \bar{\lambda}(T) v^2, \\
m_2^2(T, v) &= \bar{m}_S^2(T) + \frac12 \bar{a}_2(T) v^2, \\
\label{eq:mGsq}
m_G^2(T, v) &= \bar{m}_\phi^2(T) + \bar{\lambda}(T) v^2,
\end{align}
where the barred parameters are understood to be those of the effective 3D theory and are therefore temperature dependent. $m_1, m_2$ correspond to masses of a Higgs-like and a singlet-like excitation, and $m_G$ is the mass of unphysical Goldstone-like modes in $R_{\xi = 0}$ gauge.

We show the $Z_2$ symmetric potential as function of $v$ for three temperatures in Fig.~\ref{fig:veff-IR-div}.\footnote{The ``one-loop'' potential shown in Fig.~\ref{fig:veff-IR-div} actually includes $\mathcal{O}(g^4)$ corrections from hard thermal modes, ie. the barred parameters in Eqs.~(\ref{eq:m1sq})-(\ref{eq:mGsq}) are the same as for the two-loop $\Veff$. Inclusion of these effects goes beyond the $\mathcal{O}(g^3)$ accuracy of one-loop $\Veff$ and is done here in order to facilitate direct comparison of IR effects. In the main numerical analysis we do not include these corrections in the one-loop potential.}
For values of $v$ that result in a negative mass-square, $m^2(T, v) < 0$, for any of the scalar eigenstates, the effective potential has an imaginary part. This happens already at one-loop level due to the aforementioned $(m^2)^{3/2}$ terms. The one-loop potential also develops a visible kink at field values for which $m^2(T, v) = 0$. At two-loop the situation at vanishing $m_1^2$ or $m_2^2$ is much more severe: $m_1^2 = 0$ produces to a very narrow ``spike'', visible in second and third plots, in the real part of $\Veff$ caused by the logarithmic divergence discussed around Eq.~(\ref{eq:sunset-integral}), while $m_2^2 = 0$ leads to a wider but finite spike. For $m_G^2 = 0$ the potential remains finite because no diagram of type (\ref{eq:sunset-diagram}) with three Goldstone propagators exists. The reason $m_2^2 = 0$ does not produce a genuine divergence is because in the $Z_2$ symmetric and $x = 0$ limit used here, the diagram with three singlet-like ($h_2$) propagators is absent as well. A variation of the $H_3$ diagram exists, however, with one $h_1$ and two $h_2$ propagators that produces $\sim \ln(\bar{\mu} / (\sqrt{m_1^2} + 2\sqrt{m_2^2}))$. The logarithm can become unnaturally large and start dominating the $\Veff$ when $m_1^2$ and $m_2^2$ are both very small, causing the wider spikes seen in all three plots. We refer to \cite{Niemi:2021qvp} and the supplementary material of \cite{Niemi:2020hto} for details of diagrammatic calculations.

We emphasize that these complications posed by non-analytic IR behavior have been known already for a long time \cite{Laine:1994zq, Fodor:1994bs}. It also needs to be emphasized that the effective potential defined by (\ref{eq:Veff-def}) only carries physical meaning at field values that satisfy Eqs. (\ref{eq:min_cond}), ie. in the extremal points \cite{Nielsen:1975fs,Fukuda:1975di}. Indeed, all thermodynamical quantities can be extracted from extremal values of $\Veff$ (corresponding to free energies of different phases), and in the present context any spurious effects outside local minima only become an issue if they prevent the computation of said minima.

A self-consistent way for avoiding non-analytic behavior near a first-order phase transition was outlined in \cite{Arnold:1992rz} and recently formulated in \cite{Ekstedt:2022zro, Gould:2023ovu,Ekstedt:2024etx} (c.f. also \cite{Ekstedt:2020abj,Lofgren:2023sep}). The method relies on the observation that in models where one expects a large discontinuity in the location of the minimum (ie. a strong first-order transition), there are field-dependent one-loop contributions to the potential that cannot be treated as perturbations around the tree-level solution. Hence a further, background-field dependent resummation of specific diagrams is required, effectively absorbing the potential barrier generated at one-loop into a new ``leading order'' potential. One can then perform an ``$\hbar$ expansion'' around the newly resummed ``leading order'' potential that properly describes a first-order phase transition and thus gives well-defined results for $\Veff(v_\text{min}, x_\text{min})$.
The downside of the formalism in \cite{Ekstedt:2022zro, Gould:2023ovu} is that identifying a consistent resummation scheme relies on a (background-field dependent) hierarchy of scales among fields in the theory. A self-consistent application of the method can therefore be cumbersome in models with many scalar fields, especially if performing large parameter-space scans where different types on phase transitions can occur.

For the present paper we have not attempted these field-dependent resummations, and use instead the simpler potential where propagators are resummed with thermal mass corrections only, and we include all two-loop diagrams in the 3D EFT. We also solve for the minima using a direct numerical algorithm (details given in the next section), instead of using an order-by-order $\hbar$ expansion. As demonstrated in \cite{Gould:2023ovu}, a direct minimization can give good results even for phase transitions for which a field-dependent resummation would formally be required. This is not a coincidence: in a power counting sense, implementing the resummation only modifies the potential at order $\mathcal{O}(g^{4.5})$ in the formal expansion parameter.
The direct method is also known to agree well with nonperturbative lattice results when the EWPT is strongly first order, as demonstrated by a recent lattice study in the singlet-extended model \cite{Niemi:2024axp}. 

The final hurdle for our numerical analysis are then the IR sensitive ``spikes'' in $\Veff$ caused by large logarithms at specific field values, point (ii) above. Our minimization algorithm, described in detail in the next section, generally does a good job at avoiding these spikes. However, in a large parameter scan it is practically impossible to always ensure that spurious effects do not occur. As a consistency check, we always verify that imaginary parts of $\Veff$ are absent in what the algorithm thinks is the global minimum, and require well-behaved derivatives at the critical temperature. The latter is checked by measuring the latent heat $L$ and $\langle \he\phi\phi \rangle$ at $T_c$ from derivatives of the potential (see section \ref{sec:quantities}), and discarding points where these evaluate to nonsensically large values.

\section{Details of the numerical analysis}
\label{sec:analysis}

\begin{figure*}[t]
	\centering
	\begin{subfigure}[b]{0.49\textwidth}
		\centering
		\includegraphics[width=1.0\textwidth]{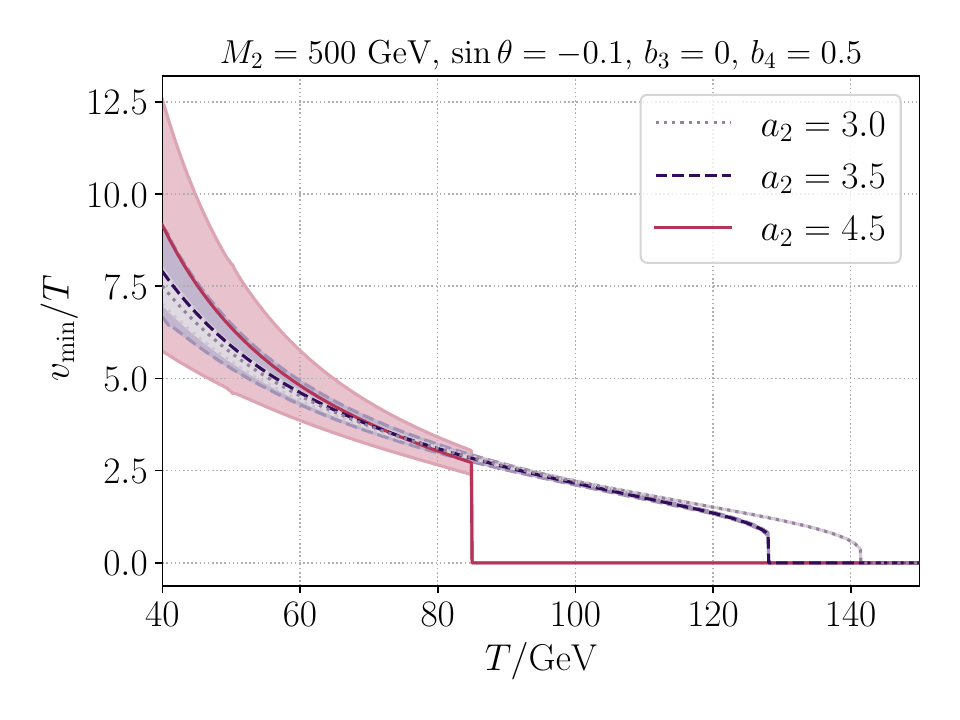}
	\end{subfigure}
	\hfill
	\begin{subfigure}[b]{0.49\textwidth}
		\centering
		\includegraphics[width=1.0\textwidth]{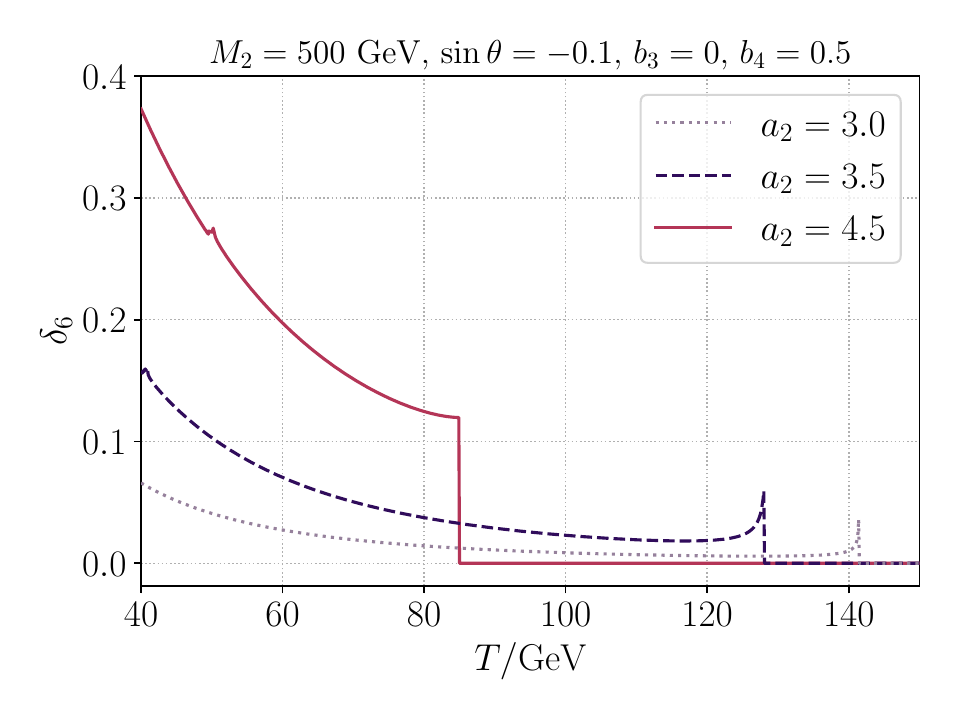}
	\end{subfigure}
	\caption{
		Left: evolution of the Higgs background field (``VEV'') in the global minimum of the two-loop effective potential, at three values of $a_2$. The critical temperature is where the field jumps from $v_\text{min} = 0$ to $v_\text{min} \neq 0$. The shaded bands depict systematic error in $v_\text{min}$ due to neglect of higher-dimensional operators from the high-$T$ effective theory as described in the text. Right: temperature-dependence of the error estimator $\delta_6$, Eq.~(\ref{eq:delta6}).
	}
	\label{fig:dim6-illustration}
\end{figure*}

As discussed in section~\ref{sec:model} the free parameters for our model are the mass of the BSM excitation $M_2$, the mixing angle (or $\sin\theta$) and the couplings $a_2, b_3, b_4$. 
We have performed a uniform scan over the free parameters as follows ($\Delta$ refers to the grid spacing):
\begin{align}
\begin{tabular}{@{}lll@{}}
	$M_2$ & $\in [100, 800] \, \mathrm{GeV}$, & \, $\Delta M_2 = 5 \, \mathrm{GeV}$ \\
	$\sin\theta$ & $\in [-0.3, 0.3]$, & \, $\Delta (\sin\theta) = 0.02$ \\
	$a_2$ & $\in [0, 8]$, & \, $\Delta a_2 = 0.1$ \\
	$b_3$ & $\in [-150, 150] \, \mathrm{GeV}$, & \, $\Delta b_3 = 50 \, \mathrm{GeV}$ \\
	$b_4$ & $\in [0.25, 2.0]$, & \, $\Delta b_4 = 0.25$
\end{tabular}
\end{align}
for a total of $19.8$ million points.
This choice of parameter scanning ranges originates from earlier one-loop parameter scans in the singlet extension \cite{Profumo:2007wc,Profumo:2014opa,Kurup:2017dzf}, although we have chosen relatively conservative upper bounds on $|b_3|$ and $b_4$. We will not study masses lighter than $100$ GeV as strong transitions in the small mass region are mostly associated with exponentially small couplings \cite{Kozaczuk:2019pet}. Hence, for light $M_2 \lesssim 100$ GeV one can expect higher-order effects from the singlet to be small.

We note that $|\sin\theta|$ is experimentally bounded from above by Higgs signal strength measurements and electroweak precision studies \cite{Profumo:2014opa,Lopez-Val:2014jva,Robens:2015gla,Robens:2016xkb}. The bound from signal strengths is independent of $M_2$ for $M_2 \gtrsim 150$ GeV and was reported in Refs.~\cite{Robens:2015gla,Robens:2016xkb} to be $|\sin\theta| \lesssim 0.36$. A more recent study with full data from LHC run 2 found $|\sin\theta| \lesssim 0.23$ \cite{Niemi:2024axp}. Furthermore, a mass-dependent upper bound can be obtained from precision $W$ boson mass calculations and rules out $|\sin\theta| \gtrsim 0.25$ for $M_2 \gtrsim 500$ GeV \cite{Lopez-Val:2014jva,Robens:2016xkb}. A caveat here is that the renormalization scheme (and thus the mixing angle) used to obtain this result \cite{Lopez-Val:2014jva} differs from our \MSbar scheme (see section \ref{sec:model}), and a careful conversion between the schemes would be required to robustly apply the constraint. Nevertheless, some of the parameter space towards the upper end of our $|\sin\theta|$ scanning range may already be ruled out by experimental data.

For a given set of input parameters we search for the EWPT by starting from temperature $T= 40$ GeV 
and increasing $T$ in steps of $\Delta T = 0.1$ GeV until the global minimum of $\Veff(v,x)$ changes from $v \neq 0$ to $v=0$.
In practice the analysis proceeds as follows.
\begin{enumerate}
	\item From the inputs we compute the \MSbar renormalized parameters (scale $\bar\mu = M_Z$) at next-to-leading order as described in section~\ref{sec:model} and appendix A of Ref.~\cite{Niemi:2021qvp}. 
	
	\item We require absolute stability of the electroweak vacuum at the input scale, ie. the global potential minimum should lie at $v \approx 246$ GeV (more specifically, the minimum used in step 1. above). This check is performed at one-loop level by numerically minimizing the $T=0$ Coleman-Weinberg potential. It should be noted that in principle, the EW vacuum could be a long-lived metastable state instead of the global energy minimum. We do not consider this special case here.
	
	\item The parameters are renormalization group (RG) evolved to scale $\bar\mu = 700$ GeV using one-loop RG equations (see eg.~\cite{Brauner:2016fla}). The reasoning behind this choice of scale is as follows. In high-$T$ calculations a common choice is $\bar\mu = \bar\mu_T \approx 7T$ as this is optimal for minimizing temperature-dependent logarithms that arise from thermal loops \cite{Farakos:1994kx}. However, the critical temperature can differ a lot between one- and two-loop analyses, and using a $T$-dependent scale complicates comparisons of the two calculations. The specific choice $\bar\mu = 700$ GeV is not too far from the ``optimal'' scale of $\bar\mu_T \approx 7T$ if the critical temperature is of order $100$ GeV, which we find to be the case at two loops.
	
	It is possible that some coupling enters nonperturbative regime during RG-running, indicating breakdown of perturbation theory at the scale relevant for high-$T$ physics. To check for this, we impose naive perturbative bounds on the dimensionless couplings at the scale $\bar\mu = 700$ GeV, ie. the cost of a four-point vertex in Feynman diagrams should not exceed the $4\pi$ suppression associated with loop integration. This requires 
	\begin{align}
	\label{eq:perturbativity-bound}
	|\lambda|, |b_4| < \frac{2\pi}{3}, \quad |a_2| < 4\pi.
	\end{align} 
	Points that do not satisfy these conditions are discarded. Requiring perturbative unitarity is known to put more stringent bounds on the running couplings \cite{Robens:2015gla}, but we do not consider unitarity constraints in this work.

	\item Matching to the effective theory (\ref{eq:EFT-action}) is performed following Ref.~\cite{Niemi:2021qvp}. This includes RG evolution down from the matching scale $\bar\mu = 700$ GeV to scale $\bar\mu = T$ at which our analysis of $\Veff$ is carried out.
	
	\item We obtain the two-loop effective potential from \cite{Niemi:2021qvp} and find its global minimum $(v_\text{min}, x_\text{min})$ as function of the temperature, using a numerical minimization algorithm whose details are described below. The temperature loop proceeds linearly from low to high temperature and is terminated once the global minimum changes to $v_\text{min} = 0$ and the $T$ at which this happens is identified as the critical temperature. Our upper bound for the temperature was $1$ TeV, and we give up if the critical temperature is higher than this.
	
\end{enumerate}
This procedure gives the full $\mathcal{O}(g^4)$ result for the effective potential in the high-$T$ approximation, while optimizing a subset of higher-order logarithms using the renormalization group. For comparison with one-loop calculations we have repeated the scan using one-loop accuracy in steps 4 and 5 (see section~\ref{sec:Veff}). 

The main challenge in the analysis is the reliable determination of the global $\Veff(v, x)$ minimum. We do this by choosing a set of ``sensible'' $(v, x)$ pairs that are passed as starting points for a \textit{local} minimization algorithm.%
\footnote{We performed the local minimization using the derivative-free \textsc{BOBYQA} algorithm \cite{bobyqa} and its \textsc{C++} implementation in the \textsc{dlib} library \cite{dlib09}.}
In our testing this was both faster and more reliable than global minimization algorithms, when the set of starting points was chosen as follows:
\begin{enumerate}[(i)]
	\item Each local extremum of the tree-level (resummed, $T$-dependent) $\Veff(v, x)$ is included as a starting point. At tree-level, the extrema conditions (\ref{eq:min_cond}) can be solved analytically with the cubic formula. This often has solutions where both $v$ and $x$ are non-vanishing.
	
	\item We ensure that the combinations $(0, 0)$, $(v, 0)$, $(0, x)$, $(0, -x)$ are included, where nonzero values of a field refer to a large value ($\gtrsim T$). Note that we may restrict to $v \geq 0$ without loss of generality.
		
\end{enumerate}
These starting points are typically sufficient for finding all local minima of the full $\Veff$, from which the global minimum can be deduced. Outside its minima, the effective potential defined by (\ref{eq:Veff-def}) can develop an imaginary part (see \cite{Weinberg:1987vp} for a physical interpretation). Here we always minimize the real part only.

It should be pointed out that the effective potential constructed from Eqs.~(\ref{eq:field-shifts}) and (\ref{eq:Veff-def}) describes a first-order transition at any finite order of perturbation theory (with the possible exception of the tree level potential). The reason is that one-loop corrections from gauge field zero-modes produce cubic terms of form $g^3 v^3$ that create a potential barrier between minima at $v=0$ and $v\neq 0$. In many models, including the familiar SM case, a weak EWPT ($\Delta v_\text{min}/T \ll 1$) as predicted by the effective potential is actually a crossover when studied with nonperturbative methods \cite{Kajantie:1995kf, Kajantie:1996mn,Csikor:1998eu,Niemi:2020hto,Gould:2022ran,Niemi:2024axp}. Given that weak transitions are not particularly interesting for cosmology in the first place, we will only report transitions with $\Delta v_\text{min} / T > 0.7$.%
\footnote{
In order to find a boundary in parameter space between first-order and crossover regions, one can often integrate out the heavy singlet, and utilize the fact that the phase diagram of the ${\rm SU}(2)$ + Higgs EFT is known nonperturbatively, see \cite{Andersen:2017ika, Niemi:2018asa, Gould:2019qek,Niemi:2020hto, Friedrich:2022cak}. We have not attempted this approach here as our focus is on strong transitions, which can be reliably found using perturbation theory \cite{Niemi:2024axp}, assuming of course that the 3D EFT approach remains valid.
}

We have verified that the procedure described here correctly reproduces the results for benchmark points presented in \cite{Niemi:2021qvp}. Despite this, our scanning program occasionally found minima where the potential had an imaginary part, and in other cases we found transitions with spuriously large latent heat (see next section for definitions). These suggest a failed minimization, possibly due to the IR issues described in section \ref{sec:IR-div}. We also found spurious transitions where the Higgs field jumps back to $v_\text{min} = 0$ at a low temperature, an effect that can likely be attributed to a failure of our high-$T$ EFT at low $T$.
In total, points where these problems arise amount to roughly $8\%$ of all our points where a phase transition is observed (according to the aforementioned $\Delta v_\text{min} / T > 0.7$ criterion). The overwhelming majority ($82\%$) of failed points have a large mixing angle $|\sin\theta| > 0.2$ and as such, are partly ruled out by collider experiments (see discussion above).
All these spurious transitions have been discarded from our results below. We will also focus mostly on parameter space with $|\sin\theta| \leq 0.2$ where the success rate of our scanning algorithm is good.

\section{Quantities of interest}
\label{sec:quantities}

Once the critical temperature $T_c$ is found, at $T = T_c$ we calculate the latent heat
\begin{align}
\label{eq:latent}
L = -T_c \Delta \der[\Veff(v_\text{min}, x_\text{min})]{T},
\end{align}
where $\Delta$ denotes the difference of low- and high-$T$ phases. We also compute jumps in the background fields $\Delta v_\text{min}$ and $\Delta x_\text{min}$,
and discontinuity of the Higgs condensate $\langle \he\phi\phi \rangle = d \Veff / d\bar{m}_\phi^2$. Here $\bar{m}^2_\phi$ is the resummed version of $m^2_\phi$ as it appears in the EFT (\ref{eq:EFT-action}): $\bar{V}(\phi,S) \supset \bar{m}^2_\phi \he\phi\phi$. The value of $\Delta v_\text{min}$ is gauge dependent by construction but expected to be close, 
in $R_{\xi=0}$ Landau gauge,
to the gauge-invariant quantity \cite{Laine:2017hdk, Niemi:2021qvp}
\begin{align}
\label{eq:vphys}
\Delta \vphys \equiv \sqrt{2 \Delta \langle \he\phi\phi \rangle}.
\end{align}
Our parameter-space scans verify this expectation, see Fig.~\ref{fig:field-jumps} below. 

For later purposes we note that the latent heat is related to expectation values of local operators in the 3D action, and their jumps across the phase transition, as
\begin{align}
\label{eq:latent-condensates}
L = -\Tc \Delta \Big( \pder[\bar{m}_\phi^2]{T} \langle\he\phi\phi\rangle + \pder[\bar{b}_1]{T} \langle S \rangle + \pder[\bar{m}_S^2]{T} \langle S^2 \rangle + \dots \Big)
\end{align}
where the barred parameters are those appearing in the 3D EFT, cf.~\eqref{eq:potential-3d}, and we show only terms with largest temperature derivatives. This equation follows directly by applying the chain rule on Eq.~\eqref{eq:latent}, see \cite{Niemi:2024axp} for details.
The importance of Eq.~\eqref{eq:latent-condensates} lies in the observation that expectation value jumps are a feature of the high-$T$ effective theory, while temperature derivatives of action parameters arise solely from UV physics, ie. the 4D $\rightarrow$ 3D matching relations.

\def\figscale{0.4075}
\def\smallLift{\vspace{-0.75cm}}
\begin{figure*}[!t]
	\centering
	\includegraphics[width=\figscale\textwidth]{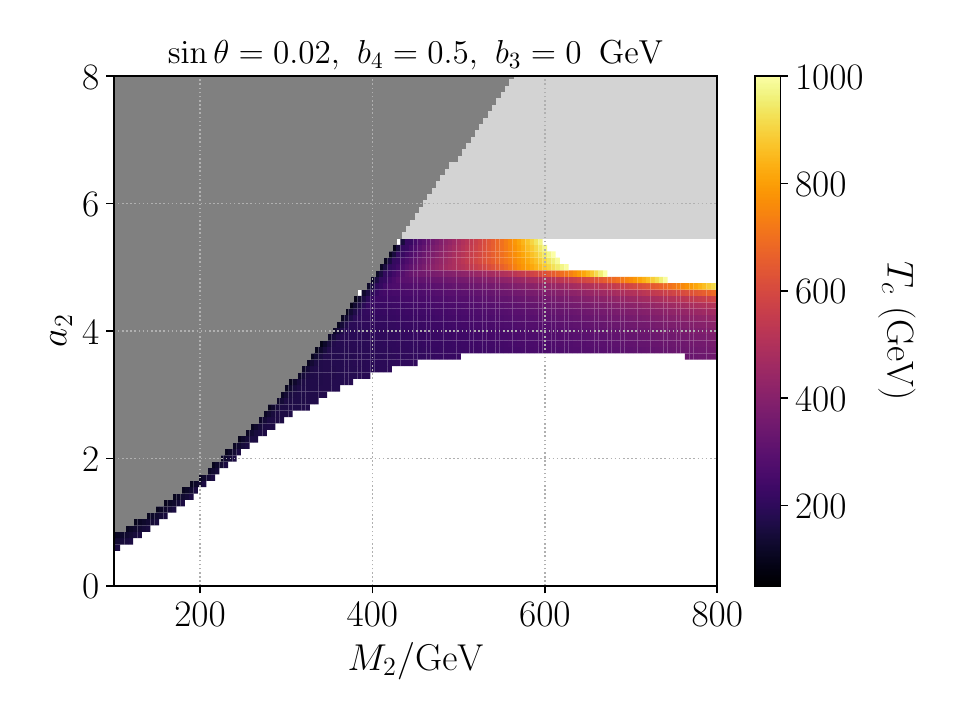}
	\includegraphics[width=\figscale\textwidth]{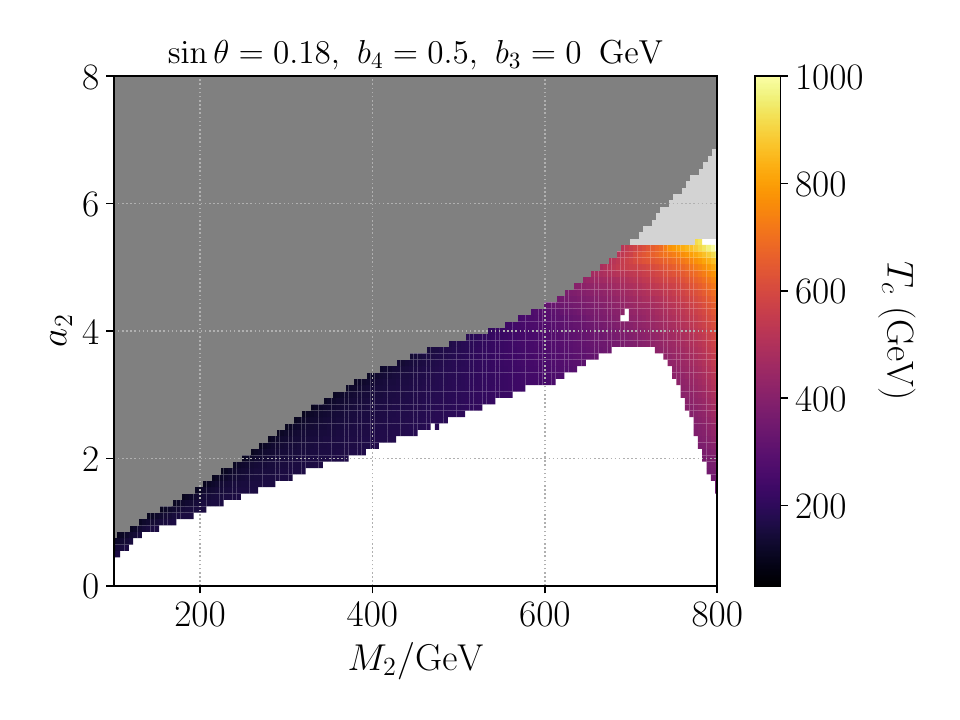}
	
	\smallLift
	\footnotesize{One-loop scan} \\
	\includegraphics[width=\figscale\textwidth]{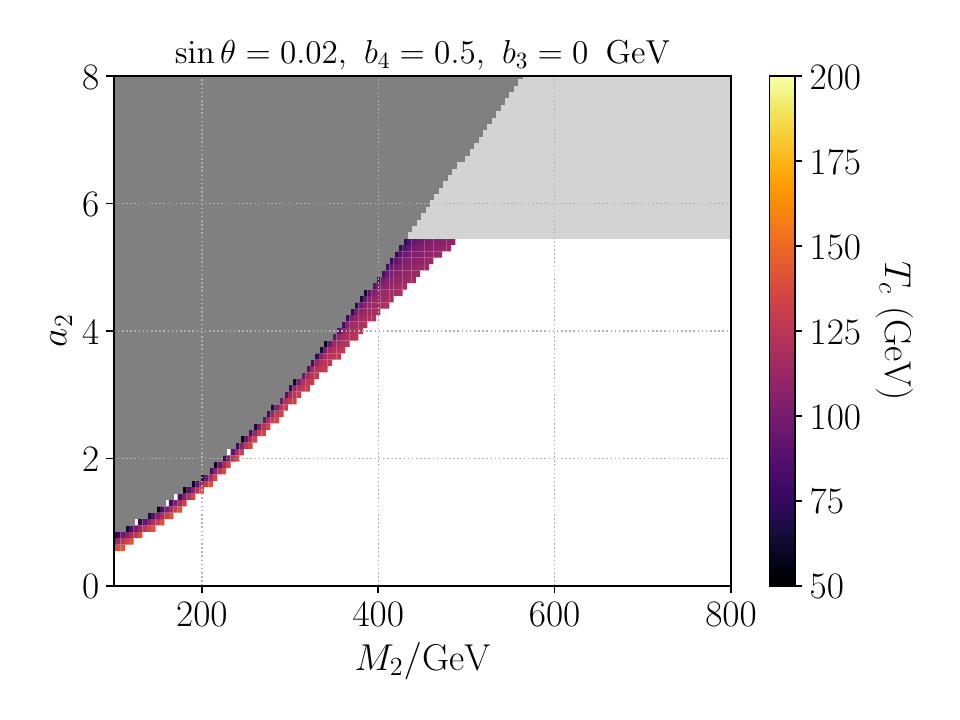}
	\includegraphics[width=\figscale\textwidth]{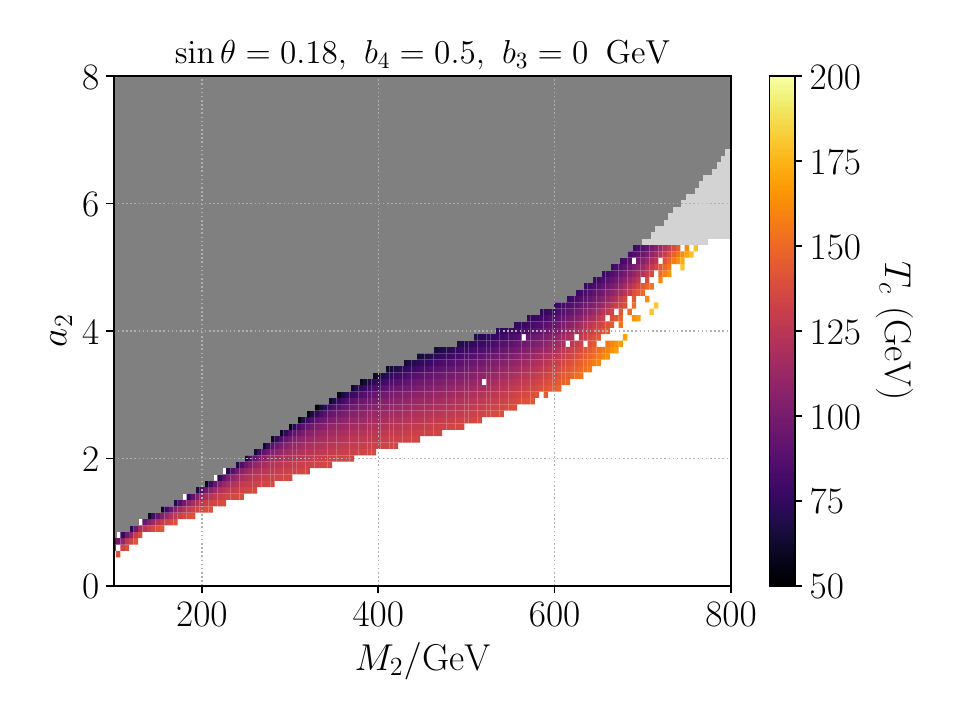}
	
	\smallLift
	\footnotesize{Two-loop scan}
	\caption{Heat maps of the critical temperature in $(M_2, a_2)$-plane, at one-loop (top row) and two-loop (bottom row). Dark and light gray denote regions of metastable $T=0$ vacuum and nonperturbatively large couplings, respectively. We only show transitions with $\Delta v / T > 0.7$.}
	\label{fig:Tc-comparison-st}
	\includegraphics[width=\figscale\textwidth]{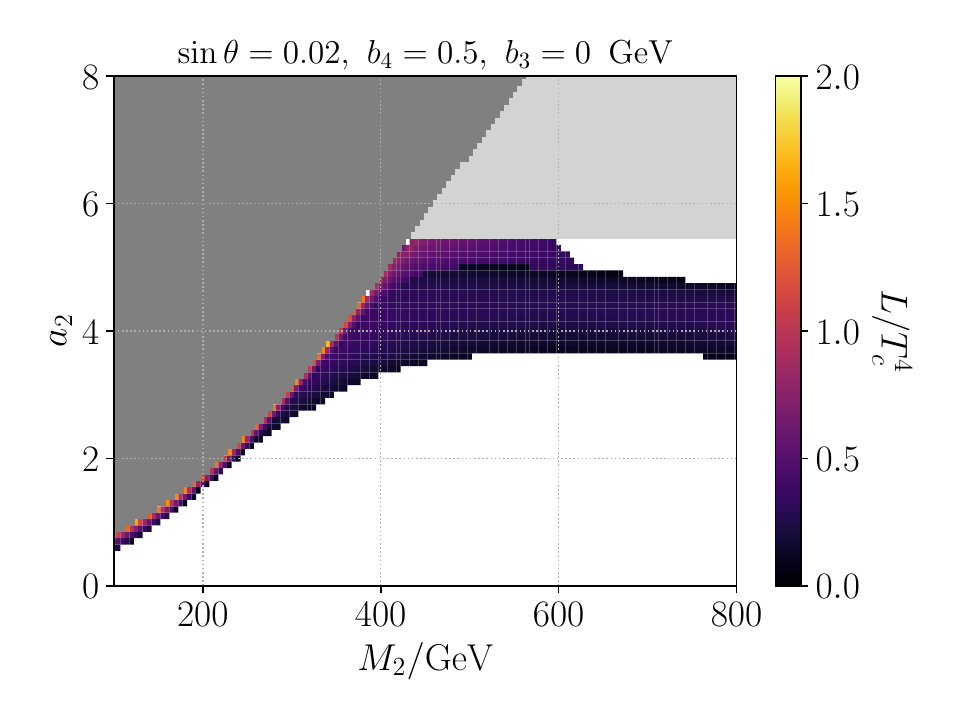}
	\includegraphics[width=\figscale\textwidth]{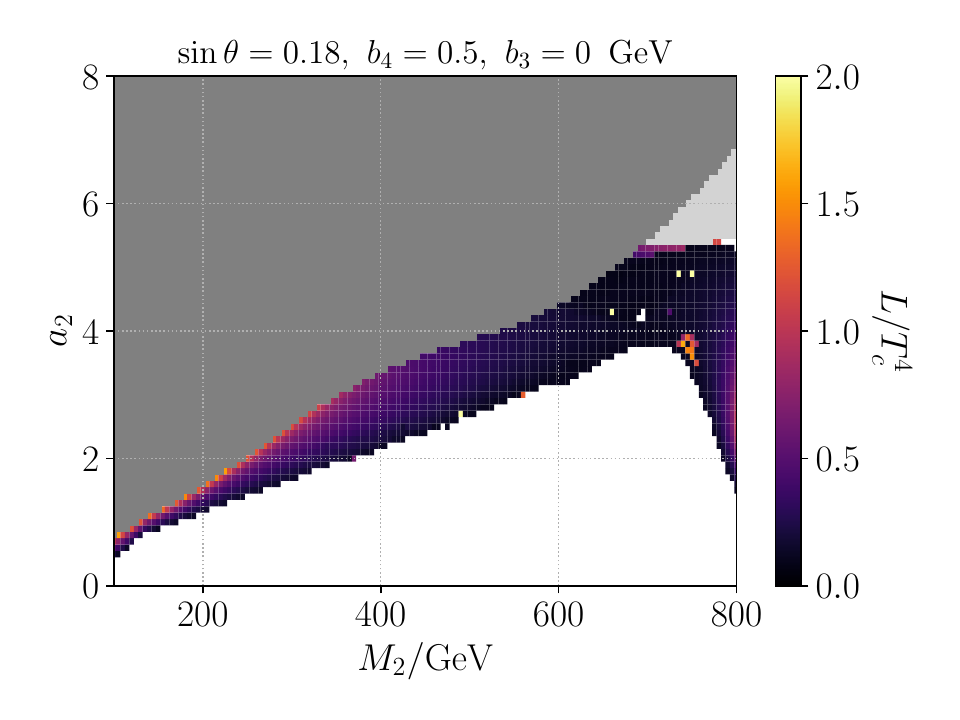}
	
	\smallLift
	\footnotesize{One-loop scan} \\
	\includegraphics[width=\figscale\textwidth]{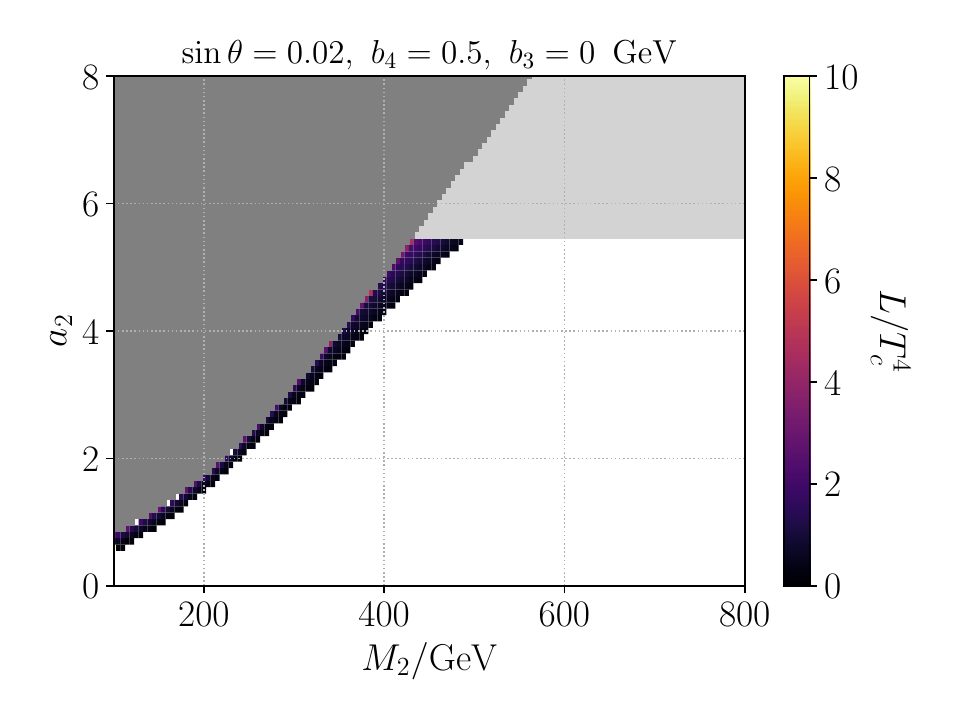}
	\includegraphics[width=\figscale\textwidth]{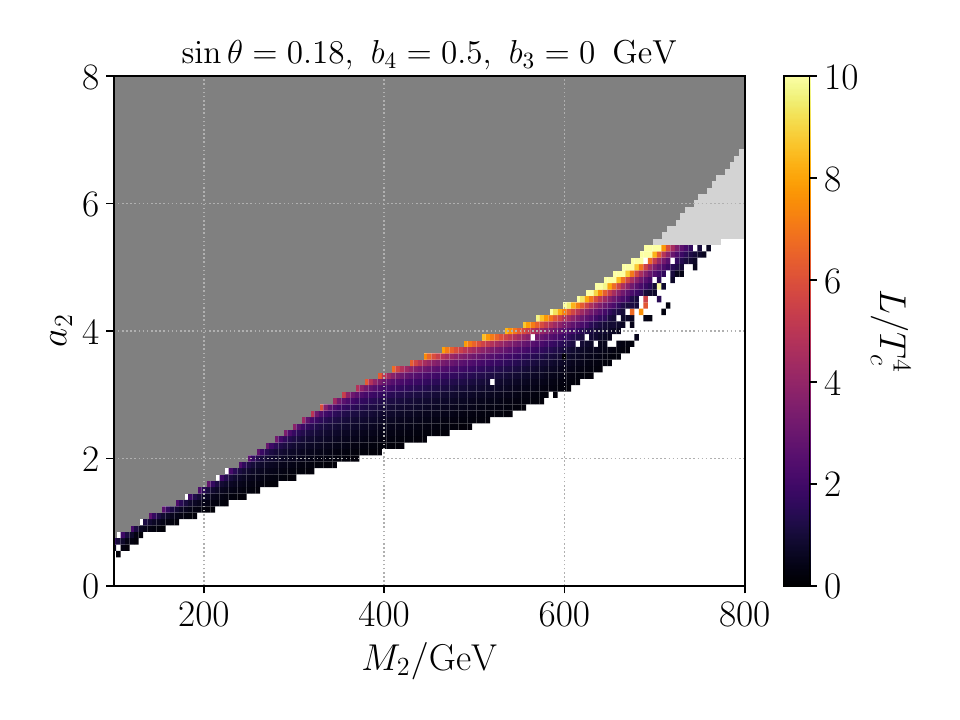}
	
	\smallLift
	\footnotesize{Two-loop scan}
	\caption{As Fig.~\ref{fig:Tc-comparison-st} but showing the latent heat $L/T^4_c$. In the two-loop case we have capped the heatmap scale at $L/T^4_c = 10$ to better show color variations in other parts of the phase transition region.}
	\label{fig:latent-comparison-st}
\end{figure*}

The dimensionless ratios $L/T^4$ and $\Delta \vphys / T$ are often used as measures of the transitions strength. The former gives a lower bound on the energy available for gravitational wave production in the phase transition \cite{Espinosa:2010hh}, and the latter is related to the suppression of Chern-Simons diffusion rate in the broken phase and is thus relevant for electroweak baryogenesis \cite{DOnofrio:2014rug}. In the minimal SM, the latent heat and Higgs condensate jump are approximately proportional \cite{Farakos:1994xh}, but in the singlet extension the latent heat obtains contributions also from discontinuities in condensates involving the singlet  (cf. \cite{Niemi:2024axp}). We nevertheless observe $L/T^4$ and $\Delta \vphys / T$ to be positively correlated.

Following \cite{Gould:2019qek}, we may relate the latent heat at $T_c$ to a lower bound of another measure of transition strength often used in gravitational-wave studies, the so-called $\alpha$ parameter. Here, $\alpha$ is defined as the ratio of (pseudo-) trace anomaly and enthalpy of the symmetric phase, c.f. \cite{Giese:2020znk, Tenkanen:2022tly}:
\begin{align}
\label{eq:alpha-bound}
\alpha(T_*) \gtrsim \frac{30}{4 \pi^2 g_*} \frac{L(T_c)}{T^4_c},
\end{align}
where $T_* < T_c$ is a temperature close to the percolation temperature (meaning that more than 2/3 of the universe is in the low-temperature phase) and $g_* = 107.75$ is the number of relativistic degrees of freedom at $T_*$.\footnote{
We have corrected here a misprint in Eq.~(6) of \cite{Gould:2019qek} related to missing factor 4 in the denominator.
}
See eg. \cite{Athron:2023rfq} for details on the meaning of $T_*$.
The lower bound of Eq.~\eqref{eq:alpha-bound} holds in many cases where $\alpha$ turns out to be monotonically decreasing function of temperature.
As an example, the condition $\alpha > 0.1$ is commonly quoted as a heuristic lower bound for transitions strong enough to be observable at LISA \cite{Caprini:2019egz}. In the absence of significant supercooling this corresponds to $L/T^4_c \gtrsim 10$.
We use this value as an upper bound in range of $L/T_c^4$ shown in bottom row of Fig.~\ref{fig:latent-comparison-st}.  

We also define an error estimator for our high-$T$ approximation, ie. the EFT (\ref{eq:EFT-action}), following \cite{Kajantie:1995dw, Niemi:2018asa, Gould:2019qek, Niemi:2021qvp}. Let $v_\text{min}(T)$ be value of the Higgs background field in the broken phase at temperature $T \leq T_c$. This is found from the effective potential as described above. We then modify the $\Veff$ by explicitly including scalar operators of dimensions five and six (those in Eq.~(\ref{eq:dim6}) in the tree level part only, and repeat the minimization. This gives $v_\mathrm{min, 6}(T)$. Let $\delta_6(T)$ denote the relative shift in the location of the broken phase caused by the higher-dimensional operators:
\begin{align}
\label{eq:delta6}
\delta_6(T) = \Big| \frac{v_\text{min}(T) - v_\text{min,6}(T)}{v_\mathrm{min}(T)} \Big|.
\end{align}
The subscript emphasizes that this error estimate arises from operators up to field dimension six.

The estimator $\delta_6(T)$ can be evaluated at any temperature where the broken phase exists as a (meta-)stable minimum of the potential. The expectation is that $\delta_6$ becomes large deep in the broken phase where $v_\text{min} / T \gg 1$. This is the regime where masses generated by the Higgs mechanism are large compared to the temperature, thus invalidating the high-$T$ EFT approach. We illustrate this behavior in Fig.~\ref{fig:dim6-illustration} at three values of $a_2$, which contributes to Wilson coefficients of the higher-dimensional operators. For $a_2 = 3.0, 3.5$ the error in $v_\text{min}/T$ is at the percent level and the truncated EFT performs well even at rather large $v_\text{min} / T \approx 2.5$. In contrast, the $a_2 = 4.5$ case always has $\delta_6$ of at least $0.1$ and the convergence of high-$T$ approximations is questionable. For comparison, in the minimal SM one has $\delta_6 \approx 0.01$ at $v_\text{min} / T = 1$ and $\delta_6$ is dominated by top-quark contributions to the Wilson coefficient of the $(\he\phi\phi)^3$ operator \cite{Kajantie:1995dw}.

The quantity $\delta_6(T)$ may not be well defined in the immediate vicinity of the critical temperature because the inclusion of higher-dimensional operators may shift the critical temperature enough that the broken phase is no longer metastable, and the shift $v_\text{min} - v_\text{min,6}$ cannot then be calculated. This is the reason for the apparent ``spiking'' in $a_2 = 3.0, 3.5$ curves in fig~\ref{fig:dim6-illustration}. A better error estimator near the critical temperature could be eg. the shift in $T_c$, but we have not implemented this in our scans. In the results below we always report $\delta_6$ evaluated below the critical temperature, at $T = 0.8T_c$.

As mentioned above, $v_\text{min}$ itself is not a physical quantity, but the relative shift defined here is still a useful probe of the error inherent in EFT computations of physical quantities due to the close relationship between $\Delta v_\text{min}$ and the gauge-invariant $\Delta \vphys$ defined in (\ref{eq:vphys}). An analogous error estimator can be defined also for the singlet field $x_\text{min}$, see Fig~\ref{fig:delta6}.

\section{Results of parameter space scan}
\label{sec:results}

\def\figscale{0.34}
\def\shiftLeft{\hspace{-0.25cm}}
\begin{figure*}[!t]
	\centering
	\shiftLeft
	\includegraphics[width=\figscale\textwidth]{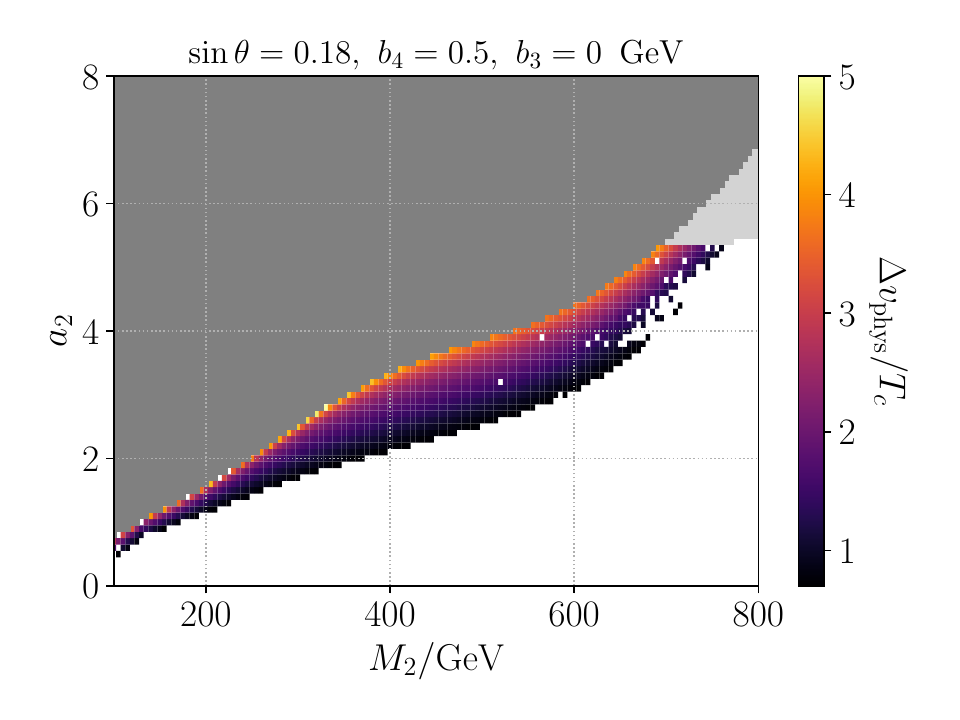}
	\shiftLeft
	\includegraphics[width=\figscale\textwidth]{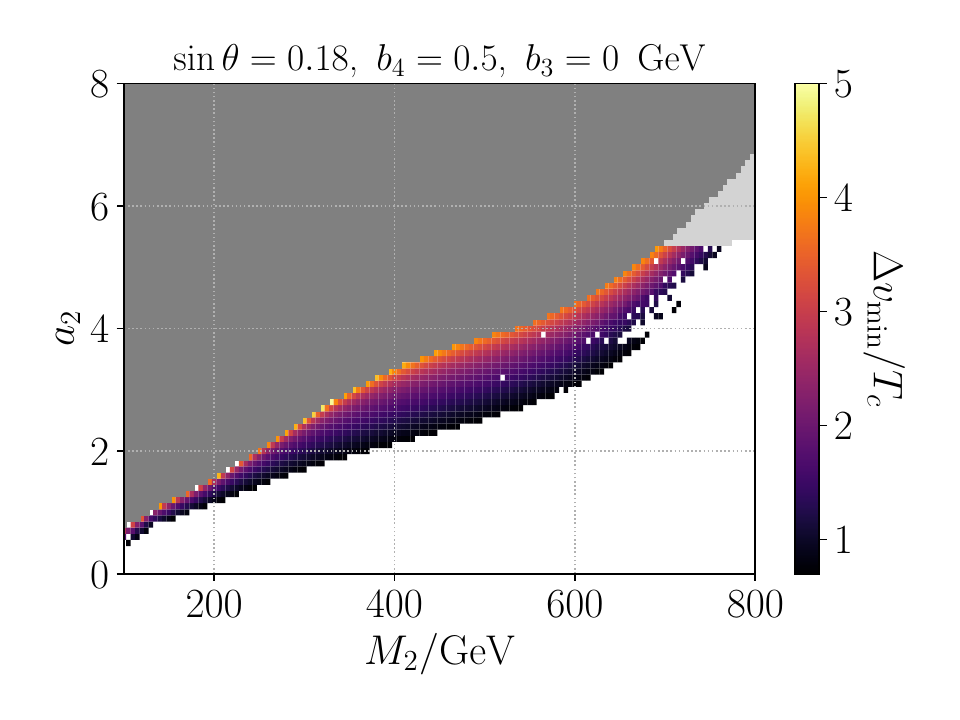}
	\shiftLeft
	\includegraphics[width=\figscale\textwidth]{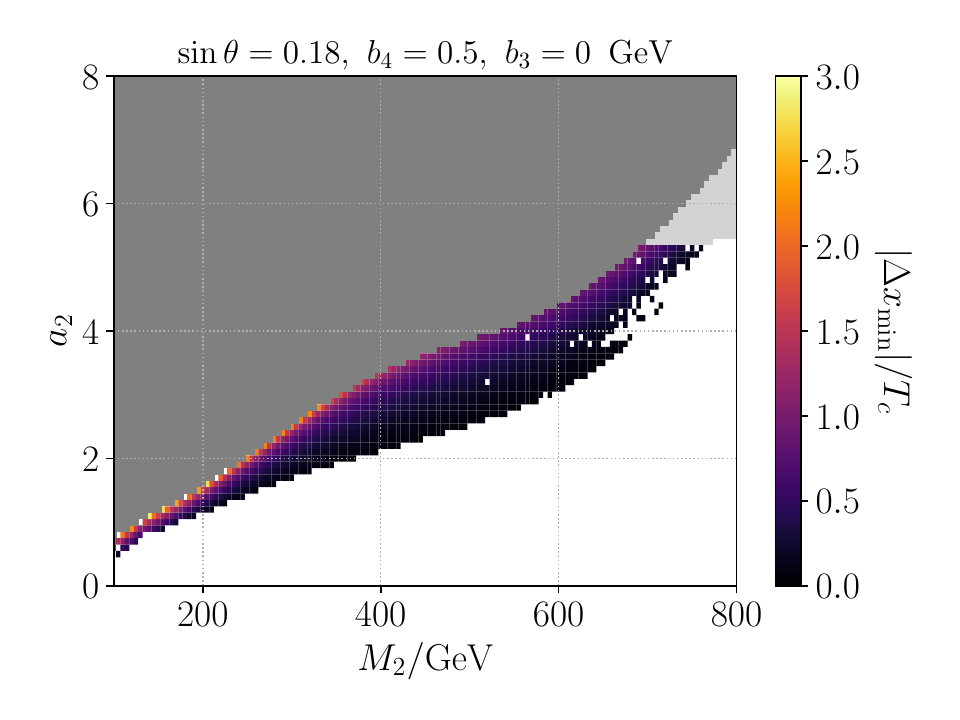}
	\caption{
		Discontinuities of the Higgs condensate $\Delta \vphys = \sqrt{2\Delta\langle \he\phi\phi\rangle}$ (left) and the Higgs and singlet background fields (middle and right) across the EWPT. Plots are at relatively large mixing angle $\sin\theta = 0.18$, and only two-loop results are shown. Gray regions are as described in the caption of Fig.~\ref{fig:Tc-comparison-st}.
	}
	\label{fig:field-jumps}
\end{figure*}

\def\figscale{0.45}
\begin{figure*}[!t]
	\centering
	\includegraphics[width=\figscale\textwidth]{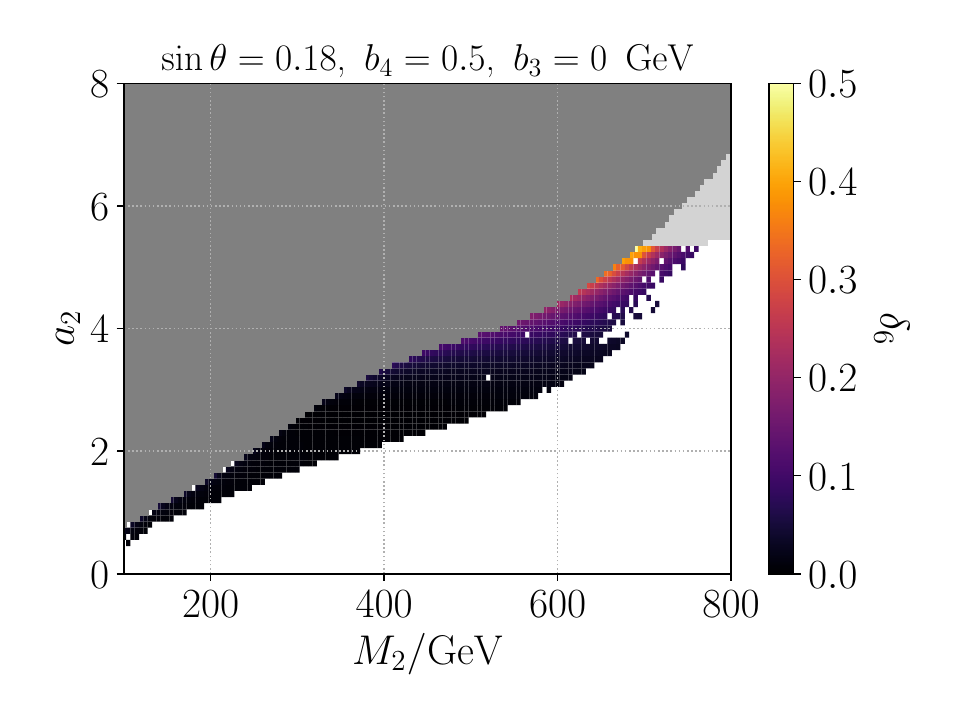}
	\includegraphics[width=\figscale\textwidth]{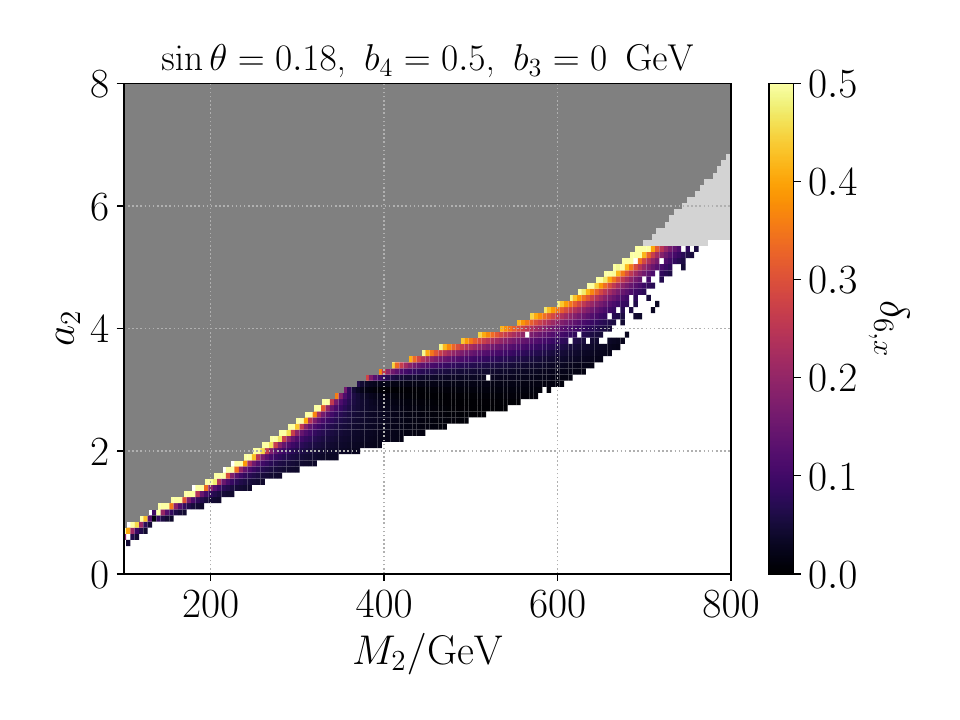}
	\caption{
		Error estimator $\delta_6$ as defined in Eq.~\eqref{eq:delta6}, and an analogous quantity for the singlet background field $x$ (right plot). We find $\delta_6 < 0.1$ in most of the first-order EWPT region, suggesting that the high-$T$ EFT approach is a valid tool for computing phase transition parameters while keeping the error at ten-percent level. The error grows large towards the large-$M_2$ and large-$a_2$ region, suggesting breakdown of the EFT in this region. $\delta_{6,x}$ is seen to consistently be larger than $\delta_6$, suggesting that the singlet VEV can be determined less precisely than the Higgs analogue.
	}
	\label{fig:delta6}
\end{figure*}

Fig.~\ref{fig:Tc-comparison-st} displays heatmaps of the critical temperature in $(M_2,a_2)$-plane for two values of $\sin\theta$, with one-loop (two-loop) results on the top (bottom) row.
Dark gray regions denote where the electroweak vacuum is metastable, ie. not the global energy minimum at zero temperature (see section~\ref{sec:analysis}), and in light gray regions the perturbativity bound of Eq.~\eqref{eq:perturbativity-bound} is not satisfied. In white regions at small $a_2$ we find no strong transitions according to our requirement $\Delta v/T_c > 0.7$. In the one-loop plots there is a further white region at large $a_2$ and $M_2$ where we do not have phase transition data because the critical temperature exceeded our upper limit of $T = 1$ TeV. The occasional white points seen within colored regions correspond to data that was discarded due to the scanning program failing to give good results at those points, see section~\ref{sec:analysis} for description of possible causes.

Comparing one- and two-loop heatmaps, regions of strong transitions are seemingly narrower at two loops.
Notably, for small $\sin\theta$ (left column) the region of strong transition at one loop extends to large $M_2$ and is associated with high critical temperatures, yet this feature is missing in the two-loop results. For a large mixing angle (right column), the feature is somewhat less pronounced but clearly visible for $M_2 \gtrsim 700$ GeV. The main quantitative feature between one- and two-loop analyses is that the one-loop approximation consistently predicts a much higher critical temperature. This result appears fairly general as we observe it for other values of $\sin\theta, b_3, b_4$ as well (plots not shown for the sake of brevity). We investigate possible reasons for this effect in Appendix~\ref{sec:impact}.
A rough qualitative agreement between one- and two-loops is seen at small $a_2, M_2$, particularly near the boundary of metastable $T=0$ vacuum, where both analyses find strong first-order transitions. 

To assess EWPT strength, we turn to the dimensionless latent heat $L/T_c^4$, the Higgs condensate discontinuity $\Delta \vphys$ (see Eq.~\eqref{eq:vphys}) and jumps in the background fields $v,x$.
Comparison of one- and two-loop $L/T_c^4$ is shown in Fig.~\ref{fig:latent-comparison-st} again at two values of $\sin\theta$. One immediately sees that the region of strong transitions is significantly larger at $\sin\theta = 0.18$ compared to the $\sin\theta = 0.02$ case.
This is to be expected, since a larger mixing angle leads to larger $|a_1|$ cubic coupling in the potential which in turn produces a tree-level barrier between the $v=0$ and $v \neq 0$ minima.
Two-loop $\Delta \vphys$ and field jumps are shown in Fig.~\ref{fig:field-jumps} for the $\sin\theta = 0.18$ case.
For the singlet field jump we always plot its absolute value.
One immediate takeaway from Fig.~\ref{fig:field-jumps} is that the results for the ``physical'' $\Delta \vphys$ and the $R_{\xi=0}$ value of $\Delta v$ are practically indistinguishable, as already anticipated in the discussion around Eq.~\eqref{eq:vphys}.
Therefore we do not distinguish between $\Delta \vphys$ and $\Delta v$ in the following discussion.
Similar observations have been made recently in \cite{Gould:2023ovu}, c.f. Fig.~14 therein, when working at two-loop order. See also related discussion in \cite{Lofgren:2023sep}.

\def\figscale{0.355}
\def\smallLift{\vspace{-0.25cm}}
\def\shiftLeft{\hspace{-0.50cm}}
\begin{figure*}[!t]
	\centering
	\shiftLeft
	\includegraphics[width=\figscale\textwidth]{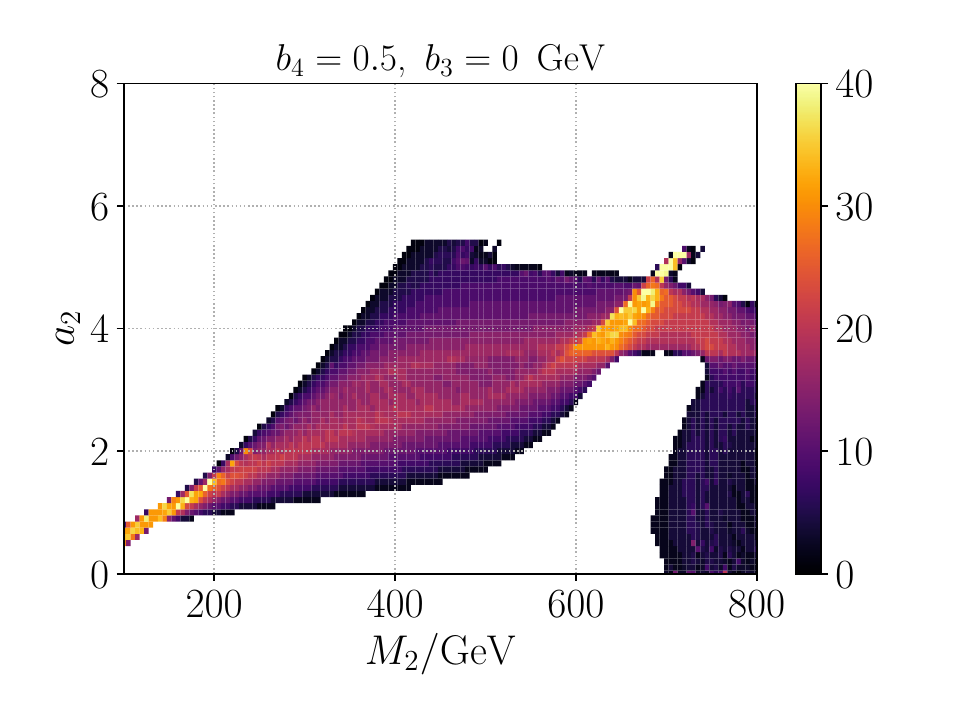}
	\shiftLeft
	\includegraphics[width=\figscale\textwidth]{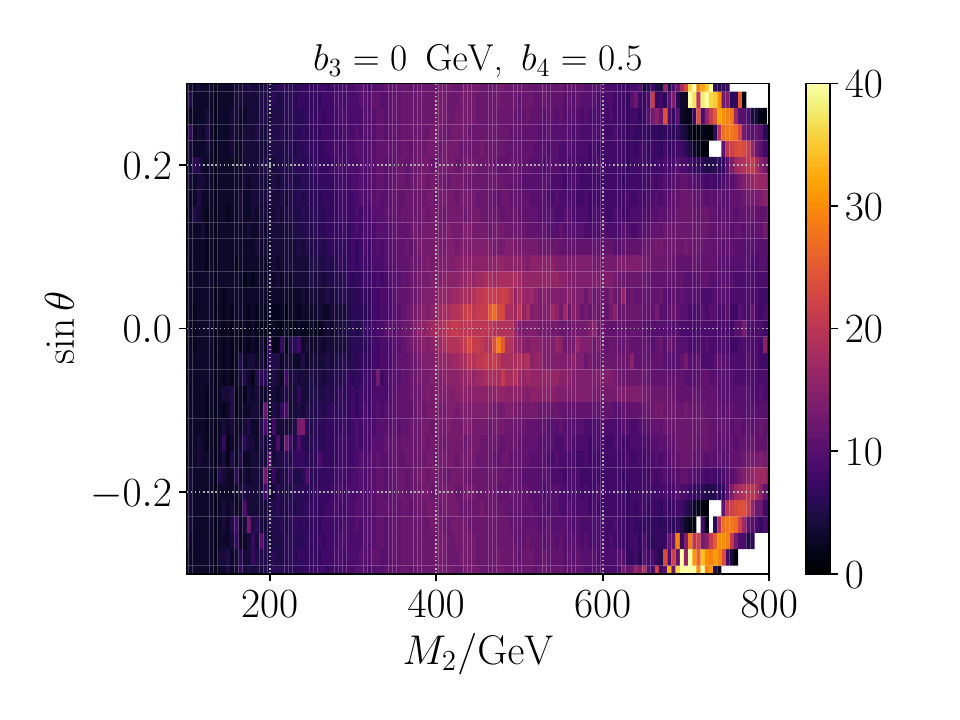}
	\shiftLeft
	\includegraphics[width=\figscale\textwidth]{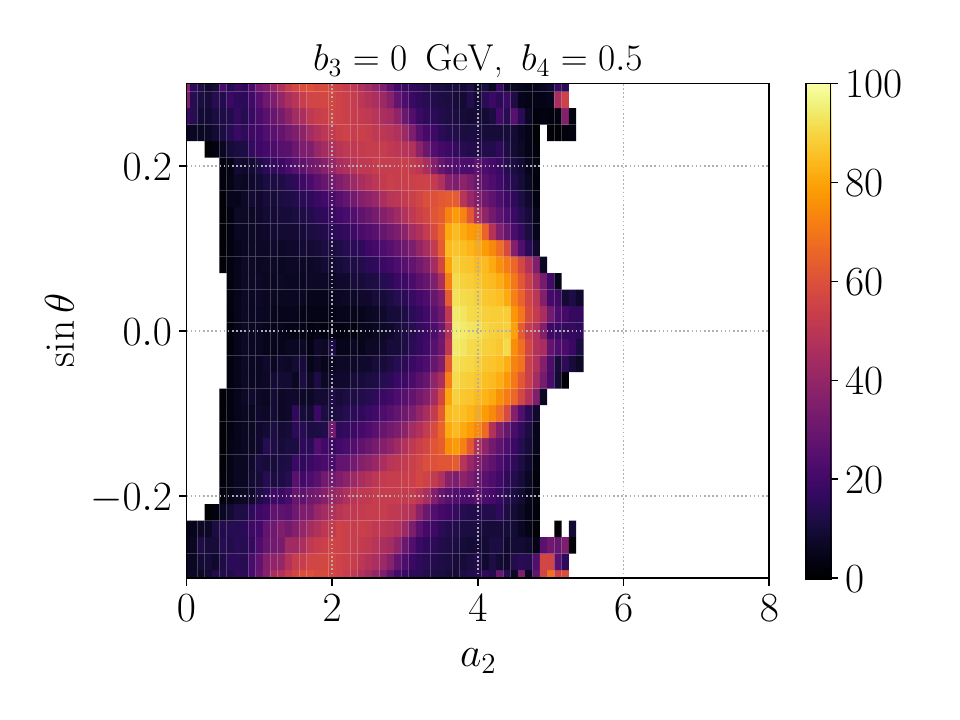}
	
	\smallLift
	\footnotesize{One-loop scan} \\
	\shiftLeft
	\includegraphics[width=\figscale\textwidth]{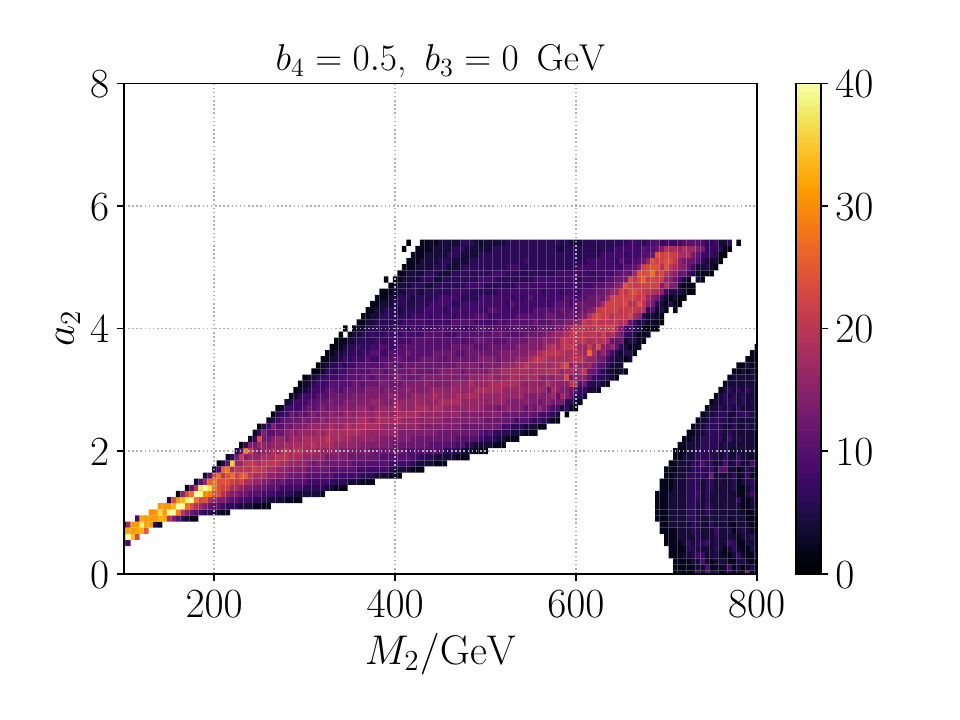}
	\shiftLeft
	\includegraphics[width=\figscale\textwidth]{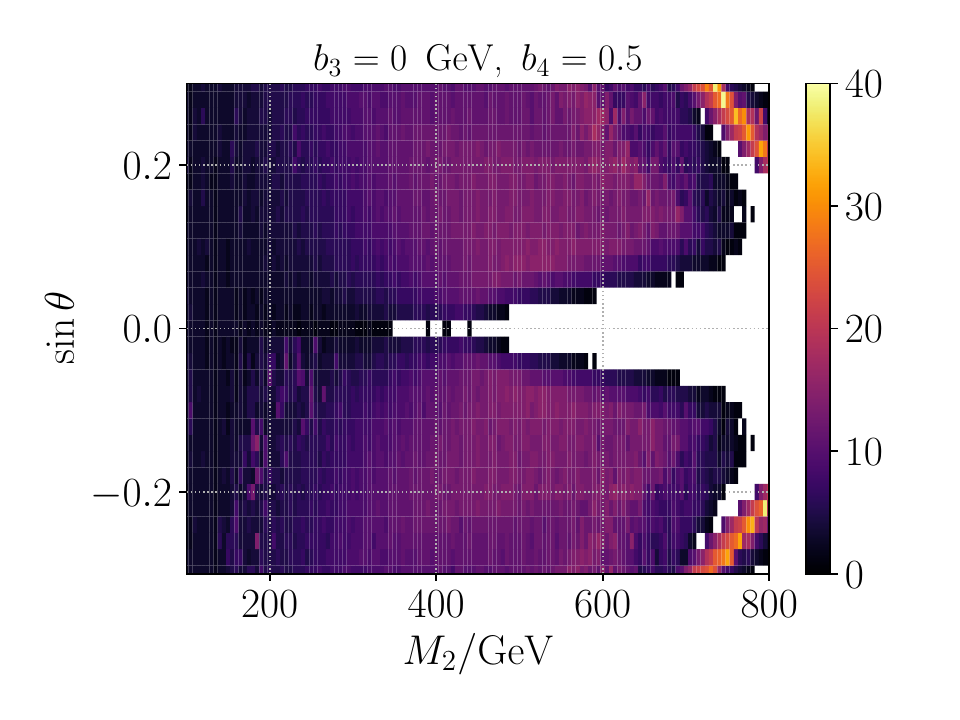}
	\shiftLeft
	\includegraphics[width=\figscale\textwidth]{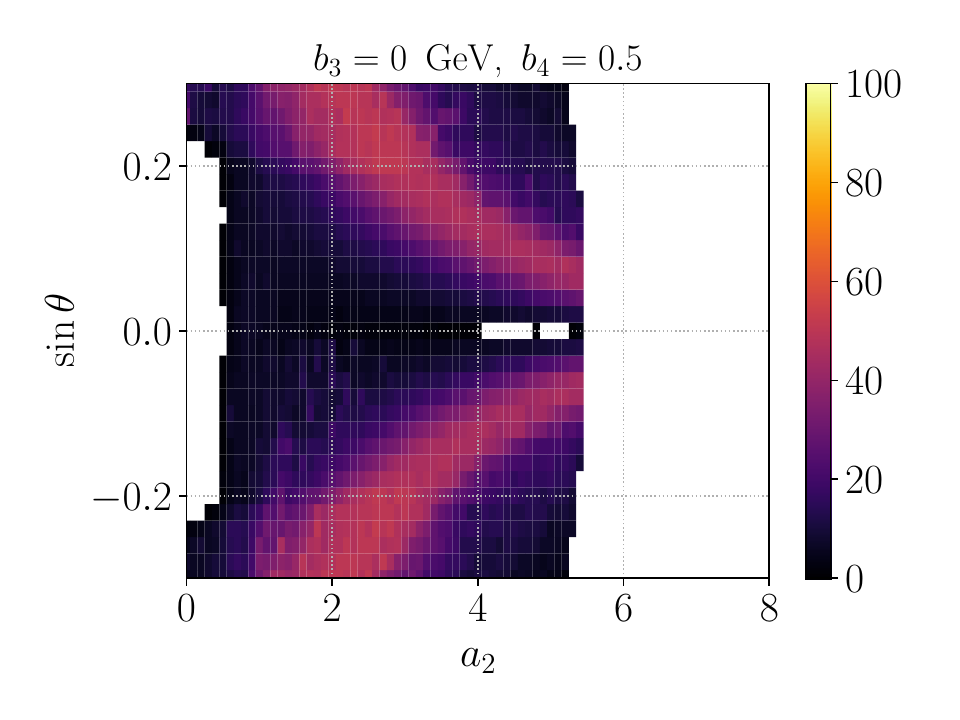}
	
	\smallLift
	\footnotesize{Two-loop scan}
	\caption{
		Two-dimensional projections of the three-dimensional $(M_2,a_2,\sin \theta)$ parameter space, such that the total number of points giving a first-order EWPT is plotted as a heatmap. Top row is one-loop scan, bottom row is two-loop.
	}
	\label{fig:FOPT-projected}
\end{figure*}

In terms of both $\Delta v/T$ and $L/T_c^4$, strongest transitions are found in the immediate vicinity of the region where $T=0$ electroweak vacuum becomes metastable (dark gray region).
This narrow band of strong transitions is precisely where one expects to find two-step phase transitions, ie. the $\Veff(v, x)$ minimum changes as $(0, x_1) \rightarrow (0, x_2)$ at some high temperature,
and the actual EWPT proceeds as $(0, x_2) \rightarrow (v, x_3)$ at the $T_c$ shown here. This two-step pattern can result in very strong EWPT, particularly near the $Z_2$ symmetric limit of the model where a tree-level potential barrier
between the symmetric and broken phases is generated by the earlier, singlet-specific transition; however the region of viable two-step transitions is known to be relatively narrow \cite{Kurup:2017dzf, Lewicki:2024xan}.
In the $\sin\theta \neq 0$ case studied here, the singlet background field is always discontinuous at $T_c$ and we have not tracked possible earlier transitions in the singlet-only direction,
however the heatmap of $\Delta x/T$ in Fig.~\ref{fig:field-jumps} reveals large singlet jumps in the bottom-left quadrant of $(M_2, a_2)$ parameter space, suggesting two-step pattern in this region. 
The switch from one-step to two-step behavior also explains the sudden increase in transition strength seen especially in bottom-left quadrants of the heatmaps.
Notably, comparison of one- and two-loop heatmaps in Fig.~\ref{fig:latent-comparison-st} suggests that strong two-step transitions are slightly rarer at two-loop level.

The two-loop heatmap of $L/T_c^4$ in the $\sin\theta = 0.18$ case (bottom right-hand plot of Fig.~\ref{fig:latent-comparison-st}) deserves further scrutiny.
At large $a_2 \gtrsim 4$, $M_2 \gtrsim 500$ GeV we find the latent heat to grow very rapidly with even small changes to the input parameters.%
\footnote{The heatmap scale is capped at $L/T_c^4 = 10$ to facilitate presentation. Near the region of nonperturbatively large couplings the values of $L/T_c^4$ range between 10 to 30. 
}.
To understand why this behavior appears peculiar, recall the relation between latent heat and field operator jumps, Eq.~\eqref{eq:latent-condensates}.
In Fig.~\ref{fig:field-jumps} that shows jumps of background fields, no rapid increase in $\Delta v/T_c$ or $\Delta x/T_c$ is observed that would explain the corresponding growth of $L/T_c^4$.
This suggests the effect originates from the other part of Eq.~\eqref{eq:latent-condensates}: the mapping between 4D and 3D descriptions, the quality of which depends on high-$T$ expansions.

We show in Fig.~\ref{fig:delta6} the error estimator $\delta_6$, defined in Eq.~\eqref{eq:delta6}, for the $\sin\theta = 0.18$ case. We see that the error due to neglecting higher orders of high-$T$ expansion stays small in most of the region with first-order transitions, less than $10\%$, and our results concerning the phase transition should be reliable there. However, for $a_2 \gtrsim 4$ the $\delta_6$ quickly grows to $0.2 - 0.5$, suggesting an error of several tens of percents in the location of $\Veff$ minimum. Unsurprisingly, the region of large $\delta_6$-error mostly overlaps with the region of peculiarly large latent heat, and we can conclude that the ``very strong transitions'' seen in Fig.~\ref{fig:latent-comparison-st} at large $M_2$ are likely not reliable. In order to realistically probe this, and any other region where $\delta_6$ is large, alternative approaches are needed that do not rely on high-$T$ approximations. This could be an interesting direction for future research.

It is also worth emphasizing that $\delta_6$ only becomes large when approaching the light-gray region, which is where perturbativity is lost already at $T=0$. A different pattern, however, is observed in the analogous error estimate for the singlet background field, $\delta_{6,x}$; this is defined as in \eqref{eq:delta6} but for $x_\text{min}$ instead of $v_\text{min}$. 
A heatmap of $\delta_{6,x}$ in the same $\sin\theta = 0.18$ plane is shown on the right-hand side of Fig.~\ref{fig:delta6}. The singlet-specific quantity $\delta_{6,x}$ is large also in the two-step region at small mass, $M_2 \lesssim 300$ GeV, overlapping with the region where the singlet VEV is seen to develop a large discontinuity at the EWPT (Fig.~\ref{fig:field-jumps}). We have not attempted to identify if any individual higher-dimensional operator is responsible for this effect. It is also possible that this behavior changes if higher-dimensional operators involving derivatives are included in the analysis, see \cite{Chala:2024xll} for a recent related work.
Additional plots similar to Figs.~\ref{fig:latent-comparison-st} and \ref{fig:delta6} are collected in Appendix \ref{sec:more-plots}.

Figs.~\ref{fig:Tc-comparison-st} and \ref{fig:latent-comparison-st} demonstrated that the two-loop potential predicts narrower regions of parameter space with strong transitions in these parameter planes. In fact, this result holds more generally: Fig.~\ref{fig:FOPT-projected} shows two-dimensional projections of the $(M_2,a_2,\sin \theta)$ parameter space, each subfigure being a heatmap of the total number of points that have a first-order EWPT in that plane.%
\footnote{
In all plots so far, we have fixed singlet self-couplings $b_3,b_4$ to specified values. This is because our scan over these parameters was considerably less dense than over $M_2, a_2, \sin \theta$. Figs.~\ref{fig:FOPT-projected} does not qualitatively change even if $b_3, b_4$ are also projected over.
} 
Top (bottom) row shows one-loop (two-loop) results, and in each plane the one-loop scan finds strong transitions in regions that are absent in the two-loop scan. The contrast is most stark around $M_2 = 750$ GeV and $a_4 = 4$ in the left-hand plot, as well as near the $\sin\theta = 0$ limit in the middle and right-hand plots. These results demonstrate that one-loop parameter scans may predict completely incorrect ``hot spots'' for strong EWPT. Regarding the $\sin\theta \approx 0$ region, it should be pointed out that a region of two-step transitions could still be present here (see eg.~\cite{Lewicki:2024xan}), but this region is too narrow to be visible in our scans.

\section{Discussion}
\label{sec:conclusions}

We have studied the electroweak phase transition in the non-$Z_2$ symmetric singlet-extended Standard Model, performing large-scale parameter space scans over free parameters of the model. 
This study is among the first parameter-space studies utilizing a two-loop thermal effective potential in any model of new physics, earlier works including \cite{Croon:2020cgk,Friedrich:2022cak,Gould:2023jbz,Lewicki:2024xan}. 
We have found that previous one-loop studies vastly overestimate results for critical temperature, 
and can lead to overly optimistic predictions for the number of parameter points that admit strong transitions.
The strongest transitions are found near parameter space regions where the EW minimum is metastable, and in these regions the two-loop analysis predicts significantly stronger transitions compared to one-loop.%
\footnote{ 
Our discussion in appendix~\ref{sec:impact} provides support that in the limit of heavy singlet, this is a result of
two-loop corrections contributing with opposite sign to the effective Higgs thermal mass and effective self-coupling, compared to one-loop terms.
This can lead to significantly smaller prediction for the critical temperature, and enhanced dimensionless ratios $L/T^4_c$ and $\Delta v_{\text{min}}/T_c$.
}

A central ingredient in computing the two-loop thermal potential (and also in a recent nonperturbative study of the model \cite{Niemi:2024axp}) is the high-temperature effective field theory (EFT) approach and related high-$T$ expansions. One of our goals in this work was to investigate the validity of this method in detail and provide generic error estimates for the EFT approach. This is especially important now that the modeling of cosmological phase transitions is entering a new era of precision study, largely in preparation for LISA-generation gravitational-wave detectors.

To study the numerical uncertainty inherent in the EFT method, we studied the effect of leading higher dimensional operators that are commonly neglected from the EFT. Our results indicate that plenty of strong transitions can be studied with only very small, percent-level inherent uncertainty if higher-dimensional operators are neglected. However, for the very strongest transitions found in our scans the corresponding uncertainty can reach several tens of percent. A proper two-loop study in these regions of the parameter space would require a new method of thermal resummation that does not rely on high-$T$ approximations. Complications of such a procedure have been discussed in \cite{Laine:2017hdk}.

In regards to model phenomenology of the singlet extension, it could be interesting to accompany our observation that one-loop scans seem to overestimate the number of valid strong-EWPT scenarios with proper statistical measures regarding the likelihood of such transitions in experimentally-probable regions of the parameter space, see eg.~\cite{Dorsch:2014qja,Chen:2017qcz,Biekotter:2023eil,Zhang:2023jvh}. Methodology presented in the paper at hand could be used to bring phenomenological phase-transition studies to two-loop level also in other BSM models. Recent such one-loop studies include  refs.~\cite{Kazemi:2021bzj,Cline:2021iff,Zhou:2022mlz,Anisha:2023vvu,Ma:2023kai,Ahriche:2023jdq,vonHarling:2023dfl,Goncalves:2023svb,Zhang:2023mnu,Addazi:2023ftv,Abe:2023zja,Biermann:2024oyy}.

\begin{acknowledgments}
We thank Andreas Ekstedt, Oliver Gould, Johan L{\"o}fgren, Michael Ramsey-Musolf, Kari Rummukainen, Philipp Schicho, Jasmine Thomson-Cooke, Van Que Tran, Yanda Wu, Guotao Xia and Jiang Zhu for discussions. This work was supported in part by National Natural Science Foundation of China grant 11975150. LN was additionally supported by Academy of Finland grants 320123, 345070 and 354572.
\end{acknowledgments}


\appendix

\section{Anatomy of the thermal effective potential}
\label{sec:anatomy}

Technology to compute two-loop thermal effective potential in high-temperature expansion has existed for a long time \cite{Farakos:1994kx}, yet its use is still not common practice. Hence, we give here a pedagogic review how to compute it, paying special attention to power counting, to ensure consistent use of perturbation theory \cite{Arnold:1992rz,Lofgren:2023sep}%
\footnote{Recent \cite{Ekstedt:2024etx} has pushed similar computations in gauge-Higgs theories to three loops, which defines the current state of the art.}
In order to speedrun through different constitutients of the thermal effective potential, we assume here the minimal case of just one scalar 
field, with mass $m^2$ and self-coupling $\lambda$. For adding gauge fields and fermions, see \cite{Farakos:1994kx,Kajantie:1995dw}. 

We denote corresponding scalar background field $\varphi$ and mass squared $M^2_\varphi = m^2 + 3 \lambda \varphi^2$ for the quantum fluctuation.
Formally, the effective potential reads
\begin{align}
\label{eq:Veff-formal}
V^{}_{\text{eff}} &= V^{}_{\text{tree}} + V^{\text{1-loop}}_{\text{eff}} + V^{\text{2-loop}}_{\text{eff}},
\end{align}
where the one-loop part has the renowned structure
\begin{align}
\label{eq:Vloop1-formal}
V^{\text{1-loop}}_{\text{eff}} &= V_{\text{CT}} + V_{\text{CW}} + V_{\text{T}} + V_{\text{daisy}},
\end{align} 
where different terms correspond to ultraviolet (UV) counterterms, Coleman-Weinberg potential at zero temperature, one-loop thermal function and daisy resummation, respectively.
Explicitly, in high-$T$ expansion in $M^2_\varphi/T^2$ (assuming $m^2 \sim \lambda T^2$ and $\varphi \sim T$)
\begin{align}
\label{eq:Vloop1}
V_{\text{CT}} + V_{\text{CW}} + V_{\text{T}} &= -\frac{T}{12\pi} (M^2_\varphi)^{\frac32} + \bigg( -\frac{\pi^2}{90}T^4 + \frac{T^2}{24} M^2_\varphi \nonumber \\
& - \frac{M^4_\varphi}{4(4\pi)^2} L_b + \frac{\zeta(3)}{3(4\pi)^4} \frac{M^6_\varphi}{T^2}  \bigg) + \mathcal{O}(\frac{M^8_\varphi}{T^4}),
\end{align} 
where $1/\epsilon$ poles at UV cancel between counterterms and the Coleman-Weinberg potential, and we use shorthand notation $L_b \equiv \ln\Big( \frac{\bar{\mu}^2}{(4\pi e^{-\gamma} T)^2}  \Big)$ for logarithm of the renormalisation scale (working in \MSbar scheme). 
Here $\gamma$ is the Euler–Mascheroni constant. 
The non-analytic $(M^2_\varphi)^{\frac32}$-term originates solely from the zero Matsubara modes \cite{Laine:2016hma}, while all the other temperature dependent terms come from non-zero Matsubara modes, and are analytic in $M^2_\varphi/T^2$ expansion. The daisy resummation is most often implemented as 
\begin{align}
\label{eq:daisy}
V_{\text{daisy}} &\equiv \frac{T}{12\pi} \Big( (M^2_\varphi)^{\frac32} -(\bar{M}^2_\varphi)^{\frac32} \Big),
\end{align} 
where $\bar{M}^2_\varphi \equiv M^2_\varphi + \Pi_1$ with one-loop thermal mass $\Pi_1 = T^2 \lambda/4 $. In this expression, the first term simply removes corresponding term in Eq.~\eqref{eq:Vloop1}, so that the mass in the zero mode propagator is effectively replaced by resummed propagator with thermally corrected mass $\bar{M}^2_\varphi$.

To see the reason behind such resummation, let us consider the scalar two-point correlator, for vanishing background field. 
At one-loop, at leading order in high-$T$ expansion this correlator reads%
\footnote{
Our definitions for sums over Matsubara frequencies and the integral measure coincide with \cite{Gorda:2018hvi}.   
}
\begin{align}
& \quad \quad \diagramA + \diagramB + \diagramC \nonumber \\
&= m^2 + 3 \lambda \bigg( \sum_{n \neq 0} T \int_p \frac{1}{(2\pi n T) + p^2 + m^2} + T \int_k \frac{1}{p^2 + m^2} \bigg) \nonumber \\
&= \underbrace{m^2 + \frac{T^2}{4}\lambda}_{\bar{m}^2} + \mathcal{O}(\lambda m^2) - \frac{T}{4\pi} (m^2)^{\frac32}.
\end{align}
Here we have highlighted the appearance of the one-loop corrected thermal mass $\bar{m}^2$ that contributes at leading order $\mathcal{O}(\lambda T^2)$. 
Other two terms are further suppressed,  $\mathcal{O}(\lambda^{2} T^2)$ and $\mathcal{O}(\lambda^{\frac32} T^2)$, respectively.
Since one-loop result from non-zero Matsubara modes contributes at same order as the tree-level mass, we need to resum this leading order contribution to the propagator of the zero mode when we compute loops with the zero mode. Diagrammatically, the zero mode propagator becomes
\begin{align}
\label{eq:resummed-propagator}
\diagramAA \equiv \diagramA + \diagramB,
\end{align}
where we denote resummed propagator, with mass $\bar{M}_\varphi$, by a dashed line.

Without this resummation, one would run into problems in accounting contributions of the zero Matsubara mode:
consider a following class of $(l+1)$-loop diagrams, where a zero Matsubara mode loop with momentum $K=(0,k)$ is attached by $l$ non-zero Matsubara mode loops each with momenta 
$Q^2 = (2\pi n T)^2 + q^2$, where $n \neq 0$. Diagrammatically
\begin{align}
\daisydiagram
\end{align}
where progator with momentum $k$ is denoted by a solid line, and loops with momenta $Q$ by double lines (and we have merely illustrated the $l=5$ case).
In general, this kind of $(l+1)$-loop \textit{daisy diagram} factorizes to one-loop topologies, and we can easily sketch its structure at leading order in high-$T$ expansion: (we omit overall numerical factors for simplicity)
\begin{align}
& \bigg(\lambda \sum_{n\neq 0} T \int_q \frac{1}{(2\pi n T)^2 + q^2 + M^2_\varphi}\bigg)^l T \int_k \frac{1}{(k^2+M^2_\varphi)^l} \nonumber \\
& \approx  (\lambda T^2)^l T (M^2_\varphi)^{\frac32 -l}/\pi = \frac{(M^2_\varphi)^{\frac32}}{\pi} T \Big( \frac{\lambda T^2}{M^2_\varphi} \Big)^l \sim \frac{\lambda^{\frac32}}{\pi} T^4.
\end{align}  
Here we utilized the fact that $n\neq 0$, so all $q$-integrals are safe at infrared (IR)
-- ie. Matsubara frequencies $2\pi n T$ regulate the propagator at IR  -- and hence analytic in $M^2_\phi$. We can therefore expand in $M^2_\varphi/T^2$ and simply keep the leading $M^2_\varphi$-independent part. Integral over $k$, however, is IR sensitive and leads to non-analytic fractional powers of $M^2_\varphi$. Furthermore, the base of power $l$ scales as $\lambda T^2/M^2_\varphi \sim \mathcal{O}(1)$ so irrespective of number of loops $l$, this class of diagrams contributes as $\mathcal{O}(\lambda^{\frac32} T^4)$ at high-$T$ regime $M^2_\varphi \sim \lambda T^2$. All these $(l+1)$-loop daisy diagrams are less suppressed than the one-loop Coleman-Weinberg potential which scales as $\mathcal{O}(\lambda^{2} T^4)$, and contribute at next-to-leading order, compared to the leading order potential, ie. tree-level potential with thermally corrected mass $(m^2+\Pi_1) \phi^2/2 + \lambda \phi^4/4 \sim \mathcal{O}(\lambda T^4)$.  

Using the resummed propagator of Eq.~\eqref{eq:resummed-propagator} for the zero mode in computation of one-loop vacuum diagram,
one includes an infinite Dyson series of $(l+1)$-loop daisy diagrams, formally
\begin{align}
\label{eq:daisies}
\sum_{l=0}^\infty \daisydiagram \rightarrow \diagramX = -T(\bar{M}^2)^{\frac{3}{2}}/(12\pi).
\end{align}
This is exactly the daisy resummed term in Eq.~\eqref{eq:daisy}, 
and bears the name of our favorite flower. While it technically contains an infinite series of infinite-order diagrams, all such diagrams factorize to the simple one-loop structure. Hence, despite the resummation, it is convenient to label this contribution as ``one-loop'' in Eq.~\eqref{eq:Vloop1-formal}, especially since it is parametrically larger than the Coleman-Weinberg potential.      

Often in the literature, accounting for Eq.~\eqref{eq:daisies} is referred as resumming ``the most dangerous'' diagrams. Simply put, this is to account leading -- one-loop -- contribution from the zero Matsubara mode, which is screened by excitations of much heavier non-zero Matsubara modes at high temperature regime. Hence, the effect from the zero mode is enhanced. This resummation can be made systematic, and generalised to all-orders, using effective field theory, as discussed in sec.~\ref{sec:EFT}.

Returning to the effective potential, we obtain after a slight re-organization
\begin{align}
V^{}_{\text{tree}} + V^{\text{1-loop}}_{\text{eff}} &= \frac{1}{2}\Big( m^2(\mu) + \Pi_1(\mu) -\frac{3 \lambda m^2}{(4\pi)^2} L_b \Big) \phi^2 \nonumber \\
&+ \frac{1}{4} \bar{\lambda} \phi^4 - \frac{1}{12\pi} (\bar{M}^2)^{\frac{3}{2}} + \frac{\zeta(3)}{3(4\pi)^4} \frac{M^6}{T^2} \nonumber \\
& + \mathcal{O}(\frac{M^8}{T^4}), 
\end{align} 
where $\bar{\lambda} \equiv \lambda \Big( 1 - 9 \lambda L_b/(4\pi)^2 \Big)$,
and we have for simplicity omitted field independent terms. 
In terms of power counting
\begin{align}
V^{}_{\text{tree}} + V^{\text{one-loop}}_{\text{eff}} &\sim T^4 \bigg(\lambda + \frac{\lambda^{\frac{3}{2}}}{\pi} + \frac{\lambda^2}{(4\pi)^2} + \mathcal{O}\Big(\frac{\lambda^3}{(4\pi)^4} \Big) \bigg),
\end{align}
As we have already seen,
the daisy contribution is enhanced due to IR behaviour and is parametrically larger than one-loop contribution at zero temperature, 
hence describing the leading thermal corrections at high temperature.  
The crux of this discussion, however, is realization that the one-loop effective potential is \textit{still incomplete} at $\mathcal{O}(\lambda^2 T^4)$. 
To demonstrate this, we add effects of two-loop thermal mass and two-loop contributions from the zero mode, ie. the generalization of the daisy term. 
This results \cite{Schicho:2021gca}
\begin{align}
\label{eq:Vloop2}
V^{\text{2-loop}}_{\text{eff}} &= \diagramD + \diagramE + \diagramY + \diagramZ \nonumber \\
&= \frac{\phi^2 T^2}{2(4\pi)^2} \bigg( -\frac{9}{4}\lambda^2 L_b - 6 \bar{\lambda}^2 \Big[c + \ln\Big(\frac{3T}{\bar{\mu}_3} \Big) \Big] \bigg) \nonumber \\
&+ \frac{T^2}{(4\pi)^2} \frac{3}{4}\bigg( \bar{\lambda} \bar{M}^2 -2 \bar{\lambda}^2 \phi^2 \Big[ 1 + 2 \ln\Big( \frac{\bar{\mu}_3}{3 \bar{M}} \Big) \Big] \bigg) \nonumber \\
& \sim \mathcal{O}(\lambda^2 T^4),
\end{align}
where 
$c \equiv \frac12 \Big( \ln \frac{8\pi}{9} + \frac{\zeta'(2)}{\zeta(2)} -2 \gamma \Big)$ \cite{Kajantie:1995dw} (here $\zeta$ is the Riemann zeta function)
and the second line equals $\Pi_2 \phi^2/2$, where $\Pi_2$ is the two-loop contribution to the thermal mass $\bar{m}^2_{\text{2-loop}} = m^2 + \Pi_1 -3 \lambda m^2 L_b/(4\pi)^2 + \Pi_2$. Note that the external lines describe insertions of the background field $\varphi$.
The term $-3 \lambda m^2 L_b/(4\pi)^2$ is typically absorbed to $\Pi_2$ despite it appears at one-loop, but next-to-leading order in high-$T$ expansion, making it same order as the actual two-loop piece. 
It is also worth mentioning that the automated code \DRALGO \cite{Ekstedt:2022bff} provides a slightly different form $\Pi_{2,\DRALGO} \equiv T^2 \Big(\frac{3}{4}T^2 \lambda^2 (-4 \gamma + L_b + 48 \ln A) - 6 \bar{\lambda}^2 \ln\Big(\frac{\bar{\mu}}{\bar{\mu}_3} \Big)/(4\pi)^2$, where $A$ is the Glaisher–Kinkelin constant. 
This form agrees with $\Pi_2$ defined through Eq.~\eqref{eq:Vloop2}
when substituting leading order relation $\bar{\lambda} \rightarrow \lambda$, ie. the difference $\Pi_2 - \Pi_{2,\DRALGO}$ is of higher order than target accuracy $\mathcal{O}(\lambda^2 T^2)$.%
\footnote{
In $\Pi_2$, we have added formally higher order corrections using the renormalization running, as explained in \cite{Kajantie:1995dw,Schicho:2021gca}.  
Note also that the appearing constants are related through $-\zeta'(2) = \frac{\pi^2}{6}(12 \ln A - \gamma - \ln 2\pi)$.
} 
The third line in Eq.~\eqref{eq:Vloop2} comes from the two-loop potential within the EFT \cite{Farakos:1994kx}.
The dependence of the EFT renormalization scale $\bar{\mu}_3$ drops out due to cancellation of running of the mass in second line and explicit logarithm on the third line.

Final result up to and including two-loop reads
\begin{align}
\label{eq:Vfinal}
V^{}_{\text{eff}} &= V^{}_{\text{tree}} + V^{\text{1-loop}}_{\text{eff}} + V^{\text{2-loop}}_{\text{eff}} \nonumber \\
&=  \frac{1}{2} \bar{m}^2_{\text{2-loop}} \varphi^2 + \frac{1}{4} \bar{\lambda} \varphi^4 - \frac{1}{12\pi} (\bar{M}^2_\varphi)^{\frac{3}{2}} \nonumber \\
&+ \diagramY + \diagramZ
+ \frac{1}{8} \bar{c}_6 \frac{\varphi^6}{T^2},
\end{align}
where two-loop diagrams represents the third line of Eq.~\eqref{eq:Vloop2}, and
$\bar{c}_6 \equiv 72 \zeta(3) \lambda^3/(4\pi)^4$.
The interpretation of $\bar{\lambda}$ and $\bar{c}_6$ is clear: 
they are parameters within the EFT, that receive thermal corrections from the UV scale of non-zero Matsubara modes, in exact analogy to thermal mass, via 
\begin{align}
\diagramXX + \diagramYY,
\end{align}
diagrammatically. We note that two-loop corrections to $\bar{\lambda}$ is of higher order than our target accuracy.

Eq.~\eqref{eq:Vfinal} is the complete result for the effective potential at order $\mathcal{O}(\lambda^2 T^4)$.
The contribution from leading marginal operators,  the term with coefficient $\bar{c}_6$ in Eq.~\eqref{eq:Vfinal}, is $\mathcal{O}\Big( \frac{\lambda^3}{\pi^4} T^4 \Big)$ and kept  
merely to scrutinize our use of high-$T$ expansion; it is not the complete result at such high order.
First corrections to Eq.~\eqref{eq:Vfinal} appear at three loops, and are parametrically $\mathcal{O}\Big( \frac{\lambda^{\frac{5}{2}}}{\pi^3} T^4 \Big)$ \cite{Rajantie:1996np, Gould:2021dzl, Gould:2023jbz}.

We have thus observed that each new order in the expansion of the effective potential at high temperature is suppressed by factor $\lambda^{\frac12}/\pi$. 
Indeed, the convergence of the perturbative expansion at high temperature is \textit{slower} compared to a similar expansion at zero temperature, where each new order is suppressed by factor $\lambda/\pi^2$. 

Using renormalization group evolution in terms of $\beta$-functions
$\bar{\mu} \frac{d}{d\bar{\mu}} m^2 = 6\lambda m^2/(4\pi)^2$ and $\bar{\mu} \frac{d}{d\bar{\mu}} \lambda = 18\lambda m^2/(4\pi)^2$
it is straightforward to check that Eq.~\eqref{eq:Vfinal} is independent of RG scale $\bar{\mu}$ at target accuracy $\mathcal{O}(\lambda^2 T^4)$:
\begin{align}
\bar{\mu} \frac{d}{d\bar{\mu}} V^{}_{\text{eff}} = \mathcal{O}(\lambda^{\frac{5}{2}}),
\end{align}
simply because $\bar{\mu}$-dependence cancels identically in $\bar{m}^2$ and $\bar{\lambda}$. 
Here it is crucial that the running inside the one-loop thermal mass is cancelled by the explicit logarithm from two loops, ie. 
\begin{align}
\bar{\mu} \frac{d}{d\bar{\mu}} \Big( \Pi_1[\lambda(\bar{\mu})] + \Pi_2[\ln(\bar{\mu})]  \Big) = \mathcal{O}(\lambda^{\frac{5}{2}}).
\end{align}
Finally, we note that the running in terms of $\bar{\mu}_3$ inside the one-loop daisy correction is of higher order than our target accuracy.

As we have demonstrated, convenient labels one-loop and two-loop in Eq.~\eqref{eq:Veff-formal} are ambiguous when resummations and high-temperature expansions are involved. 
In our nomenclature used throughout this article, one-loop refers to a daisy-resummed 
$\mathcal{O}(\lambda_i^{\frac32})$ 
calculation in any quartic couplings $\lambda_i$ with partial 
$\lambda^2_i$ 
terms included from $T=0$ renormalization,
and two-loop is full 
$\mathcal{O}(\lambda^2_i)$.
In general models, the appeal of limiting to a one-loop thermal effective potential of Eq.~\eqref{eq:Vloop1} is a simple closed formula for it, in terms of a master formula involving a summation over mass eigenvalues for all fields \cite{Quiros:1999jp}. 
The fact that the ``daisy resummation'' corresponds to the one-loop computation in a dimensionally reduced EFT is often lost in translation.
Unfortunately, there is no similar simple recipe for a two-loop effective potential in analogy to the closed formula at one-loop level. 
However, \DRALGO code \cite{Ekstedt:2022bff} automates such two-loop computations for generic models.
Since two-loop thermal corrections are parametrically as important as one-loop zero temperature quantum corrections, ie. 
$\mathcal{O}\Big(\frac{\lambda^2}{(4\pi)^2} \Lambda^4 \Big)$ (with $\Lambda = m^4$ at zero and $\Lambda=T$ at high temperatures), 
they can be expected to be large in BSM applications, where large portal couplings to Higgs are required for first-order phase transitions.

\section{Assessing the impact of two-loop contributions}
\label{sec:impact}

In the previous section we reviewed the general structure of the thermal effective potential. Let us now turn back to singlet extended Standard Model and investigate the impact of two-loop contributions, via analytic expressions within the EFT. 
Unlike in the rest of the article, in this appendix we define scalar background fields $v_3$, $x_3$ in 3D units, ie. they have mass dimension one half.
In general, an analytic treatment is not straightforward due to the complicated, nested structure of the effective potential. However, we can gain insights by following exercise: 
let us assume that we are close to $Z_2$-symmetric limit, and that the background field of the singlet does not change significantly across the transition.
This allows us to set $x_3 = 0$.
In this approximation terms with cubic scalar couplings have minor effects and we can set $\bar{b}_{3}=\bar{a}_{1} = 0$, as well as ignore the mixing of two scalars within the EFT, which greatly simplifies analytic formulae below. 

In addition, we assume that the singlet is significantly heavier than the Higgs and the gauge bosons within the EFT, ie.
\begin{align}
M_{S,3} \gg M_{\phi,3}, M_{W,3}, M_{Z,3},
\end{align} 
where scalar masses within the EFT have simple expressions $M^2_{S,3} = \bar{m}^2_{S} + \frac12 \bar{a}_{2} v^2_3$ and $M^2_{\phi,3} = \bar{m}^2_{\phi} + \bar{\lambda}_{} v^2_3$ at tree-level, given that singlet backgroud field $x_3$ vanishes. 
With these assumptions, we can integrate out the heavier singlet, and construct an EFT where only the Higgs and gauge bosons remain.
Hence, our strategy is to
\begin{itemize}

\item[(i)] find matching relations for the final Higgs-gauge EFT,

\item[(ii)] review how thermodynamics in the final EFT depend only on two dimensless ratios $y$ and $x$ (defined below), which control critical temperature and phase transitions strength, respectively.

\item[(iii)] find out how two-loop contributions of the singlet effect $y_c = y(T_c)$ and $x_c = x(T_c)$ at critical temperature.  

\end{itemize}
This strategy is analogous to the approach used in \cite{Brauner:2016fla,Gould:2019qek}, with a following difference: while these references integrated the heavy singlet out along with non-zero Matsubara modes in dimensional reduction, in our case the zero Matsubara mode of the singlet has been kept in construction of the first 3D EFT, but is then integrated out (in analogy to Debye screened zero Matsubara modes of temporal gauge fields).  
We note that this EFT approach describes exactly same thermodynamics as the thermal effective potential of sec.~\ref{sec:Veff}, within the aforementioned approximations. 
Yet, we have found working within the EFT framework conceptually more clear than directly inspecting the effective potential. 

Starting with (i), we find following matching relations 
\begin{align}
\label{eq:soft-corrections-mass}
\widetilde{m}^2_{\phi} &= \bar{m}^2_{\phi} - \frac{\bar{a}_{2} \sqrt{\bar{m}^2_{S}}}{8\pi} \nonumber \\ 
&- \frac{1}{(4\pi)^2} \frac{1}{4} \bigg[ \bar{a}^2_{2} + 2 \bar{a}^2_{2} \ln\Big( \frac{\bar{\mu}_3}{2 \sqrt{\bar{m}^2_{S}}} \Big) + 3 \bar{a}_{2} \bar{b}_{4} \bigg]  , \\
\label{eq:soft-corrections-coupling}
\widetilde{\lambda}_3 &= \bar{\lambda}_3 - \frac{\bar{a}^2_{2}}{32 \pi \sqrt{\bar{m}^2_{S}}} + \frac{1}{(4\pi)^2} \frac14 \frac{\bar{a}^3_{2}}{\bar{m}^2_{S}} , 
\end{align}
where parameters of the final EFT are denoted with tilde. Note that the singlet does not affect the gauge coupling so $\widetilde{g}^4_{} = \bar{g}^4_{}$.
Parameters denoted by bar can be read from \cite{Niemi:2021qvp}.  
In this appendix, we do not consider the effect to the marginal operator $(\phi^\dagger\phi)^3$.
Eqs.~\eqref{eq:soft-corrections-mass} and \eqref{eq:soft-corrections-mass} can be found using \DRALGO \cite{Ekstedt:2022bff}, except the very last term, which results from the diagrams of the form
\begin{align}
\diagramZZ
\end{align}
where dashed (solid) line represents Higgs (singlet). These contributions can be found using the effective potential in \cite{Niemi:2021qvp}. 
Current version of \DRALGO cannot be used to find such two-loop contributions to effective couplings, and indeed typically such terms have not been included \cite{Gorda:2018hvi,Niemi:2018asa} (for an exception, see recent \cite{Ekstedt:2024etx}), which is a clear oversight, given that this contribution is parametrically $\mathcal{O}(g^4)$ and hence within our target accuracy. 
Furthermore, since we want to investigate the difference between one- and two-loop results, it is crucial to include this contribution.

Turning to (ii), the infrared behaviour of 3D EFT for the Higgs with ${\rm SU}(2)$ gauge field is well understood in terms of two dimensionless ratios 
\begin{align}
y \equiv \frac{\widetilde{m}^2_{\phi} }{\widetilde{g}^4_{}}, \quad
x \equiv \frac{\widetilde{\lambda}_{} }{\widetilde{g}^2_{}}, 
\end{align}
For simplicity, in our discussion we can omit the contributions related to ${\rm U}(1)$ subgroup \cite{Kajantie:1996qd}. 
Essentially, $y$ and $x$ are just dimensionless versions of the effective Higgs mass and the self-coupling, respectively. 
Qualitatively, the critical temperature corresponds to small $y \approx 0$ and the transition strength is inversely proportional to $x^2$. 

Quantitatively, the critical line $y(x_c) = y_c$ is known at non-perturbative level from lattice simulations of \cite{Farakos:1994xh,Kajantie:1995kf,Gould:2022ran}, but it is straightforward to understand it also perturbatively \cite{Arnold:1992rz,Ekstedt:2022zro,Ekstedt:2024etx} (apart from the fact that perturbation theory is unable predict the end point of the transition).
The leading order effective potential describing a first-order transition comprises tree-level potential accompanied with one-loop contribution of the gauge field
\begin{align}
\widetilde{V}_{\text{eff}} &= \frac12 \widetilde{m}^2_{\phi} v^2_3  + \frac14 \widetilde{\lambda} v^4_3 -  \frac{\widetilde{g}^3}{16 \pi} v^3_3, \nonumber \\
&= \widetilde{g}^6 \bigg( \frac12 y \varphi^2_3  + \frac14 x \varphi^4_3 -  \frac{1}{16 \pi} \varphi^3_3 \bigg),
\end{align}
where dimensionless field $\varphi_3 \equiv \widetilde{g}^{-1} v_3$.
Using extremization condition 
$ \frac{d}{d\varphi_3} \widetilde{V}_{\text{eff}} = 0$,  
the minimum in the Higgs phase reads
\begin{align}
\varphi_{3,\text{min}} = \frac{3+\sqrt{9 - 1024 \pi^2 x y}}{32 \pi x}. 
\end{align}
In the symmetric phase the background field and the potential simply vanish, so the requirement of degenerate minima at critical temperature leads to $\widetilde{V}_{\text{eff}}(\varphi_{3,\text{min}}) = 0$, which yields the critical value for each $x$
\begin{align}
y_c(x) = \frac{1}{128 \pi^2 x}. 
\end{align}
The critical temperature can be solved from $y(T_c) = y_c$. At the critical temperature $\varphi_{3,\text{min}} = 1/(8\pi x)$ and thus we see that small $x$ corresponds to a large background field at $T_c$, ie. also to large $v_{\text{min}}/T_c$.
In other words, strong transitions occur at small values of $x$, that correspond to higher barrier separating the minima. An alternative way to see this is to inspect latent heat 
\begin{align}
L &= -T \frac{d}{dT} (T \Delta \widetilde{V}_{\text{eff}} ) = -T^2 \sum_i \frac{d\kappa_i}{dT} \Delta \frac{\partial \widetilde{V}_{\text{eff}}}{\partial \kappa_i} \nonumber \\
&\approx T \widetilde{g}^6 \eta(y) \Delta l_3, 
\end{align}
where we assumed $T=T_c$ so $\Delta \widetilde{V}_{\text{eff}}$ term (without derivative) vanishes, $\kappa_i$ runs through $\widetilde{m}^2_{\phi}$, $\widetilde{\lambda}$ and $\widetilde{g}^2$ (due to chain rule) and we kept only the most dominant term proportional to $\eta(y) \equiv T dy/dT$, c.f. \cite{Kajantie:1995kf}. 

From dimensionless form of the scalar condensate $\Delta l_3 \equiv \Delta \langle \phi^\dagger \phi \rangle/\widetilde{g}^6 \equiv \widetilde{g}^{-6} \Delta d\widetilde{V}_{\text{eff}}/dy = 1/(128\pi^2 x^2)$ we observe that at leading order in perturbation theory the latent heat -- at critical temperature -- is inversely proportional to $x^2$, c.f.~\cite{Kajantie:1997hn,Gould:2019qek}.

Finally, turning to (iii) we inspect how the singlet two-loop terms affect $y$ and $x$.
As a benchmark reference, we use a point with $M_2 = 400$ GeV, $a_2 = 4$ and $b_4 = 0.5$, $b_3 = 0$ and $\sin \theta = 0.02$, for which it is straightforward to check that aforementioned assumptions hold, ie. it is possible to integrate out heavier singlet and effects of cubic terms (proportional to $a_1$ and $b_3$) are small.
We have checked that following observations hold fairly generally for other points (that satisfy the same assumptions from above).

\begin{figure*}[t]
	\centering
	\begin{subfigure}[b]{0.49\textwidth}
		\centering
		\includegraphics[width=1.0\textwidth]{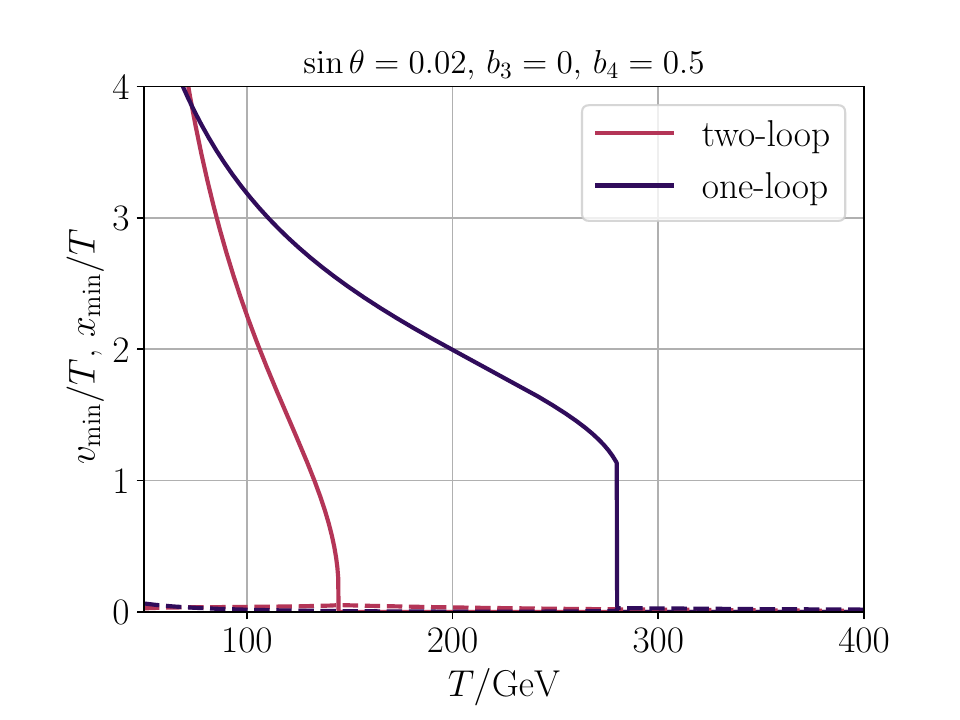}
	\end{subfigure}
	\hfill
	\begin{subfigure}[b]{0.49\textwidth}
		\centering
		\includegraphics[width=1.0\textwidth]{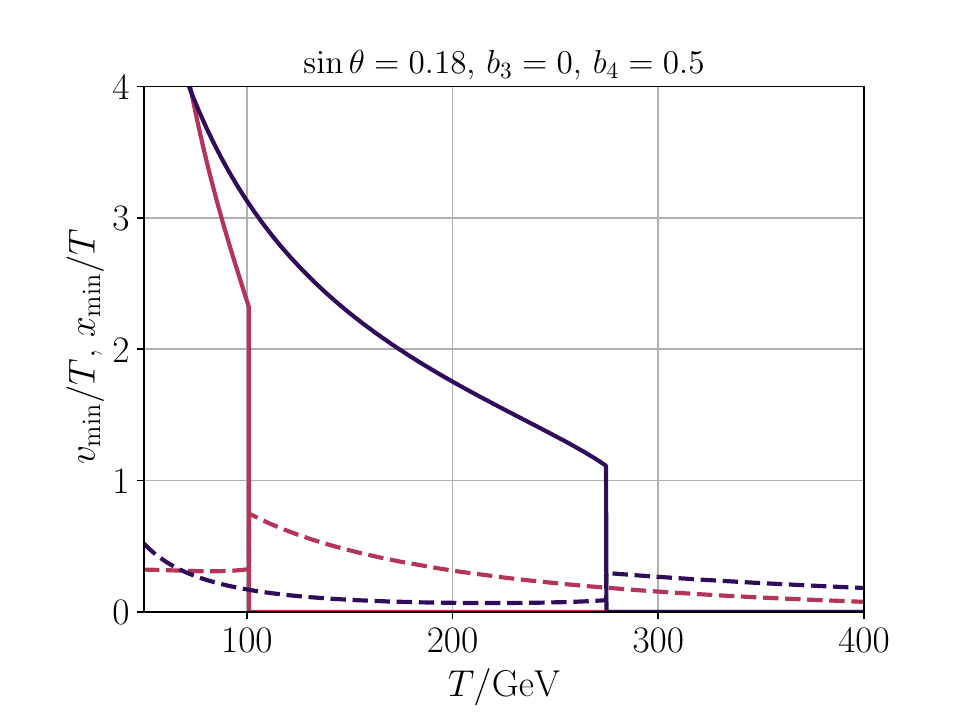}
	\end{subfigure}
	\caption{
Evolution of minima as function of temperature for benchmark points with two different $\sin \theta$ from Figs.~\ref{fig:Tc-comparison-st} and \ref{fig:latent-comparison-st}, with fixed $a_2 = 4$ and $M_2 = 600$ GeV. Solid (dashed) lines depict $v_\text{min}$ ($x_\text{min}$) at one-loop (purple) and two-loop (magenta). Discontinuity in the singlet direction is much smaller than in the Higgs direction, and is barely visible in the left panel. 
}
\label{fig:vByT-sample}
\end{figure*}

For starters, we note that the singlet mass parameter has a formal structure
\begin{align}
\bar{m}^2_{S} = m^2_{S} +  \Pi_{S,1} + \Pi_{S,2},
\end{align}
in which the vacuum mass $m^2_{S}$ dominates over thermal corrections $\Pi_{S,1}$ and $\Pi_{S,2}$ originating from non-zero Matsubara modes, at one and two loops, respectively (their expressions can be found in \cite{Niemi:2021qvp}).
This indicates that thermal corrections in $\Pi_{S,2}$ are not our main suspect for causing large difference between one- and two-loop results in sec.~\ref{sec:results}.  

Similarly, the Higgs self-coupling of Eq.~\eqref{eq:soft-corrections-coupling} has a formal structure%
\footnote{
In addition, there are contributions from temporal gauge field components, yet their effects are numerically subdominant and we can ingore them in our discussion here.
}
\begin{align}
\label{eq:tilde-lam}
\widetilde{\lambda}_{3} &= \bigg[T \lambda(\bar{\mu}) - \frac{\bar{a}^2_{2}}{32 \pi \sqrt{\bar{m}^2_{S}}} \bigg]_{\text{1-loop}} \nonumber \\
& + \bigg[ \Gamma_{1}[L_b(\bar{\mu})] + \frac{1}{(4\pi)^2} \frac14 \frac{\bar{a}^3_{2}}{\bar{m}^2_{S,3}} \bigg]_{\text{2-loop}} ,
\end{align}
where $\Gamma_1$ encodes thermal contributions from non-zero Matsubara modes.%
\footnote{
Note that $\Gamma_1 \sim \mathcal{O}(g^4)$ is counted as two-loop contribution, despite it diagrammatically composes from one-loop diagrams.  
}
Let us break down magnitudes and relative signs of different contributions in Eq.~\eqref{eq:tilde-lam}.
We emphasize that singlet affects the Higgs self-coupling (as well as other \MSbar scheme parameters) already at zero temperature, ie.
\begin{align}
\label{eq:MSbar-lam}
\lambda = \lambda_{\text{tree}} + \lambda_{\mathcal{O}(g^4)}, 
\end{align}
where tree-level term is given by Eq.~\eqref{eq:params-tree2} and the correction at $\mathcal{O}(g^4)$ can be read from appendix A in \cite{Niemi:2021qvp}.
This correction (which is counted as two-loop level, despite it appears diagrammatically at one-loop in zero temperature) \textit{reduces} $\lambda$, due to singlet contributions. On the otherhand, we have highlighted in Eq.~\eqref{eq:tilde-lam} that $\lambda = \lambda(\bar{\mu})$ is running. In our numerical study when we run parameters to scale $\bar{\mu} = 700$ GeV, we find that despite the singlet reducing the initial value, the final value for $\lambda$ at scale 700 GeV is not far off its tree-level value. Since we use same values and running for $T=0$ parameters at one and two loops, as explained in \ref{sec:analysis}, these effects cannot explain the difference of the results in the two scans. Nonetheless, we emphasize that loop corrections in Eq.~\ref{eq:MSbar-lam} are important, and need to be accounted at two-loop level.   

In thermal correction $\Gamma_1$, the singlet contribution is simply $-\frac{L_b}{(4\pi)^2} \frac{a^2_2}{4}$ \cite{Niemi:2021qvp}, and is large compared to the SM parts in $\Gamma_1$. Despite the explicit negative sign in singlet contribution, $\Gamma_1$ increases $\widetilde{\lambda}_{3}$, since $L_b = \ln\Big( \frac{\bar{\mu}^2}{(4\pi e^{-\gamma} T)^2}  \Big) < 0$, ie. is negative at relevant temperature range around $T_c \approx 144$ GeV and $\bar{\mu}=700$ GeV. 
We higlight that in Eq.~\eqref{eq:tilde-lam} the $\bar{\mu}$-dependence due to running of $\lambda(\bar{\mu})$ cancels with explicit logarithm in $\Gamma_1$ within our target accuracy at two loops. 

Overall, both terms in the two-loop level part of \eqref{eq:tilde-lam} contribute with positive sign and are of similar magnitude, while one-loop term comes with negative sign. In our numerics, we observe good convergence: two-loop terms are suppressed compared to one-loop correction, that describes largest correction to the dominant tree-level term.  

We can thus deduce how singlet corrections modify effective Higgs self coupling (and hence $x$): the SM part is large, and leads to 
$x_c \gg x_* \sim 0.1$ resulting a crossover \cite{Kajantie:1995kf,Gurtler:1997hr,Gould:2022ran}.%
\footnote{
Note that in the SM at leading order $x = \lambda/g^2$ and larger Higgs mass results in larger $\lambda$, inflating $x_c$ away from possible first-order regime. 
}
To get into the regime $0 < x_c < x_*$ where transition is of first order, singlet contributions must come with negative sign and be large enough in magnitude, and this is exactly what happens at one-loop. This is the main mechanism to induce first-order transition in the limit of a heavy BSM field that couples to the Higgs through strong enough portal: the new field effectively decreases thermal self-interaction coupling of the Higgs, allowing a first-order transition. 
Two-loop contributions increase $x$, and hence dilute the one-loop effect, seemingly weakening the transition.
This, however, is not the full story, and for that we turn to the effective Higgs thermal mass that governs the $y$ parameter.

The effective Higgs mass in Eq.~\eqref{eq:soft-corrections-mass} has the formal structure
\begin{align}
\label{eq:tilde-mass}
\widetilde{m}^2_{\phi,3} &= \bigg[ m^2_{\phi}(\bar{\mu}) + \Pi_{\phi,1}(\bar{\mu}) - \frac{\bar{a}_{2} \sqrt{\bar{m}^2_{S}}}{8\pi}  \bigg]_{\text{1-loop}} \nonumber \\
& + \bigg[ \frac{1}{(4\pi)^2} \frac{1}{4} \bigg( -\bar{a}^2_{2} - 2 \bar{a}^2_{2} \ln\Big( \frac{\bar{\mu}_3}{2 \sqrt{\bar{m}^2_{S}}} \Big) \nonumber \\
& + 3 \bar{a}_{2} \bar{b}_{4} \bigg) + \Pi_{\phi,2}[L_b(\bar{\mu})] \bigg]_{\text{2-loop}},
\end{align}
where $\Pi_{\phi,1}$ and $\Pi_{\phi,2}$ describe contributions from the non-zero Matsubara modes, at one and two loops, respectively \cite{Niemi:2021qvp}.
Again, contributions from temporal gauge field components are numerically insignificant, and regardless irrelevant for our discussion on singlet effects. 
Similar to previous discussion with the Higgs self-coupling, the zero-temperature correction \cite{Niemi:2021qvp}
\begin{align}
m^2_{\phi} =  m^2_{\phi,\text{tree}} +  m^2_{\phi, \mathcal{O}(g^4)},
\end{align}
is required at two-loop level, and receives large contributions from the singlet (singlet effects reduce the mass parameter). 
Again, the $\bar{\mu}$-dependence cancels between one- and two-loop terms at target accuracy, yet in the case of the mass parameter one needs to note that the running inside the one-loop thermal mass is cancelled by explicit logarithms in the two-loop correction.   

Apart from $m^2_{\phi} < 0$ and negative one-loop contribution from integrating out the singlet -- the third term in first line of Eq.~\eqref{eq:tilde-mass} -- all other contributions come with positive sign. This is perhaps surprising for the term in the second line in Eq.~\eqref{eq:tilde-mass}: the coefficient of the $\bar{a}^2$ term has explicit minus sign, so for this coefficient to be positive, requires that argument of the logarihmic term satisfies $\bar{\mu}_3/\bar{m}_{S} < 2/\sqrt{e}$. This is the case in our benchmark setup since $\bar{\mu}_3 = T$ and in practise $\bar{m}_{S} \gtrsim T$ in vicinity of $T_c$.%
\footnote{
Note that formally our setup requires $m_{S,3} \sim g T \ll \pi T$ as we keep the singlet in the 3D EFT after dimensional reduction, and only subsequently integrate it out. 
Magnitude of $\bar{m}_{S} \gtrsim T$ does not spoil our discussion, provided that $\bar{m}_{S} \ll \pi T$, which we have checked to hold.      
}
Again we observe good convergence, as two-loop contributions are suppressed compared to one-loop. Both at one and two loops, terms originating from integrating out the singlet are larger than contributions from non-zero Matsubara modes.

Thus, we see that two-loop contribution increases $y(T)$, and this leads to 
smaller $T_c$ compared to mere one-loop result. 
As we have seen in Sec.~\ref{sec:results}, this difference can be multiple hundreds of GeV, ie. very significant.%
\footnote{
Observation that the one-loop approximation highly overestimates the critical temperature, has been pointed out previously in \cite{Niemi:2018juv} in analogous setup in the context of the real-triplet extended SM \cite{Patel:2012pi,Niemi:2018asa,Niemi:2020hto,Gould:2023ovu}, in the limit where heavy triplet -- that couples to Higgs with sufficiently large portal interaction -- is integrated out. 
}
    
In summary so far, we have demonstrated that two-loop corrections result in smaller $T_c$ (by increasing thermal mass), but also increase $x_c$ (thermal Higgs self-coupling) which would seemingly indicate weaker transitions. The transition strength, however, is normalized by a positive power of $T_c$, ie. $v_{\text{min}}/T_c$ or $L/T^4_c$, which can be significantly larger at two loops due to drastically smaller critical temperature (see discussion below).
This shows that there is a delicate balance between different contributions affecting the properties of the transition.

Finally, let us connect the previous discussion to the full effective potential used in our analysis in sec.~\ref{sec:results}.
In Fig.~\ref{fig:vByT-sample} we plot minima of the effective potential as function of temperature for two benchmark points with $a_2=4$ and $M_2=600$ GeV present in $(M_2,a_2)$-plane of figs.~\ref{fig:Tc-comparison-st} and \ref{fig:latent-comparison-st}.
In cases of both $\sin\theta = 0.02$ (left panel) and  $\sin\theta = 0.18$ (right panel) one-loop analysis vastly overestimates $T_c$. 

In the case of $\sin \theta = 0.02$, which corresponds to our discussion so far in this appendix, 
the point $a_2=4$ and $M_2=600$ GeV is further away from the metastability region (relative to $\sin \theta = 0.18$ case) and 
one-loop analysis overestimates $v_{\text{min}}/T_c$ compared to two-loop. 
In the case of larger $\sin \theta = 0.18$ the exact \textit{opposite} happens. Similar trends are true for $L/T_c^4$, 
see table~\ref{table:benchmarks}. 
However, in both cases results for $v_{\text{min}}$ and $L$ at $T_c$ are \textit{larger} at one-loop,
while dimensionless ratios $v_{\text{min}}/T_c$ and $L/T_c^4$ are smaller at two loops for $\sin \theta = 0.02$, 
and larger at two loops for $\sin \theta = 0.18$.
\begin{table}[h!]
\centering
\begin{tabular}{|c|c|c|c|c|c|c|}
\hline
\(\sin \theta\) & order & $\frac{T_c}{\text{GeV}}$ & $\frac{v_{\text{min}}}{\text{GeV}}$ & $\frac{L}{{\text{GeV}}^4}$ & $\frac{v_{\text{min}}}{T_c}$ & $\frac{L}{T^4_c}$ \\
\hline
0.02 & 1-loop & 279.8 & 317.4 & \( 1.32 \times 10^9 \) & 1.13 & 0.21 \\
\phantom{0.02} & 2-loop & 144.4 & 37.8 & \( 1.47 \times 10^7 \) & 0.26 & 0.03 \\
\hline
0.18 & 1-loop & 274.6 & 305.6 & \( 1.06 \times 10^9 \) & 0.90 & 0.12 \\
\phantom{0.18} & 2-loop & 101.0 & 234.3 & \( 2.60 \times 10^8 \) & 2.32 & 2.50 \\
\hline
\end{tabular}
\caption{Thermodynamic quantities of interest corresponding to Fig.~\ref{fig:vByT-sample}.}
\label{table:benchmarks}
\end{table}

We have verified that these discussed features happen also more generally for other parameter values. 
Our discussion in this appendix therefore can shed some light to the features that one can observe in figs.~\ref{fig:Tc-comparison-st}, \ref{fig:latent-comparison-st}, \ref{fig:latent-comparison-a2-Mh2} and \ref{fig:FOPT-projected}.
In summary, two-loop effects

\begin{itemize}

\item increase Higgs thermal mass, resulting in significantly smaller $T_c$,

\item increase effective Higgs self-coupling, resulting in smaller $v_{\text{min}}$ and $L$. 

\end{itemize}

We emphasize that these features rely on the EFT picture of this appendix, and hence only apply for a limited part of the full parameter space. For instance, the most strong transitions that we have found occur when also singlet VEV has a significant discontinuity. Discussion herein does not govern those situations, and in such cases two-loop effects that reduce $T_c$ as well as $v_{\text{min}}$ and $L$ can be more complicated. Regardless, we observe in general that two-loop effects reduce $T_c$ and strengthen (weaken) transitions located near (further away from) region where electroweak Higgs minimum at zero temperature becomes metastable.     

\def\xx{0.32}
\begin{figure*}[t]
	\centering
	\begin{subfigure}[b]{\xx \textwidth}
		\centering
                \includegraphics[width=1.0\textwidth]{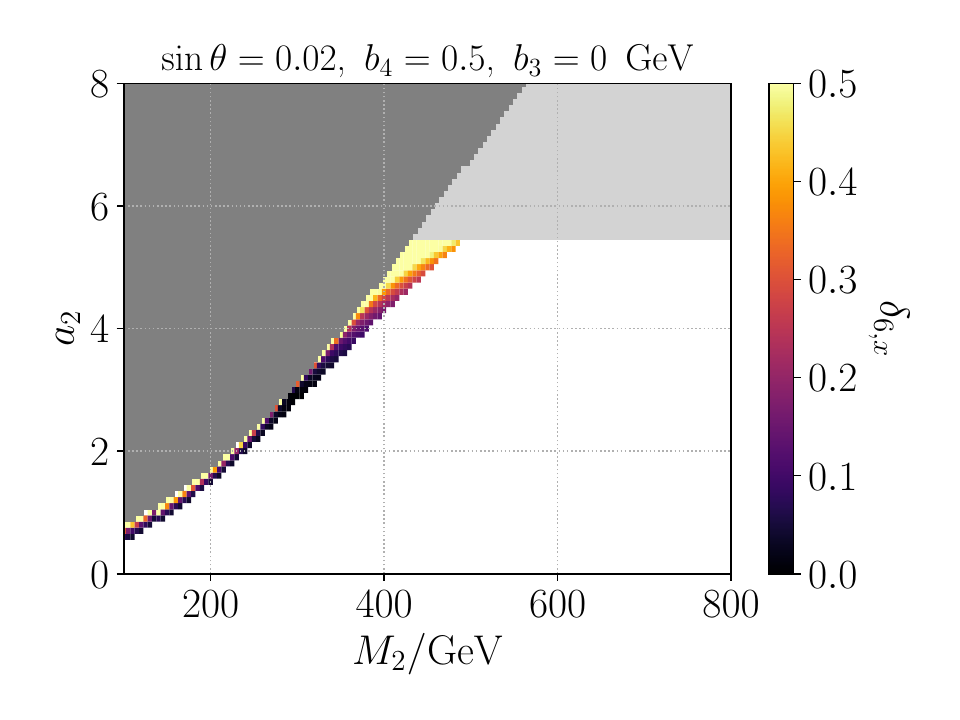}
	\end{subfigure}
	\begin{subfigure}[b]{\xx \textwidth}
		\centering
		\includegraphics[width=1.0\textwidth]{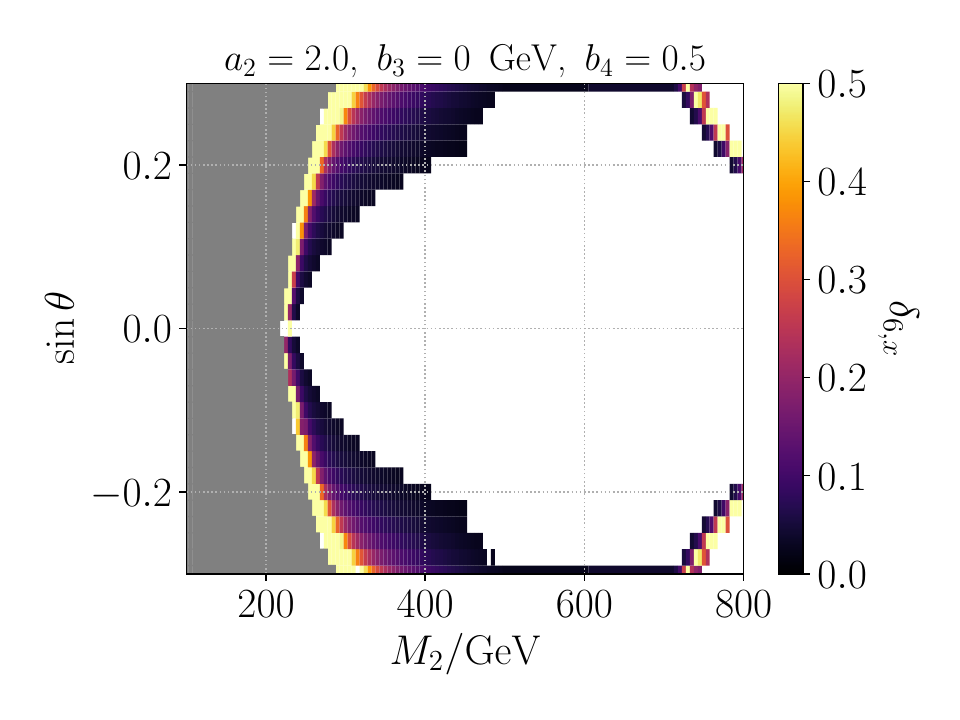}
	\end{subfigure} 
        \begin{subfigure}[b]{\xx \textwidth}
		\centering
		\includegraphics[width=1.0\textwidth]{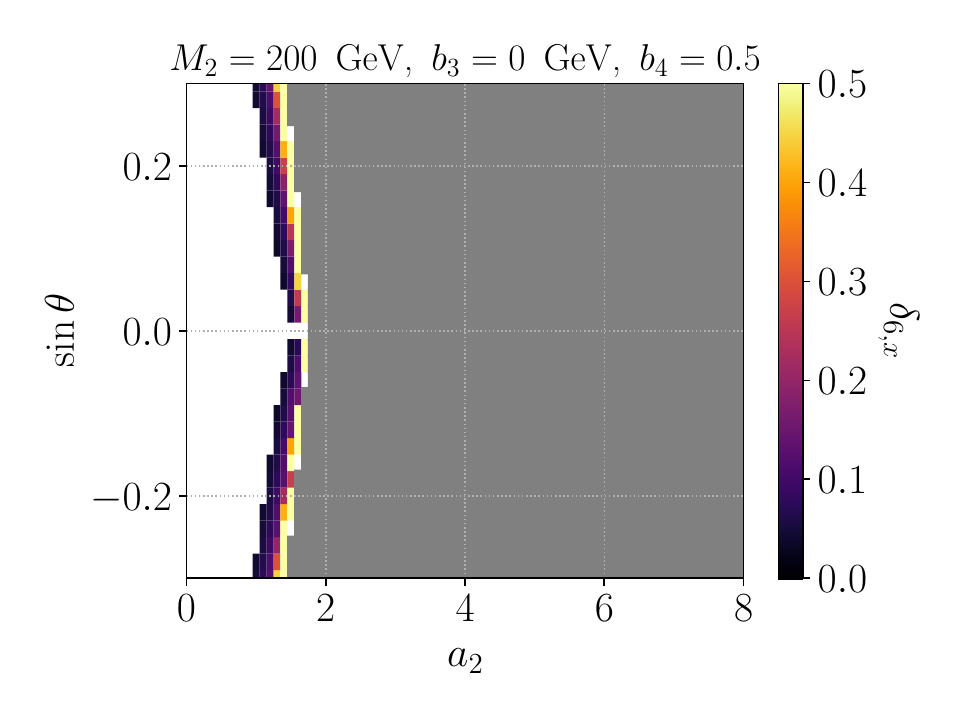}
	\end{subfigure} 
	\begin{subfigure}[b]{\xx \textwidth}
		\centering
		\includegraphics[width=1.0\textwidth]{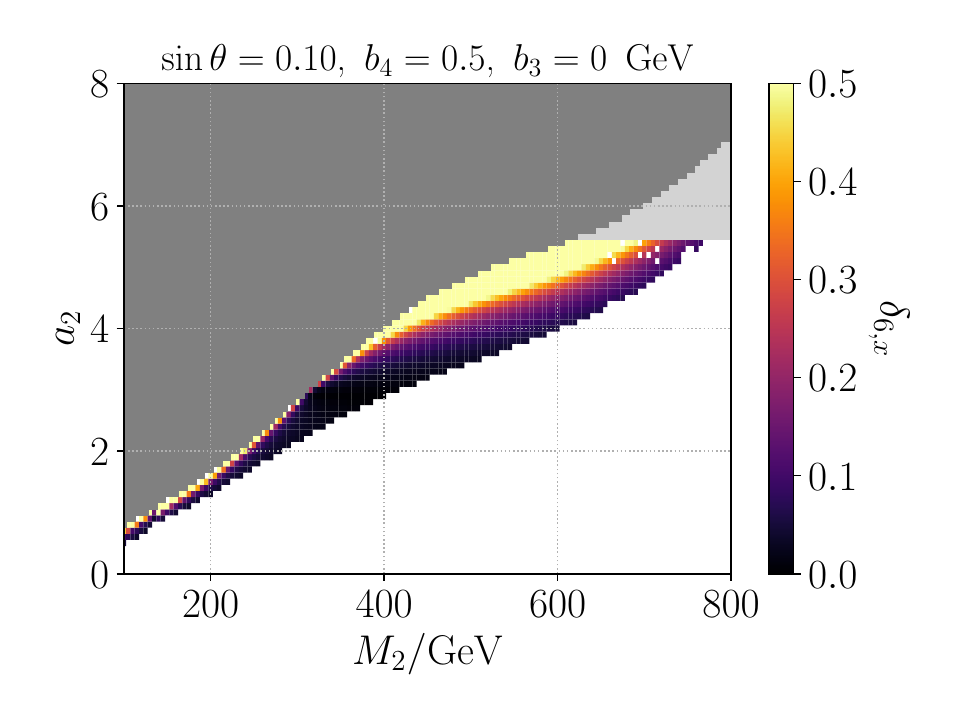}
	\end{subfigure}
	\begin{subfigure}[b]{\xx \textwidth}
		\centering
		\includegraphics[width=1.0\textwidth]{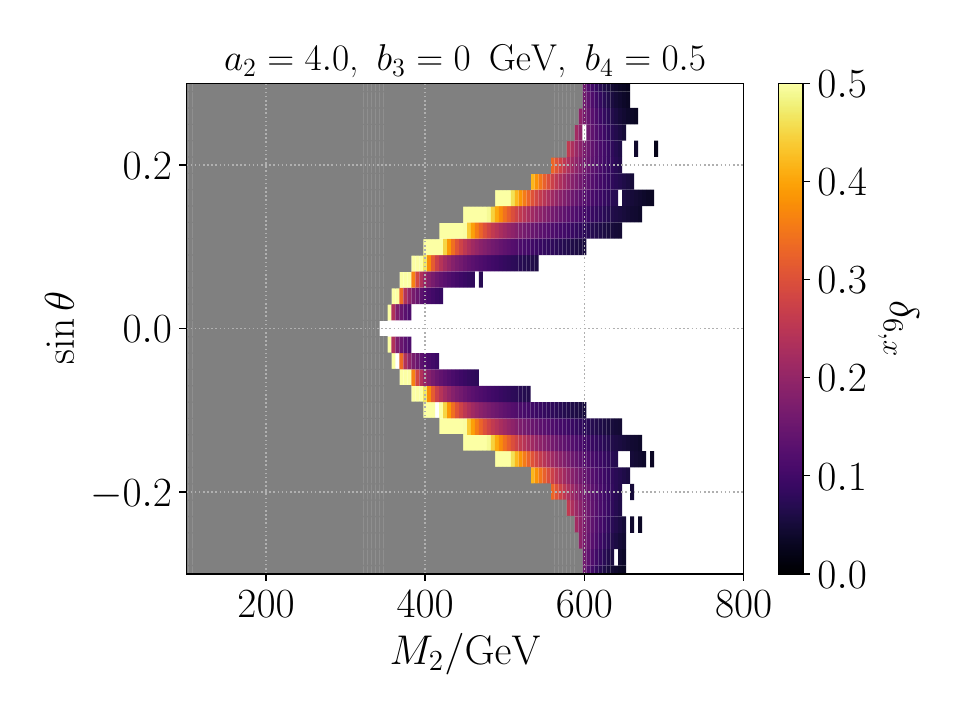}
	\end{subfigure}
        \begin{subfigure}[b]{\xx \textwidth}
		\centering
		\includegraphics[width=1.0\textwidth]{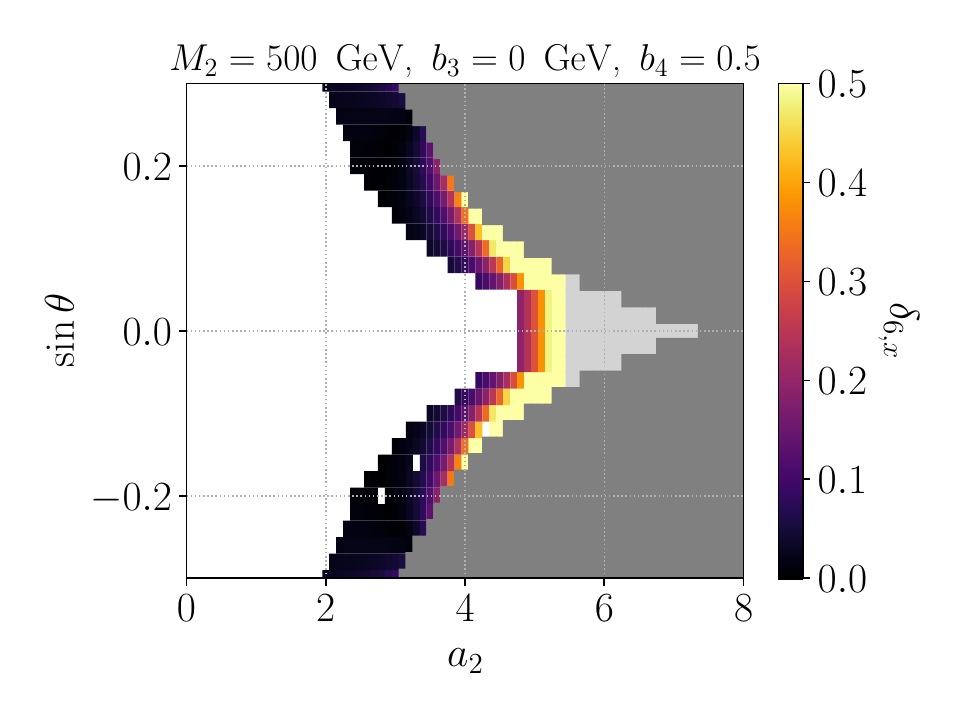}
	\end{subfigure}
	\caption{
As Fig.~\ref{fig:delta6} (right), but showing the error estimator $\delta_{6,x}$ in multiple other parameter planes.   
}
\label{fig:dim6x-error}
\end{figure*}

\section{Additional plots}
\label{sec:more-plots}

\def\xx{0.3}
\begin{figure*}[t]
	\centering
	\begin{subfigure}[b]{\xx \textwidth}
		\centering
		\includegraphics[width=1.0\textwidth]{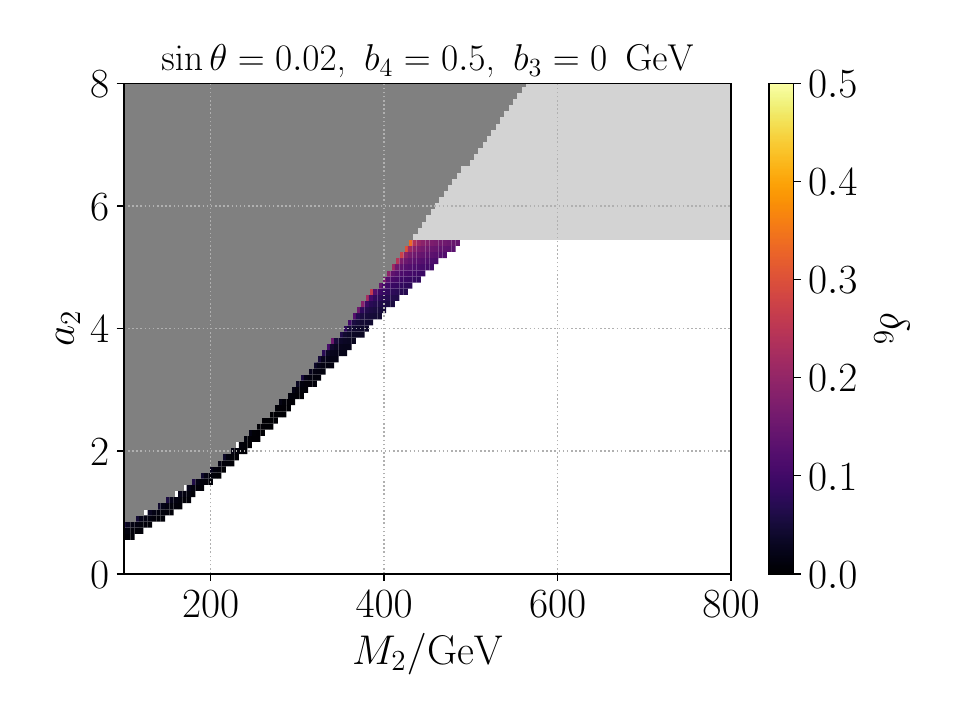}
	\end{subfigure}
	\begin{subfigure}[b]{\xx \textwidth}
		\centering
		\includegraphics[width=1.0\textwidth]{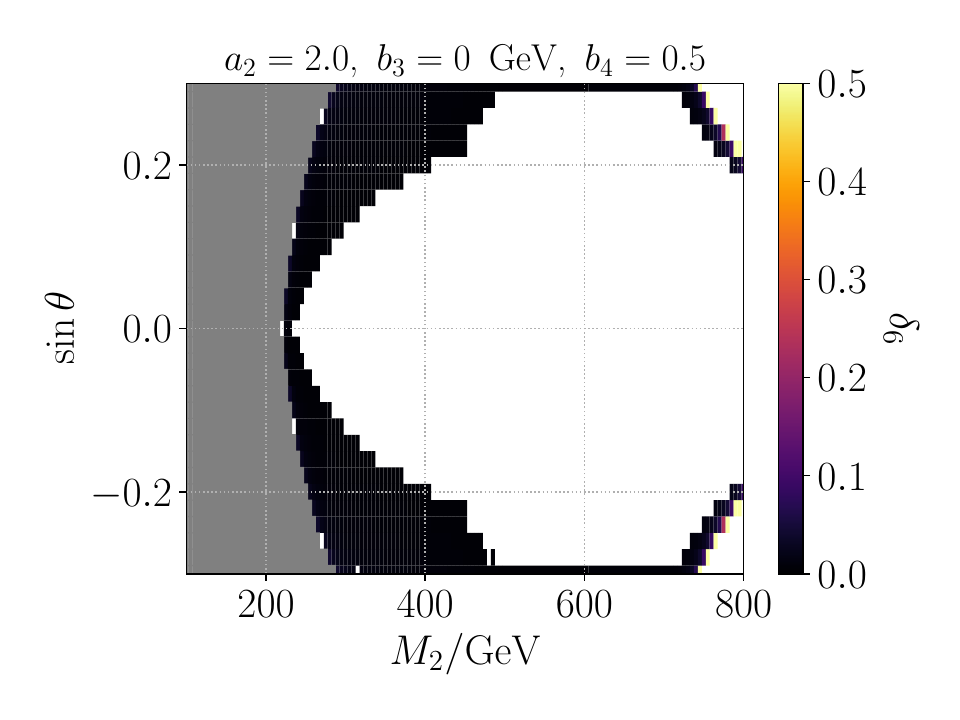}
	\end{subfigure}
	\begin{subfigure}[b]{\xx \textwidth}
		\centering
		\includegraphics[width=1.0\textwidth]{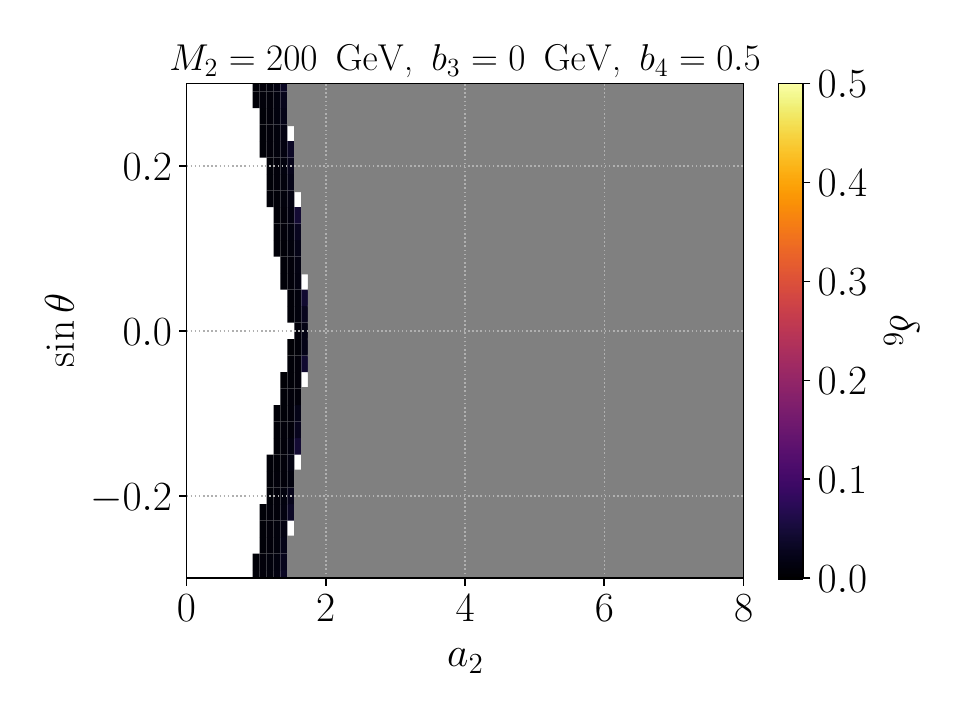}
	\end{subfigure}
	\begin{subfigure}[b]{\xx \textwidth}
		\centering
		\includegraphics[width=1.0\textwidth]{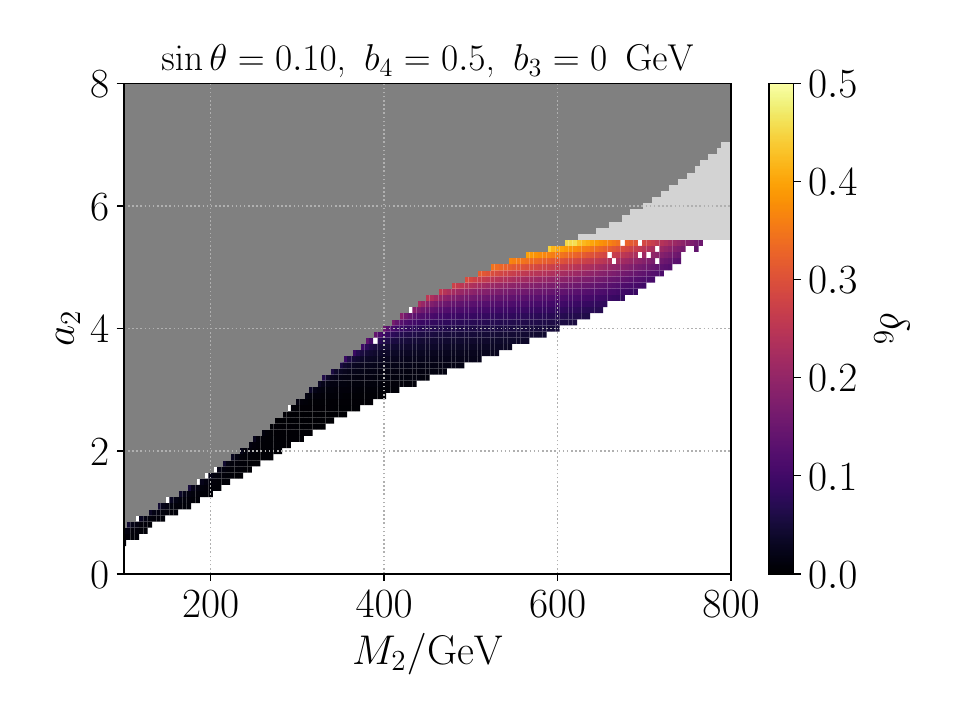}
	\end{subfigure}
	\begin{subfigure}[b]{\xx \textwidth}
		\centering
		\includegraphics[width=1.0\textwidth]{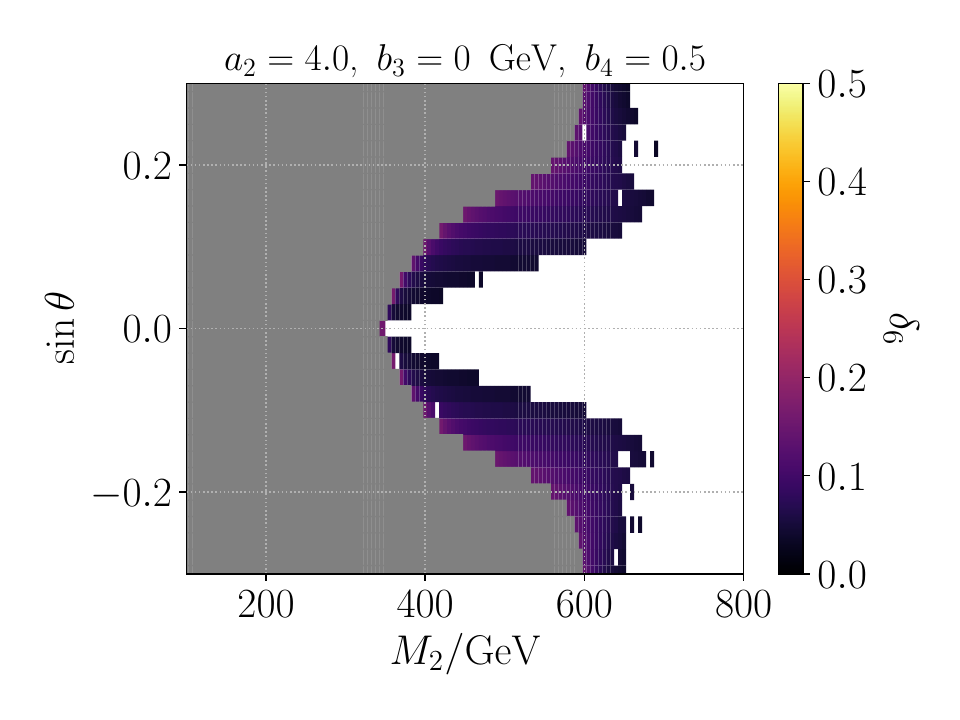}
	\end{subfigure}
	\begin{subfigure}[b]{\xx \textwidth}
		\centering
		\includegraphics[width=1.0\textwidth]{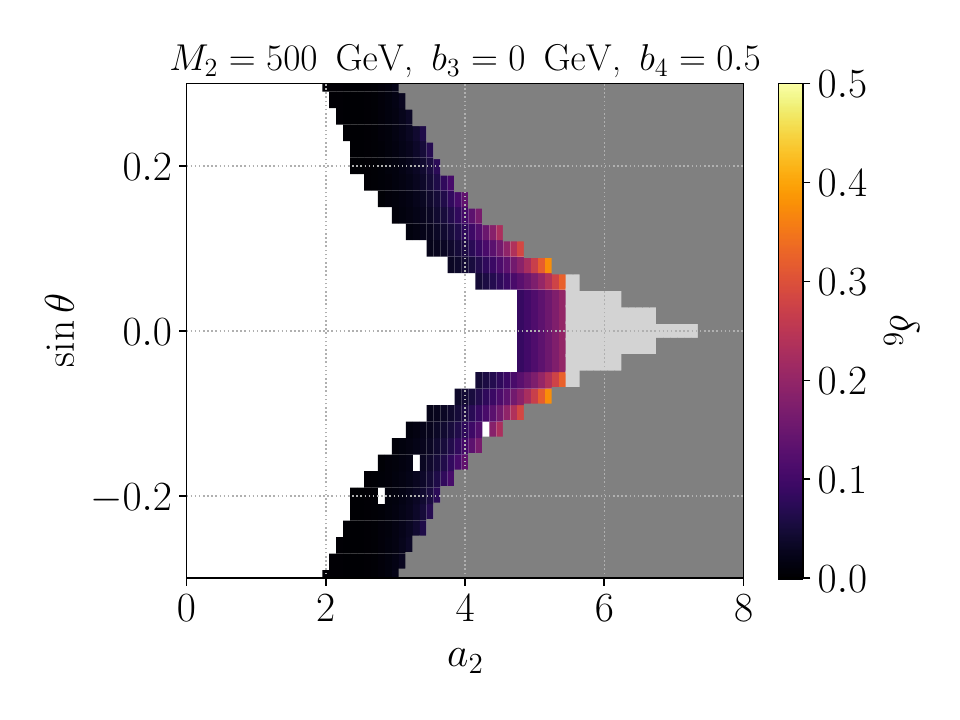}
	\end{subfigure}
	\caption{
As Fig.~\ref{fig:dim6x-error}, but showing the error estimator $\delta_{6}$  in analogy to Fig.~\ref{fig:delta6} (left). 
	}
	\label{fig:dim6-error}
\end{figure*}

In this appendix we present additional plots of the parameter space that didn't fit in the main text.
In analogy to Fig.~\ref{fig:delta6}, in Figs.~\ref{fig:dim6x-error} and \ref{fig:dim6-error}, we plot error estimators $\delta_{6,x}$ and $\delta_6$, respectively, in several other parameter planes.  
Again, we observe that the effect of marginal operators to singlet VEV is relatively much larger compared to the effect on Higgs VEV. 
We speculate this to be due to singlet being heavier than Higgs, in which case the high-$T$ expansion converges slower in general. Furthermore, the Wilson coefficient of the pure-Higgs operator $(\phi^\dagger\phi)^3$ contains a partial cancellation between BSM effects and the relatively large top quark contributions due to bosonic and fermionic loops contributing with opposite signs, see Eq.~(46) in \cite{Niemi:2021qvp}˝. Singlet operators, on the otherhand, are not effected by the top quark. 
Relatively smaller $\delta_6$ in the Higgs direction could indicate that while one could study electroweak barogenesis reliably for such transitions, the uncertainty in the singlet direction could propagate larger errors in the latent heat, and therefore to the overall prediction of transition strength.
However, we have not investigated this further at the present work at hand. 

\def\xx{0.34}
\begin{figure*}[t]
	\centering
	\begin{subfigure}[b]{\xx \textwidth}
		\centering
		\includegraphics[width=1.0\textwidth]{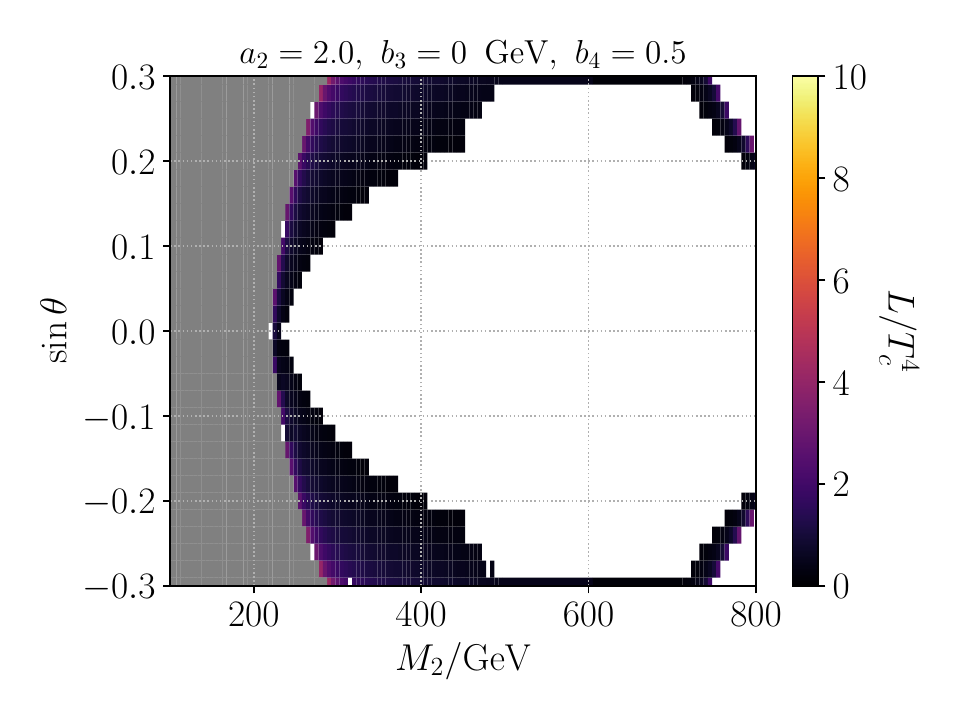}
	\end{subfigure}
	\begin{subfigure}[b]{\xx \textwidth}
		\centering
		\includegraphics[width=1.0\textwidth]{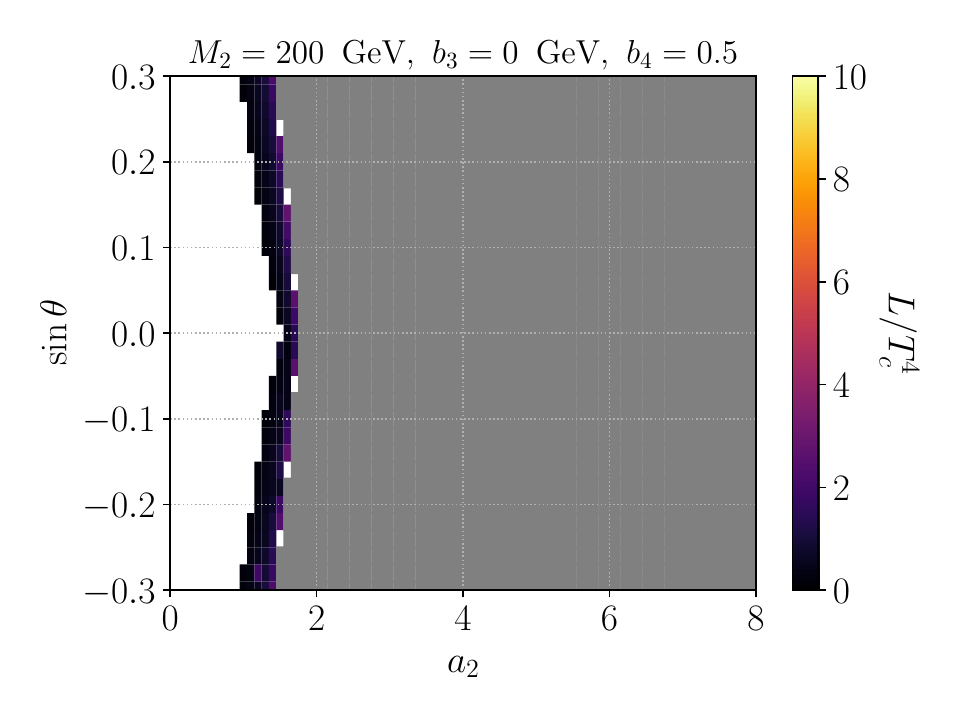}
	\end{subfigure}
	\begin{subfigure}[b]{\xx \textwidth}
		\centering
		\includegraphics[width=1.0\textwidth]{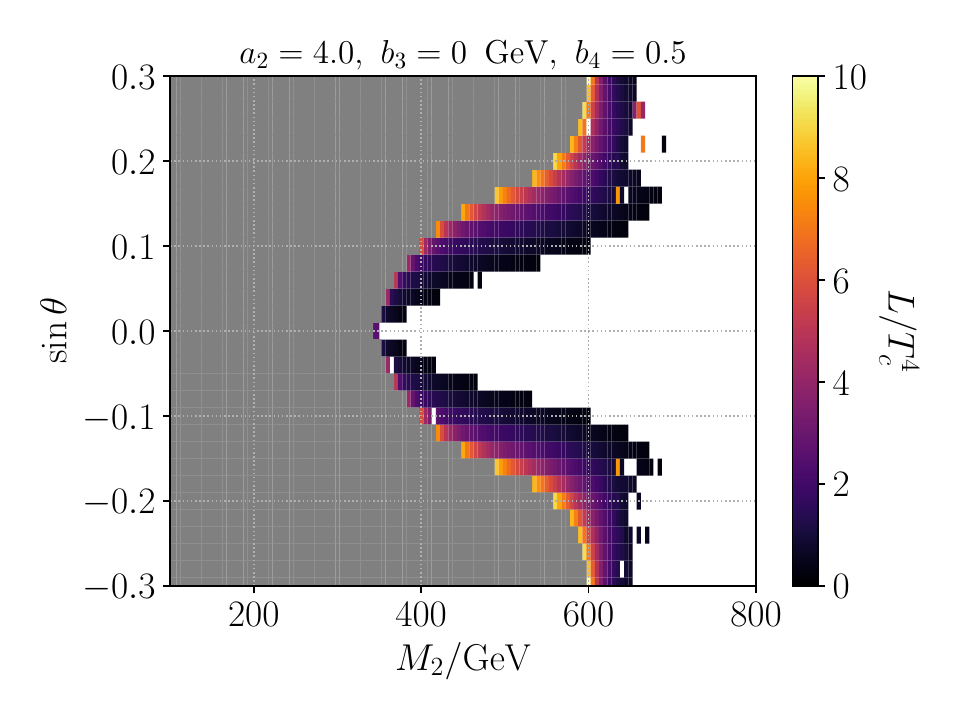}
	\end{subfigure}
	\begin{subfigure}[b]{\xx \textwidth}
		\centering
		\includegraphics[width=1.0\textwidth]{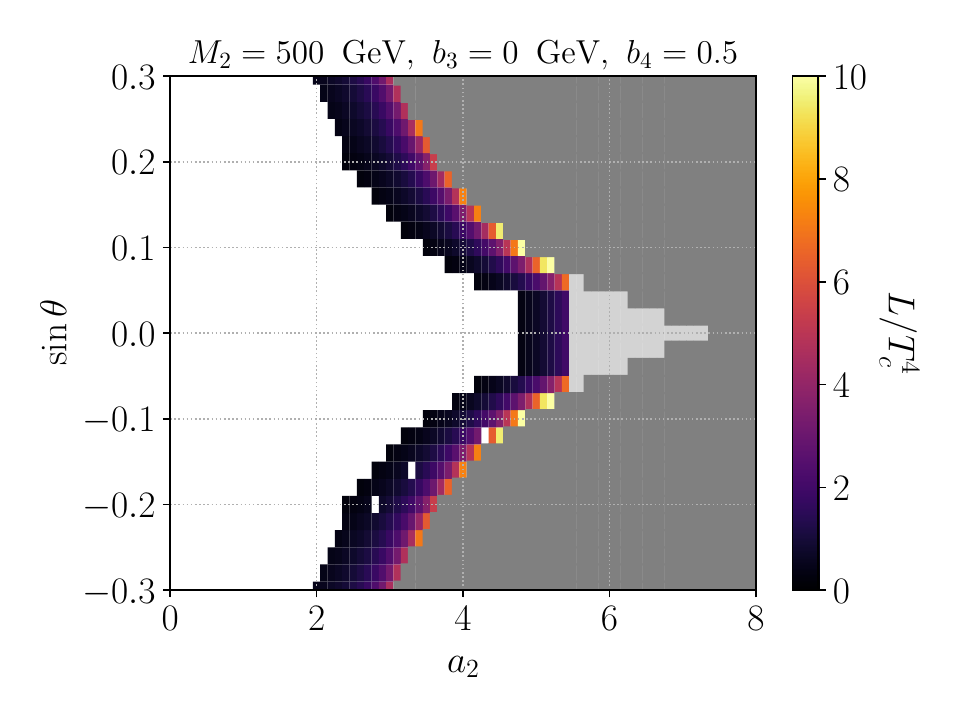}
	\end{subfigure}
	\caption{
As Fig.~\ref{fig:latent-comparison-st} (bottom row), but showing the two-loop result for the latent heat in $(M_2,\sin\theta)$ and $(a_2,\sin\theta)$ planes, for two different values of $a_2$ and $M_2$, respectively.  
	}
	\label{fig:latent-comparison-a2-Mh2}
\end{figure*}

Finally, in Fig.~\ref{fig:latent-comparison-a2-Mh2}, in analogy to Fig.~\ref{fig:latent-comparison-st} (bottom row) we plot two-loop results for the latent heat in other two parameter planes, shown in Figs.~\ref{fig:dim6x-error} and \ref{fig:dim6-error}. For larger $a_2$ and $M_2$, the transition strength increases towards the boundary
of metastable $T = 0$ vacuum (in gray).

\twocolumngrid
\bibliographystyle{apsrev4-1}
\bibliography{singletrefs}

\end{document}